\documentclass[11pt, a4paper]{article}
\usepackage{styleBuding}
\usepackage{amsmath}
\usepackage{bm}
\usepackage{amssymb}
\usepackage{listings}
\usepackage{multirow}
\usepackage{makecell}
\usepackage{array}
\usepackage{subfigure}
\usepackage{siunitx}
\usepackage{booktabs}
\usepackage{amssymb}% http://ctan.org/pkg/amssymb
\usepackage{pifont}% http://ctan.org/pkg/pifont

\usepackage{csquotes}

\newcommand{\rom}[1]{\uppercase\expandafter{\romannumeral #1\relax}}

\title{HL-LHC sensitivity to higgsinos from natural SUSY with gravitino LSP}
\author[a,b]{Jianpeng Dai}
\author[a,b,1]{,~Song Li}
\author[a,c]{,~Jin Min Yang}
\author[d]{,~Yang Zhang}
\author[a,e,1]{,~Pengxuan Zhu%
\note{Corresponding authors}}
\author[a,b,1]{,~Rui Zhu}

\emailAdd{daijianpeng@mail.itp.ac.cn}
\emailAdd{lisong@itp.ac.cn}
\emailAdd{jmyang@itp.ac.cn}
\emailAdd{zhangyangphy@zzu.edu.cn}
\emailAdd{pengxuan.zhu@adelaide.edu.au}
\emailAdd{zhurui@itp.ac.cn}
\affiliation[a]{CAS Key Laboratory of Theoretical Physics, Institute of Theoretical Physics,
                Chinese Academy of Sciences, Beijing 100190, P. R. China}
\affiliation[b]{School of Physical Sciences, University of Chinese Academy of Sciences,
                Beijing 100049, P. R. China}
\affiliation[c]{School of Physics, Henan Normal University, Xinxiang 453007,  P. R. China}
\affiliation[d]{School of Physics, Zhengzhou University, Zhengzhou 450000, P. R. China}
\affiliation[e]{ARC Centre of Excellence for Dark Matter Particle Physics, University of Adelaide, North Terrace, Adelaide SA 5005, Australia}
\abstract{In the realm of natural supersymmetric models, higgsinos are typically the lightest electroweakinos.  In gauge-mediated supersymmetry breaking models, the lightest higgsino-dominated particles decay into a $Z$-boson or a Higgs boson ($h$), along with an ultra-light gravitino ($\tilde{G}$) serving as the lightest supersymmetric particle (LSP). 
This scenario suggests a significant non-resonant $hh$ production. Basing on the recent global fitting results of the $\tilde{G}$-EWMSSM (MSSM with light electroweakinos and an eV-scale gravitino as the LSP) performed by the \textsf{GAMBIT} collaboration, which support a higgsino-dominated electroweakino as light as 140 GeV, we develop two simplified models to evaluate their detection potential at the high-luminosity LHC (HL-LHC) with $\sqrt{s} = 14~{\rm TeV}$ and an integrated luminosity of $3000~{\rm fb}^{-1}$.
The first model examines the processes where heavier higgsino-dominated states decay into soft $W/Z$ bosons, while the second focuses on direct decays of all three higgsino-dominated electroweakinos into $W/Z/h$ plus a $\tilde{G}$. 
Our study, incorporating both models and their distinct decay channels, utilizes detailed Monte Carlo simulations for signals and standard model backgrounds. We find that the HL-LHC can probe higgsinos up to 575 GeV, potentially discovering or excluding the natural SUSY scenario in the context of a gravitino LSP.
Further, we reinterpret this discovery potential using the GAMBIT global fit samples, and find that the entire parameter space of $|\mu| \leq 500~{\rm  GeV}$ with an electroweak fine-tuning measure ($\Delta_{\rm EW}$) under 100 in $\tilde{G}$-EWMSSM is accessible at the HL-LHC.
}
\arxivnumber{2311.10999}
\begin{document}
\maketitle
\section{Introduction\label{sec:intro}}
Supersymmetry (SUSY) is a unique non-trivial extension of relativistic invariance and provides a unified description for fermions and bosons. It is also an elegant framework for new physics (NP) beyond Standard Model (SM) in  particle physics~\cite{Fayet:1976cr, Witten:1981nf}. Specifically, the supersymmetric Higgs mass parameter $\mu$ should be at the electroweak scale to ensure that the Higgs fields $H_u$ and $H_d$ get the observed non-zero vacuum expectation value (vev) $v \approx 246~{\rm GeV}$ after electroweak symmetry breaking~\cite{Giudice:1988yz}. Since this $\mu$ parameter is the mass of  higgsinos, it can be measured at the Large Hadron Collider (LHC) and dark matter (DM) detection experiments~\cite{Baer:2012uy, Cao:2012fz, Kitano:2006gv, Giudice:2006sn, Allanach:2007qk}. Theoretically, if $\mu$ is much greater than electroweak scale, some other parameters need to be fine-tuned and the tuning extent is reasonably quantified by a dimensionless measure~\cite{Baer:2012uy, Hall:2011aa, Ellwanger:2011mu, Cassel:2009ps, Cassel:2010px}
\begin{equation}\label{eq:intro-natrual}
	\Delta^{\rm EW} \equiv \max\left| \frac{\partial \log m_{Z}^2}{\partial \log p_i} \right|, 
\end{equation}
where $m_Z$ is the $Z$-boson mass and $p_i$ are the SUSY parameters at electroweak scale. 

\par Of our special interest is the SUSY model with light electroweakinos and not so heavy top squarks (stops), which is called natural SUSY~\cite{Papucci:2011wy, Baer:2012uy, Hall:2011aa, Ellwanger:2011mu, Tata:2020afe, Cao:2018rix, Wang:2023suf, Yang:2022qyz, Zhao:2022pnv, Cao:2022htd, Domingo:2022pde, Chatterjee:2022pxf, Abdallah:2020yag, Wang:2019biy, Carpenter:2023agq, Baer:2023cvi, Datta:2022bvg, Cao:2019ofo, Baer:2022qqr, Baer:2022dfc, Agashe:2012zq, Kribs:2013lua, Casas:2014eca, Baer:2022wxe, Baer:2021tta, BhupalDev:2021ipu, Wu:2019hso}.
Natural SUSY is well motivated both theoretically and experimentally.
Firstly, natural SUSY is supported by the Higgs boson properties, which offer compelling perspectives for testing the effects of SUSY. Ref.~\cite{Arbey:2021jdh} reviewed the SUSY corrections to the couplings and decays of the SM-like Higgs boson and their dependence on the 19-dimensional parameter space in the phenomenological minimal supersymmetric standard model (MSSM), which showed that the SM-like Higgs state discovered at the LHC can still be described as a SUSY-like Higgs state. 
Secondly, the electroweakinos and the stops can still be light~\cite{ATL-PHYS-PUB-2023-005}. The searches at the current LHC experiment have pushed the detection limits on the colored sparticles to about 2 TeV, mostly independent on their decay modes and the mass differences with other sparticles. The bounds on third generation squark masses are somewhat weaker, with the detection limit on the stop mass is around 1.2 TeV. About the electroweakinos, the current LHC limits are rather weak. The searching limits on wino-like states are about 1 TeV for a massless lightest supersymmetric particle (LSP), while for higgsino-like states the limits are only about 550 GeV. However, the current stringent LHC constraints on the SUSY particles strongly depend on the simplified models used in the result interpretation~\cite{LHCNewPhysicsWorkingGroup:2011mji}.  A detailed survey~\cite{GAMBIT:2018gjo} of the combined impacts of DM and collider constraints on the electroweakino sector in the MSSM (EWMSSM), mostly using the $36.1~{\rm fb}^{-1}$ data,  shows that the $\mu$ parameter around 200 GeV is still allowed\footnote{It should be noted that this result is somewhat influenced by the local excess of $3\ell + E_{\rm T}^{\rm miss}$ signal of the ATLAS analysis~\cite{ATLAS:2018eui}, but this excess disappeared in the ATLAS updated report~\cite{ATLAS:2019lng}. }, mainly due to the compressed mass spectrum.  
In the framework of natural SUSY, the LHC bounds on the higgsinos are even much weaker since they are quite compressed with the higgsino-like LSP and their productions merely give missing energy \cite{Buanes:2022wgm,Han:2013usa,Guchait:2021tmh}. 

\par Although the current LHC bounds on the higgsinos (especially in nautural SUSY) are rather weak,  the parameter space predicting the observed DM relic density with a $\mu$ value at electroweak scale is highly constrained by the current DM direct detection data, as investigated in the literature~\cite{Ding:2014bqa, Kang:2017rfw, Baer:2017cck, Giudice:2017pzm, Cao:2019evo, Baer:2020kwz, Cao:2021lmj, Agashe:2022uih, Lv:2022pme, Wang:2022rfd, Carpenter:2021jbd, Ahmed:2020lua}. Roughly speaking, it is rather difficult for the neutralino DM in SUSY theories to naturally suppress the DM-nucleon scattering rate to meet the current detection limits~\cite{Cao:2019qng}. On the other hand, the recent experimental hints from the $W$-boson mass anomaly~\cite{CDF:2022hxs} and the muon $g-2$ anomaly~\cite{Muong-2:2021ojo, Muong-2:2023cdq} may indicate new particles at electroweak scale~\cite{Athron:2021iuf, Arkani-Hamed:2021xlp}. In SUSY theories these anomalies hint at sub-TeV electroweakinos (see, e.g., \cite{Arkani-Hamed:2021xlp, Yang:2022gvz, Domingo:2022pde, Wang:2021bcx, Tang:2022pxh, Li:2022eby, Bagnaschi:2022qhb}). The studies in~\cite{Chakraborti:2023pis, Chakraborti:2022wii, Heinemeyer:2022anz, Heinemeyer:2022ith, Dickinson:2022hus, Bagnaschi:2022qhb, Chakraborti:2022sbj, Chakraborti:2021squ, Chakraborti:2021dli, Chakraborti:2021kkr, Chakraborti:2020vjp} show that in order to explain muon $g-2$ anomaly, the bino DM, co-annihilating with winos (sleptons) to give the observed DM relic density, requires $m_{\rm (N)LSP} \lesssim 650 (700)~{\rm GeV}$. A detailed study shows a viable MSSM parameter space with bino DM mass $200~{\rm GeV} \lesssim m_{\rm LSP} \lesssim 600~{\rm GeV}$ and $\mu \gtrsim 400~{\rm GeV}$ in light of the DM experiments, the muon g-2 anomaly and the LHC data~\cite{He:2023lgi}. The constraints on the parameter space can be relaxed in the extensions of the MSSM. For example, in the next-to-minimal supersymmetric standard model (NMSSM)~\cite{Ellwanger:2009dp, Cao:2012fz}, the fermion part of the singlet Higgs superfield $\tilde{S}$ can naturally suppress the DM-nucleon scattering rate due to its singlet nature, while it can annihilate through various mechanisms in $\mathbb{Z}_3$-NMSSM~\cite{Ellwanger:2016sur, Ellwanger:2018zxt, Almarashi:2022iol, Cao:2021ljw, Datta:2022bvg, Guchait:2020wqn} and in the general NMSSM~\cite{Cao:2022chy, Cao:2021tuh}. Numerically, the $\mu$ parameter should be greater (less) than about $500~{\rm GeV}$ in $\mathbb{Z}_3$-NMSSM~\cite{Cao:2022htd} (general NMSSM~\cite{Cao:2022chy}). 

\par In the theoretical view, the $\mu$-parameter is indeed predicted at electroweak scale in some SUSY modles like supergravity (SUGRA). Since in SUGRA the gravitino mass $m_{3/2}$ is also expected to be above $100~{\rm GeV}$ (the gravitino is expected to acquire mass through the super-Higgs mechanism) 
\begin{equation}\label{eq:intro-mgra}
	m_{3/2} = \frac{\langle F \rangle }{\sqrt{3} M_{\rm P}},
\end{equation}
with $\langle F \rangle$ being the SUSY breaking scale and 
$M_{\rm P} = (8\pi G_{\rm N})^{-1} = 2.4 \times 10^{18}~{\rm GeV} $ being the reduced Planck mass, the higgsinos may be lighter than the gravitino and the LSP may be higgsino-like. 
Because in SUGRA the gravitino couples to every sparticle with a universal coupling in form and its interaction is of gravitational strength, the gravitino has no collider phenomenology. In contrast, in    
gauge-mediated SUSY breaking (GMSB) models, the gravitino is expected to be much lighter than other sparticles and hence serve as the LSP.  
In the presence of an approximately massless gravitino LSP, 
the \textsf{GAMBIT} collaboration reinterpreted the collider data in the EWMSSM (the MSSM with light electroweakinos),  referred as the $\tilde{G}$-EWMSSM~\cite{GAMBIT:2023yih}. In such a scenario, it is found that the $\mu$-parameter can still be about $100~{\rm GeV}$, with a value of $170~{\rm GeV}$ for the best point, indicating a small fine-tuning measure $\Delta^{\rm EW}$ \footnote{The GAMBIT collaboration fixes the Higgs mass at 125 GeV and requires the stop masses to be heavier than 3 TeV, which is different from a UV-complete SUSY model. In a UV-complete SUSY model, for fixed values of $\mu$ and $\tan{\beta}$, the soft-SUSY-breaking parameters need to be tuned to ensure the correct Higgs mass. This is crucial to the fine-tuning measure. Especially in scenarios with a low $\tan{\beta} \lesssim 5$, which typically requires stop masses well above the TeV scale to give the correct Higgs boson mass, the fine-tuning measure $\Delta^{\rm EW}$ contributed from the stop soft-breaking parameters may be quite sizable. This interplay highlights the necessity of carefully considering the fixed parameters and their effects on the overall model predictions and the fine-tuning assessments~\cite{Slavich:2020zjv}.}. 

\par In the $\tilde{G}$-EWMSSM,  the electroweakinos are produced in pair at the LHC and subsequently decay into final state of an electroweak boson $Z/W/h/\gamma$ plus a gravitino $\tilde{G}$ or plus a lighter neutralino ($\tilde{\chi}_1^0$ mostly). For these decay modes, the most difficult channel to detect is the final state of di-Higgs plus missing energy. The ATLAS report~\cite{ATLAS:2018tti, ATLAS:2021yyr, ATLAS:2021yqv, ATLAS:2022zwa, ATLAS-CONF-2023-009} shows that there exist a gap with higgsino mass $m_{\tilde{\chi}_{1,2}^0} = m_{\tilde{\chi}_1^{\pm}}$ around the top quark mass $m_{t}$, which is unreachable at LHC Run-II (see Fig.~21 in \cite{ATL-PHYS-PUB-2023-005}). 
Although the di-Higgs channel is the most difficult to detect, it is one of the most important processes to be studied at the LHC since it allows for a measurement of the Higgs trilinear self-coupling $\lambda_{hhh}$ and for the search of some NP scenario~\cite{Han:2015pwa, Etesami:2015caa,PhysRevD.70.115002}.  
For such a di-Higgs production, the decay channels are $bbWW$~\cite{ATLAS:2019vwv, ATLAS:2018fpd}, four $b$-jets~\cite{ATLAS:2023qzf, ATLAS:2020jgy}, $bb\gamma\gamma$~\cite{ATLAS:2023gzn, ATLAS:2021ifb}, $bb\tau\tau$~\cite{ATLAS:2022xzm} and four $W$-bosons~\cite{ATLAS:2018ili}. Each channel has its pros and cons. The four $b$-jets channel and the $bb\gamma\gamma$ channel are more frequently used than other channels. The former suffers from the challenge of separating it from other processes produced in the LHC collisions. Although photons are easy to pick out amongst other processes, the $bb\gamma\gamma$ channel is suppressed by its very small branching ratio. According to the current ATLAS and CMS results, the $bb\tau\tau$ is a rather promising channel, which has a higher production rate than the $bb\gamma\gamma$ channel and lower background than four $b$-jets channel. As for the benchmark NP model like $\tilde{G}$-EWMSSM, the $bb\tau\tau$ channel can also be used to probe the process such as $pp \to \tilde{\chi}\tilde{\chi} \to hh\tilde{G}\tilde{G}$, which is expected to fill the unreachable gap in the Fig.~6 of Ref.~\cite{ATLAS-CONF-2023-009}.

\par In the spirit of leaving no stone unturned, in this work we focus on the $\tilde{G}$-EWMSSM scenario and study the HL-LHC observability. 
The content is organized as follows. In Sec.~\ref{sec:2}, we briefly describe the
$\tilde{G}$-EWMSSM scenario, i.e., the MSSM electroweakino sector with a light gravitino.  In Sec.~\ref{sec:3}, we study the HL-LHC sensitivity via scrutinizing the signature of higgsino pairs in five complementary decay channels, including the final states of $ZZ$, $Zh$, $hh$, $WZ$ and $Wh$ plus missing energy $E_{\rm T}^{\rm miss}$. The prospects of HL-LHC searching limits for the simplified models are summarized in Sec.~\ref{sec:4}. Then we reinterpret the result in the framework of $\tilde{G}$-EWMSSM with the help of the global-fit data released by \textsf{GAMBIT} collaboration, and discuss the impact on the fine-tuning measure $\Delta^{\rm EW}$. In Sec.~\ref{sec:5}, we make a summary. 

\section{\label{sec:2}Theoretical preliminaries}
The gravitino, with a mass generated in Eq.~(\ref{eq:intro-mgra}), couples universally to all matter and all force fields. Its relevant interactions can be described by the goldstino field $\tilde{G}$ ~\cite{Fayet:1977vd, Fayet:1979yb, Fayet:1986zc} 
\begin{equation}\label{eq:gtino-int}
\mathcal{L}_{\rm int} = \frac{1}{\langle F \rangle} \left[ \left(m_{\psi}^2 - m_{\phi}^2   \right) \bar{\psi}_L \phi + \frac{m_{\lambda}}{4\sqrt{2}} \bar{\lambda}^{a}\sigma^{\nu \rho} F_{\nu \rho}^{a}  \right] \tilde{G} + {\rm h.c.}, 
\end{equation}
where $(\psi, \phi)$ is the scalar and fermionic component of the chiral supermultiplets, $\lambda^a$ is the gaugino associated with the gauge field with field strength $F_{\mu\nu}^a$. 
\par In the MSSM the electroweak gauginos (including bino $\tilde{B}$ and winos $\tilde{W}$) mix with the higgsinos $(\tilde{H}_u^0, \tilde{H}_u^+, \tilde{H}_d^-, \tilde{H}_d^0)$. The neutral states form four neutralinos $\tilde{\chi}_i^0$, and the charged states form two pairs of charginos $\tilde{\chi}_i^{\pm}$. In the basis of
\begin{equation}
	\psi^0 = \left(\tilde{B}, \tilde{W}^0, \tilde{H}_d^0, \tilde{H}_u^0 \right), \quad
	\psi^{\pm} = \left( \tilde{W}^+, \tilde{H}_u^+, \tilde{W}^-, \tilde{H}_d^- \right),  
\end{equation}
the neutralino and chargino mass terms are given by 
\begin{equation}
	\mathcal{L}_{\rm EWino} = -\frac{1}{2} \left(\psi^0 \right)^{T} M_{N} \psi^0 - \frac{1}{2} \left( \psi^\pm \right)^{T} M_{C} \psi^\pm + {\rm c.c.}
\end{equation}
where the symmetric neutralino mass matrix ${M}_{N}$ takes a form 
\begin{equation}
	M_{N} = \begin{pmatrix}
		M_1 & 0 & -\frac{1}{2} g^\prime v_{d} & \frac{1}{2} g^\prime v_{u} \\
			& M_2 &  \frac{1}{2} g v_{d}	&  -\frac{1}{2} g v_{u} \\
			&	& 0 & -\mu \\
			&	&	& 0 	
	\end{pmatrix},  
\end{equation}
with $M_1$ and $M_2$ being the soft breaking masses of bino and wino fields respectively, $g^\prime$ and $g$ being $SU(2)_{L}$ and $U(1)_Y$ gauge couplings, and $v_u$ and $v_d$ being the vevs of Higgs fields $H_u^0$ and $H_d^0$ after electroweak symmetry breaking, satisfying $v = \sqrt{v_u^2 + v_d^2}$. Similarly, the chargino mass matrix $M_{C}$ is given by 
\begin{equation}
 	M_{C} = \begin{pmatrix}
 		0 & X^{T} \\
 		X & 0 
 	\end{pmatrix}, \quad 
  \quad 
 	X = \begin{pmatrix}
 		M_2 & \frac{1}{\sqrt{2}} g v_u \\
 		\frac{1}{\sqrt{2}} g v_d & \mu
 	\end{pmatrix}. 
\end{equation}

\subsection{The light higgsino scenario with a light gravitino LSP}
For the naturalness of SUSY, the higgsino mass parameter $\mu$ can not be too heavy, which can be quantified by $\Delta^{\rm EW}$ defined in Eq.~(\ref{eq:intro-natrual}). Specifically,  at the one-loop level the $Z$-boson mass is given by~\cite{Arnowitt:1992qp}
\begin{equation}
	\frac{m_{Z}^2}{2} = - \mu^2 + \frac{\left(m_{H_{d}}^2 + \Sigma_d \right) - \left( m_{H_u}^2 + \Sigma_u \right)\tan^2{\beta}}{\tan^2{\beta} -1 }, 
\end{equation} 
where $m_{H_u}^2$ and $m_{H_d}^2$ are the soft SUSY breaking masses of the Higgs fields $H_u$ and $H_d$, $\Sigma_{u}$ and $\Sigma_{d}$ represent the radiative corrections. Therefore, from Eq~(\ref{eq:intro-natrual}), one can get  
\begin{equation}
	\Delta^{\rm EW} \gtrsim \Delta_{\mu}^{\rm EW} \equiv 2 {\mu^2}/{m_{Z}^2}. 
\end{equation}
The fine-tuning $\Delta_{\mu}^{\rm EW}$ of less than one percent level indicates $\mu$ in the range of $100-500~{\rm GeV}$. Here it should be noted that the fine-tuning measure can be changed by the SUSY breaking boundary condition and the radiative correction. For example, in the constrained MSSM with non-universal Higgs and gaugino masses, $\Delta_{\mu}^{\rm EW}$ can be reduced by a factor 2~\cite{Ross:2017kjc}. Moreover, the parameter space with the square root of the stop masses $\sqrt{m_{\tilde{t}_1} m_{\tilde{t}_2}}$ greater than $1.2~{\rm TeV}$ and/or the gluino mass heavier than $1.8~{\rm TeV}$ can also contribute a $5\%$ fine-tuning ($\Delta^{\rm EW} \gtrsim 20$)~\cite{Papucci:2011wy}. In this work, we focus on the light higgsinos in the simplified model defined in Sec.~2.3, and neglect the colored sparticles~\cite{Kang:2012sy} by fixing their masses at 3 TeV. 

\par In the light higgsino scenario, the lightest three electroweakinos ($\tilde{\chi}_{1,2}^0$ and $\tilde{\chi}_1^\pm$) are higgsino like, which are nearly degenerated, with masses around the value of $\mu$. In the following, the MSSM-like heavy Higgs states $H, A, H^\pm$ are also assumed much heavy, and the SM-like $h$ acts as the $125~{\rm GeV}$ Higgs state.

\par In addition to the decay modes in the MSSM, the interaction in Eq.~(\ref{eq:gtino-int}) leads to all electroweakinos decay into the final states containing a gravitino.  In the mass eigenstates, the decay widths of the neutralinos into neutral gauge bosons are~\cite{Ambrosanio:1996jn, Feng:2004mt, Covi:2009bk}
\begin{equation}\label{eq:dkN2gagra}
	\Gamma\left(\tilde{\chi}_i^0 \to \gamma \tilde{G}\right) = \frac{\kappa_{i\gamma}}{48\pi M_{\rm P}^2} \frac{m_{\tilde{\chi}_i^0}^5}{ m_{3/2}^2} ,
\end{equation}
\begin{equation}\label{eq:dkN2Hgra}
	\Gamma\left(\tilde{\chi}_i^0 \to h \tilde{G}\right) = \frac{ \kappa_{ih}}{96\pi M_{\rm P}^2} \frac{m_{\tilde{\chi}_i^0}^5}{ m_{3/2}^2} \left(1- \frac{m_h^2}{m_{\tilde{\chi}_i^0}^2} \right)^4,
\end{equation}
\begin{equation}\label{eq:dkN2Zgra}
	\Gamma\left(\tilde{\chi}_i^0 \to Z \tilde{G}\right) = \frac{2 \kappa_{iZ}^{T} + \kappa_{iZ}^{L}}{96\pi M_{\rm P}^2} \frac{m_{\tilde{\chi}_i^0}^5}{ m_{3/2}^2} \left(1- \frac{m_Z^2}{m_{\tilde{\chi}_i^0}^2} \right)^4,
\end{equation}
where 
\begin{equation}\begin{split}
	\kappa_{i\gamma} &= \big| N_{i1} \cos{\theta_{W}} + N_{i2} \sin{\theta_{W}} \big|^2 , \\
	\kappa_{ih} &= \big| N_{i3} \sin{\alpha} - N_{i4} \cos{\alpha} \big|^2,\\
	\kappa_{iZ}^{T} &= \big| N_{i1} \sin{\theta_{W}} - N_{i2} \cos{\theta_{W}} \big|^2 ,\\
	\kappa_{iZ}^{L} &= \big| N_{i3} \cos{\beta} - N_{i4} \sin{\beta} \big|^2 ,
\end{split}
\end{equation}
depict the components of bino, wino and higgsinos in the neutralino $\tilde{\chi}_i^0$. Here $N_{ij}$ is the neutralino mixing matrix element with $(i, j)$ being the (mass, gauge) eigenstate labels, $\theta_{W}$ is the Weinberg angle, and $\alpha$ is the mixing angle between the $CP$-even neutral Higgs states. Similarly, the decay width of charginos into gravitino are given by
\begin{equation}\label{eq:dkC2Wgra}
	\Gamma\left(\tilde{\chi}_i^\pm \to W^\pm \tilde{G} \right) = \frac{2 \kappa_{iW}^{T} + \kappa_{iW}^{L}}{96 \pi M_{\rm P}^2 } \frac{m_{\tilde{\chi}_i^{\pm}}^5}{m_{3/2}^2} \left( 1 - \frac{m_W^2}{m_{\tilde{\chi}_i^\pm}^2} \right)^4 , 
\end{equation}
where
\begin{equation}
	\begin{split}
		\kappa_{iW}^{T} &=\frac{1}{2} \left( \big| V_{i1} \big|^2 + \big| U_{i1} \big|^2 \right), \\
		\kappa_{iW}^{L} &=\big| V_{i2} \big|^2 \sin^2{\beta} + \big| U_{i2} \big|^2 \cos^2{\beta},
	\end{split}
\end{equation}
with $U$ and $V$ being the chargino rotation matrices defined by $UXV = m_{\tilde{\chi}^\pm}^{\rm diag}$. 

\par In terms of a typical decay width of electroweakino into gravitino
\begin{equation}\label{eq:dk-typical}
	\Gamma_{0} = \frac{1}{48\pi M_{\rm P}^2} \frac{(100~{\rm GeV})^5}{(1~{\rm eV})^2} \approx 1.12 \times 10^{-11}~{\rm GeV},
\end{equation}
then the decay widths in Eqs.~(\ref{eq:dkN2gagra}), (\ref{eq:dkN2Hgra}), (\ref{eq:dkN2Zgra}) and (\ref{eq:dkC2Wgra}) can be written as 
\begin{small}
\begin{equation}\begin{split}
		\Gamma\left(\tilde{\chi}_i^0 \to \gamma \tilde{G}\right) &= \kappa_{i\gamma} \left( \frac{m_{\tilde{\chi}_i^0}}{100~{\rm GeV}} \right)^5 \left( \frac{m_{3/2}}{1~{\rm eV}} \right)^{-2} \times ~ \Gamma_{0}, \\
		\Gamma\left(\tilde{\chi}_i^0 \to h \tilde{G}\right) &= \frac{ \kappa_{ih}}{2} \left( \frac{m_{\tilde{\chi}_i^0}}{100~{\rm GeV}} \right)^5 \left( \frac{m_{3/2}}{1~{\rm eV}} \right)^{-2} \left(1- \frac{m_h^2}{m_{\tilde{\chi}_i^0}^2} \right)^4 \times ~ \Gamma_{0}, \\
		\Gamma\left(\tilde{\chi}_i^0 \to Z \tilde{G}\right) &= \left({ \kappa_{iZ}^{T} + \frac{\kappa_{iZ}^{L}}{2}}\right) \left( \frac{m_{\tilde{\chi}_i^0}}{100~{\rm GeV}} \right)^5 \left( \frac{m_{3/2}}{1~{\rm eV}} \right)^{-2} \left(1- \frac{m_Z^2}{m_{\tilde{\chi}_i^0}^2} \right)^4 \times ~ \Gamma_{0}, \\
		\Gamma\left(\tilde{\chi}_i^\pm \to W^\pm \tilde{G} \right) &= \left(\kappa_{iW}^{T} + \frac{\kappa_{iW}^{L}}{2}\right) \left( \frac{m_{\tilde{\chi}_i^\pm}}{100~{\rm GeV}} \right)^5 \left( \frac{m_{3/2}}{1~{\rm eV}} \right)^{-2}\left( 1 - \frac{m_W^2}{m_{\tilde{\chi}_i^\pm}^2} \right)^4 \times ~ \Gamma_{0}.
\end{split}
\end{equation}
\end{small}
For a heavier electroweakino $\tilde{\chi}_i^0$  its decay width to LSP $\tilde{G}$ is enhanced by the quintic factor of the electroweakino mass; while for a heavier LSP $\tilde{G}$, the decay width is suppressed by $m_{3/2}^2$. 

\subsection{The electroweak MSSM with a light gravitino LSP\label{mod:ewmssmgraino}}

The scenario of $\tilde{G}$-EWMSSM has been encoded in the package \textsf{GAMBIT}, which provides a global fit for the parameter space~\cite{GAMBIT:2023yih, GEWMSSM:sup}.
Here we show the result in light of current experiments, focusing on the survived samples\footnote{Actually, the survived sample in our discussion is an ill-defined concept, so we use the points with the capped profile likelihood in the $2\sigma$ region, which means that the capped likelihood can be understood as a measure of how worse agreement of the signal with data than the background-only prediction.}. 

\begin{figure}[th]
	\centering
	\includegraphics[width=0.495\linewidth]{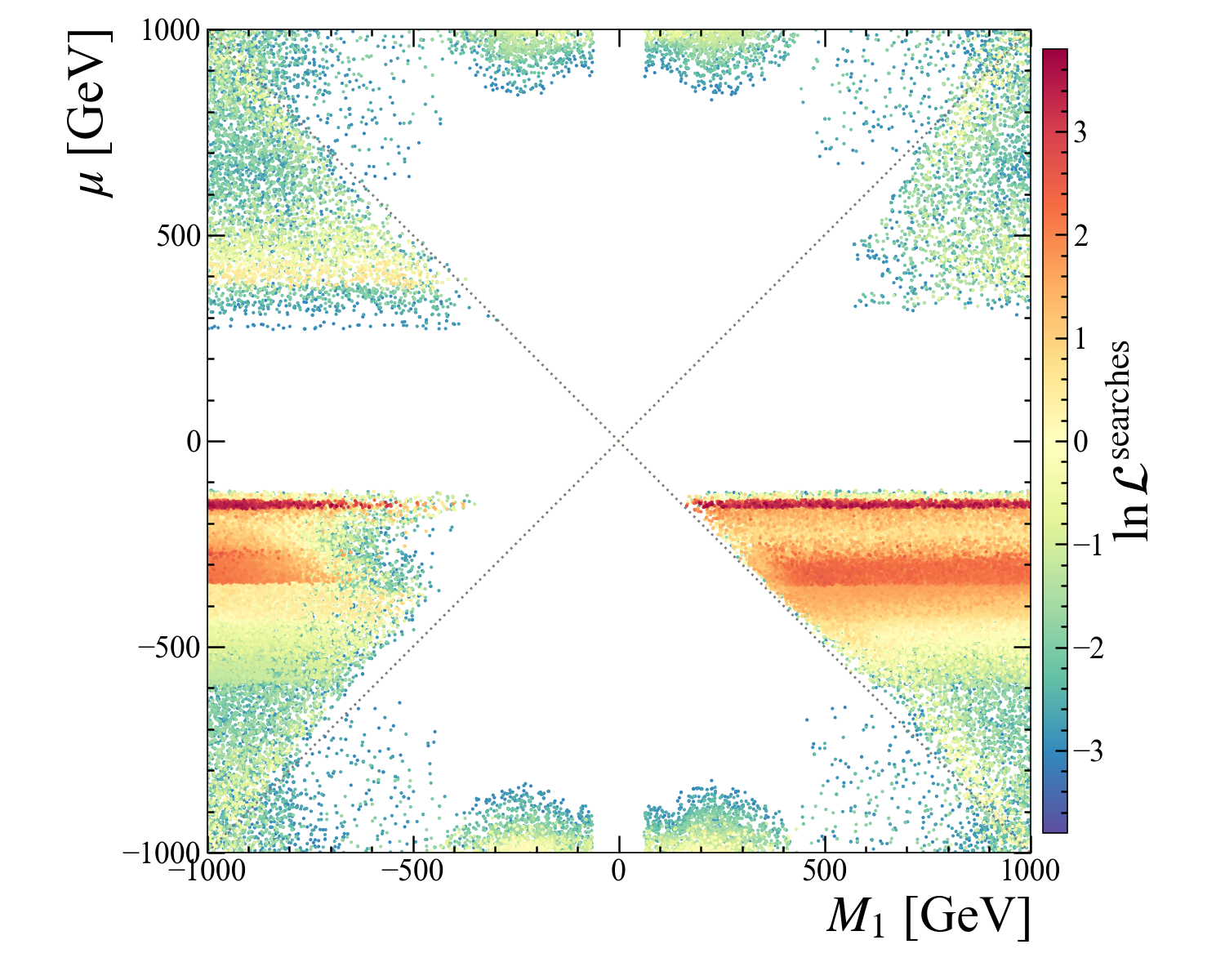}
	\includegraphics[width=0.495\linewidth]{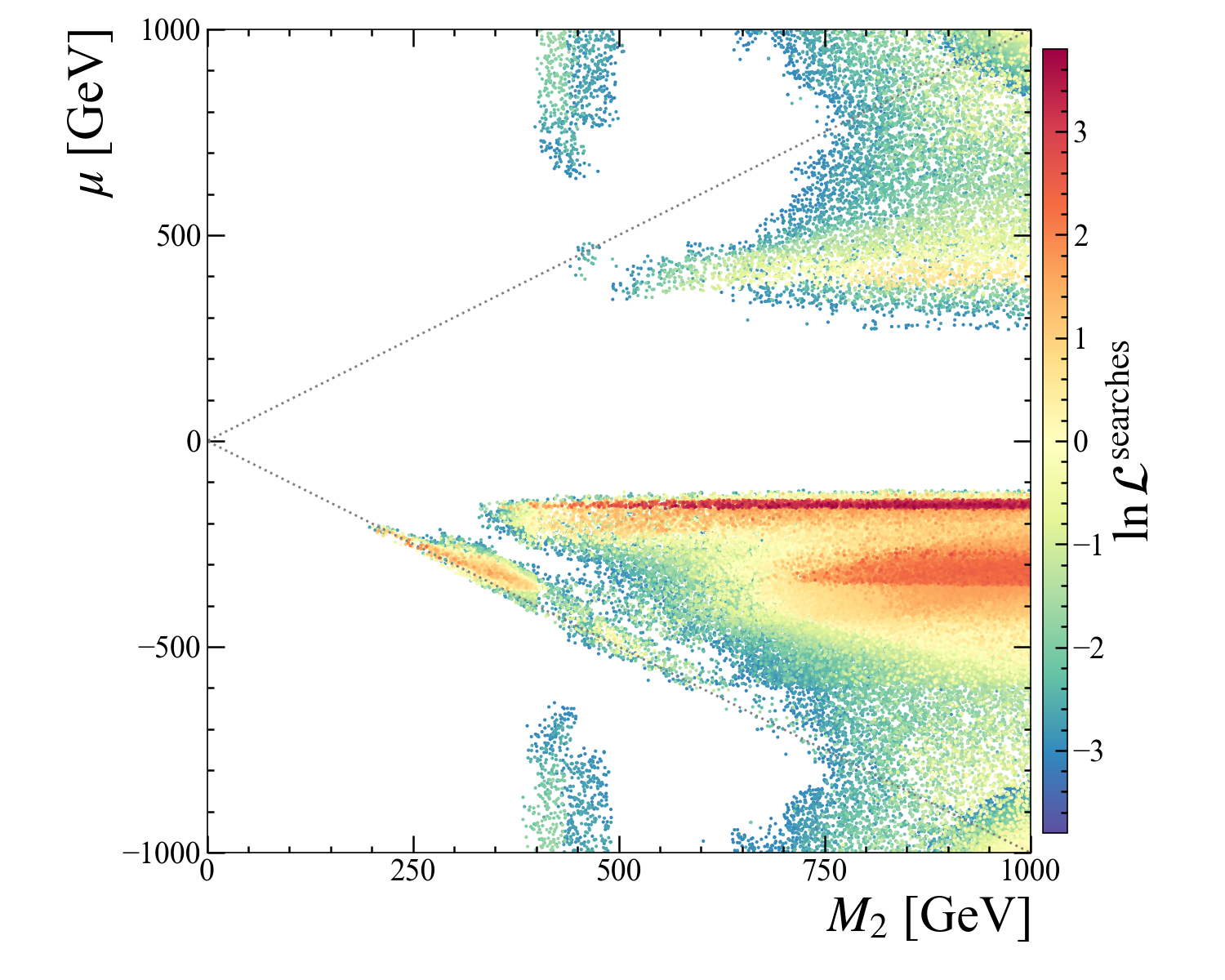}
	\caption{\label{fig:GBfit-paras} The $\tilde{G}$-EWMSSM samples in light of current experimental data, plotted in the $M_1 - \mu$ plane (left) and on $M_2 - \mu$ plane (right). The color indicates the log-likelihood ratio. The data is taken from Ref.~\cite{GAMBIT:2023yih, GEWMSSM:sup}. }
\end{figure}

In the \textsf{GAMBIT} work, the input parameters of the $\tilde{G}$-EWMSSM electroweak sector are usually chosen as 
\begin{equation}
	M_1, \quad M_2, \quad \mu, \quad \tan{\beta},
\end{equation}
while all other soft SUSY breaking parameters are fixed at $3~{\rm TeV}$, except $m_{3/2}$ which is fixed at $1~{\rm eV}$. 
Fig.~\ref{fig:GBfit-paras} shows the samples in light of current experimental data, plotted in the $M_1-\mu$ and $M_2-\mu$ planes with color coded in $\ln{\mathcal{L}} $. From Fig.~\ref{fig:GBfit-paras} one can find that the best-fit region (the points marked with red or yellow colors) favors $|\mu|$ to be smaller than about $500~{\rm GeV}$ and smaller than $|M_1|$ and $|M_2|$. 
%%%fig.1 
%%%fig.2 
\begin{figure}[t]
	\centering
	\includegraphics[width=0.45\linewidth]{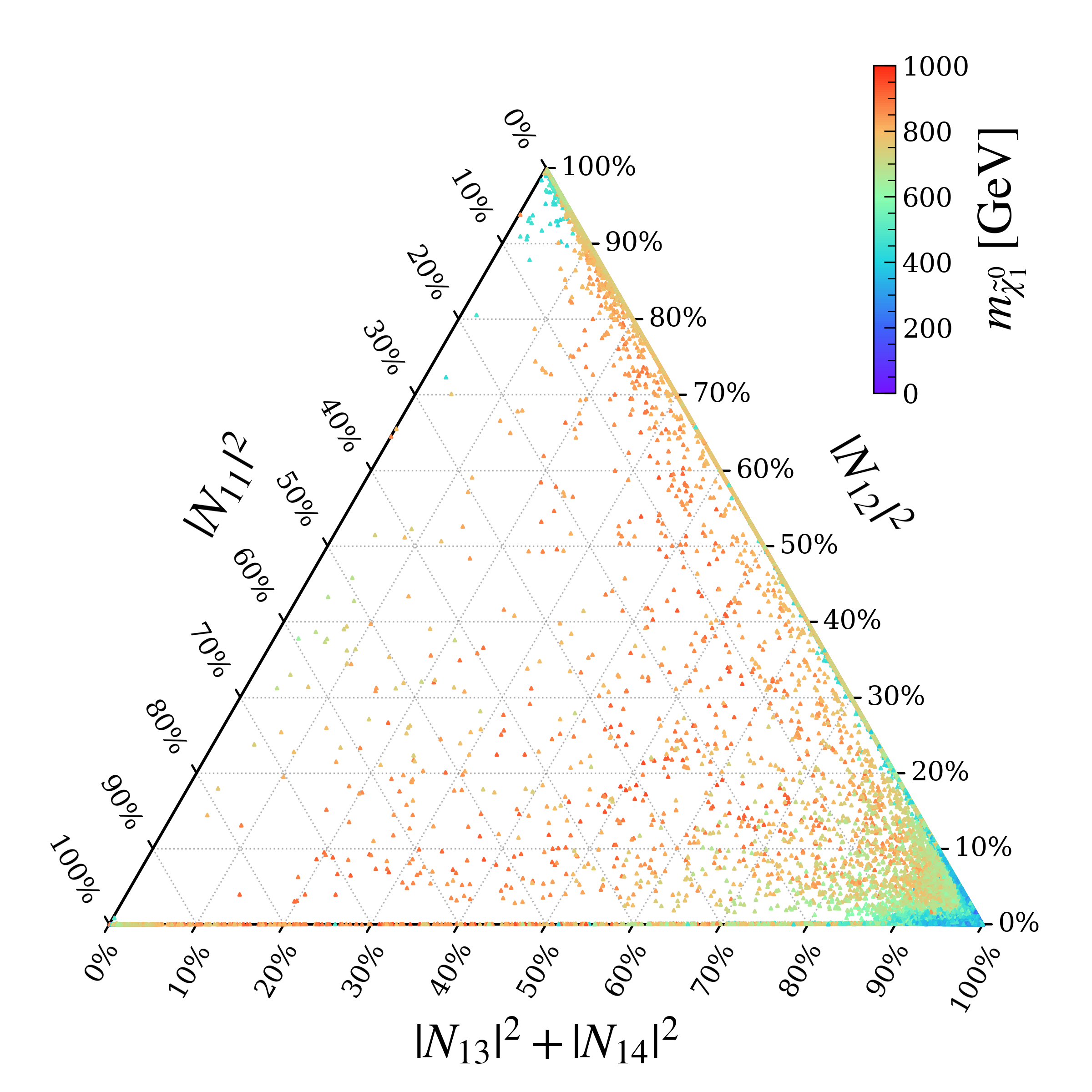}\hspace{1cm}
	\includegraphics[width=0.45\linewidth]{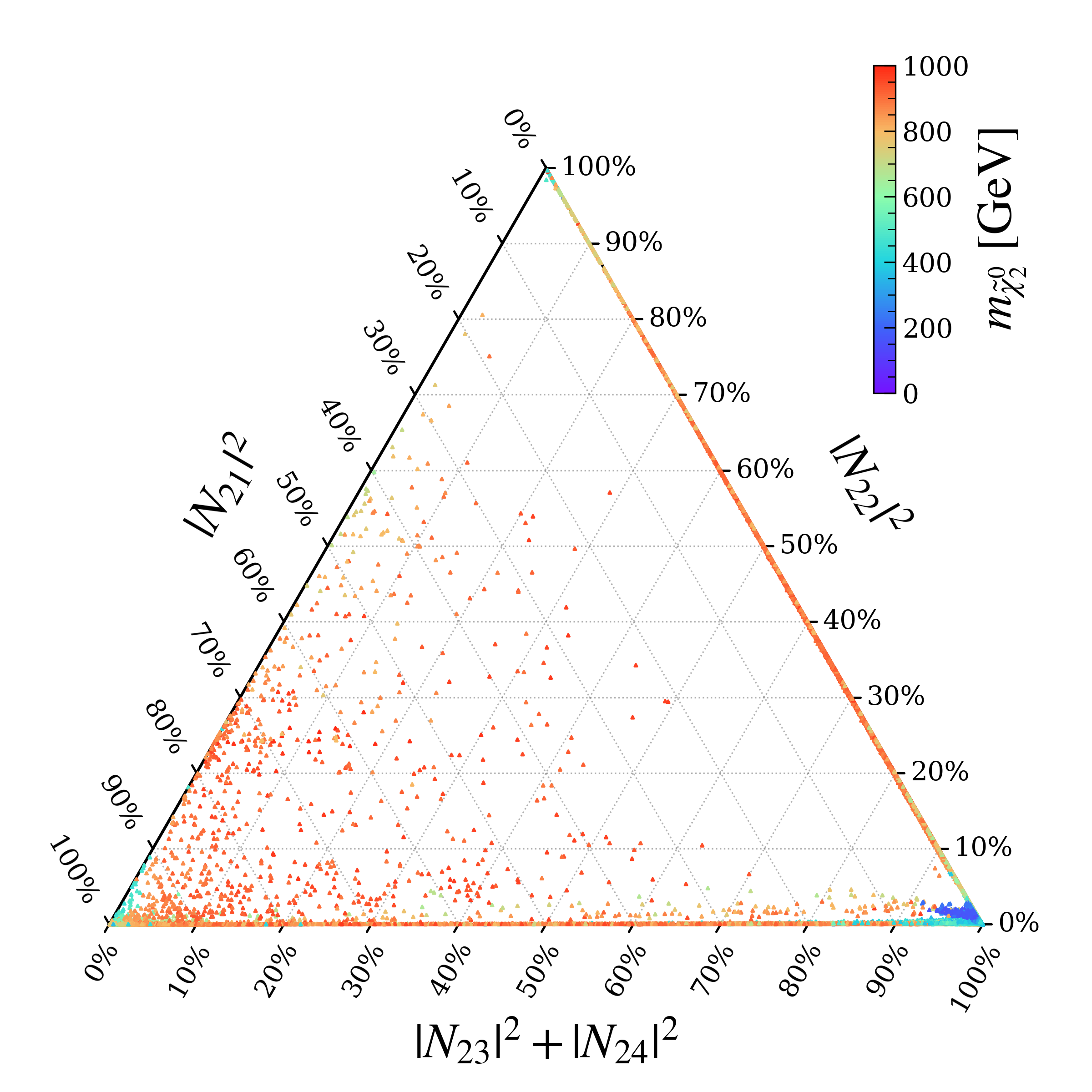}
	\caption{\label{fig:N1N2_CTs}Triangular presentations of the compositions of  $\tilde{\chi}_1^0$ (left) and $\tilde{\chi}_2^0$ (right) with color coded with their masses. The data is taken from Ref.~\cite{GAMBIT:2023yih, GEWMSSM:sup}}
\end{figure}

\begin{figure}[t]
	\centering
	\includegraphics[width=0.495\linewidth]{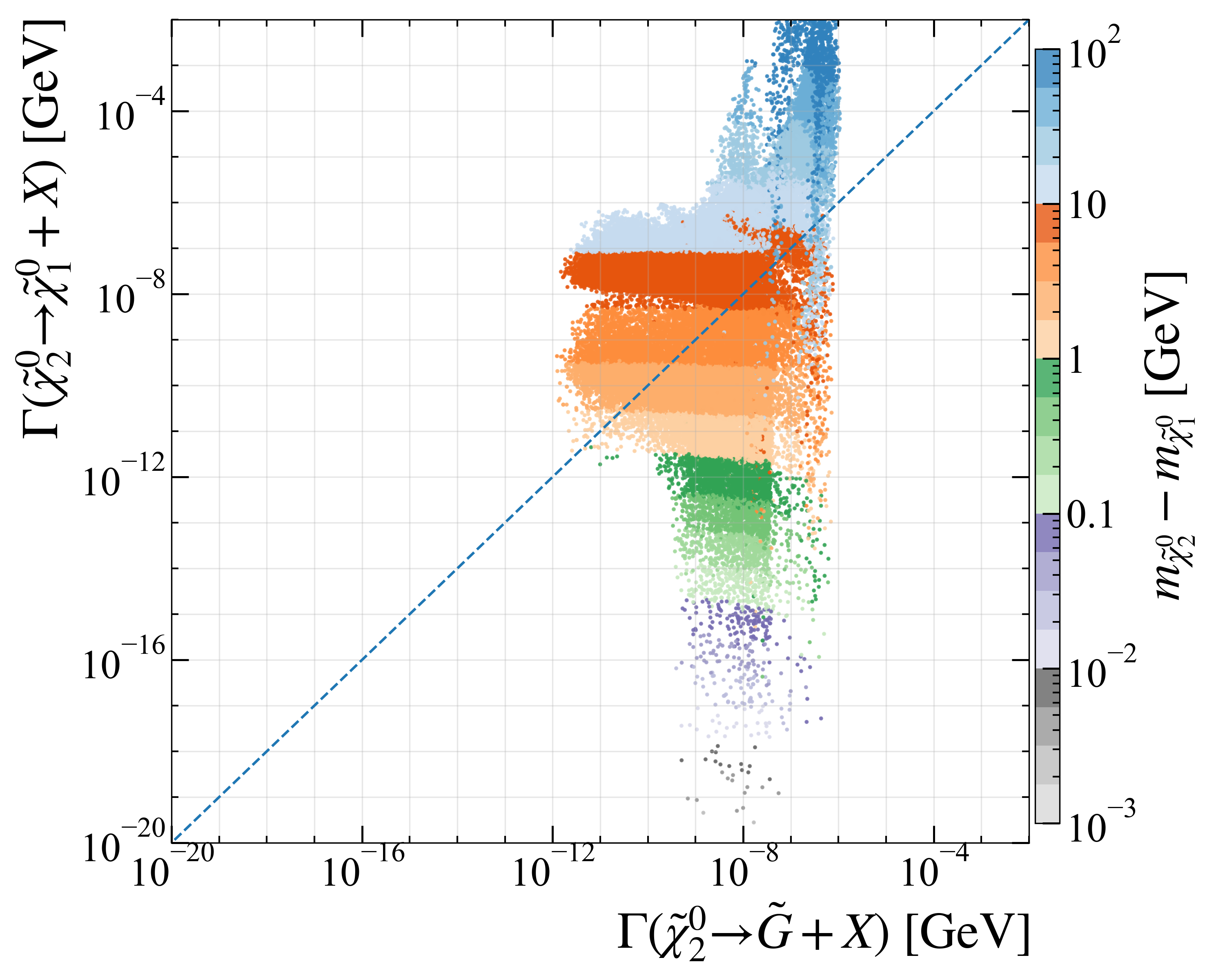}
	\includegraphics[width=0.495\linewidth]{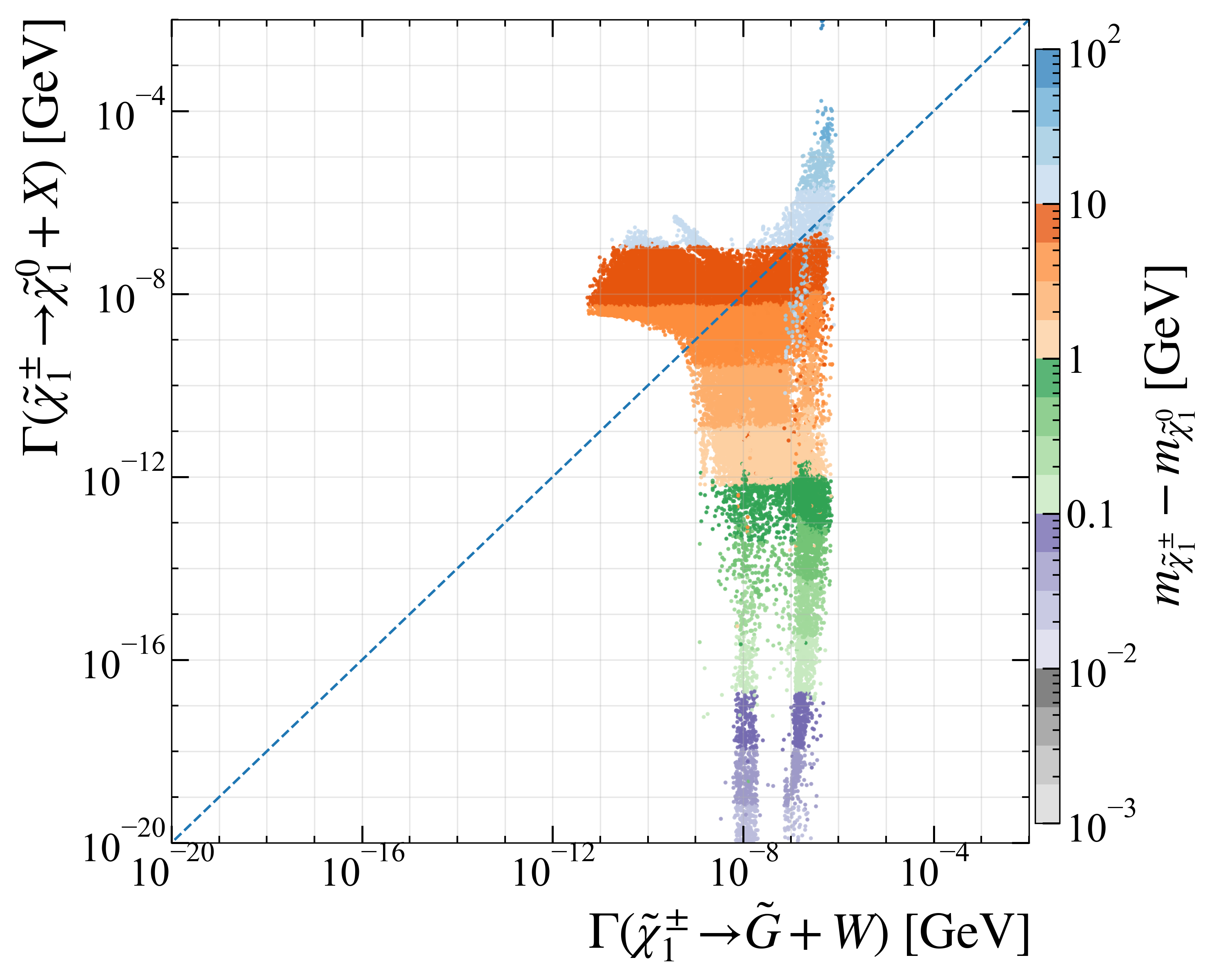}
	\caption{\label{fig:N2C2_DKs}The decay widths of 
 $\tilde{\chi}_2^0 ~(\tilde{\chi}_1^{\pm}) \to \tilde{\chi}_1^0+X$ 
 versus $\tilde{\chi}_2^0 ~(\tilde{\chi}_1^{\pm}) \to \tilde{G}+X$.  
 The color represents the mass splittings $m_{\tilde{\chi}_2^0} - m_{\tilde{\chi}_1^0}$ and $m_{\tilde{\chi}_1^\pm} - m_{\tilde{\chi}_1^0}$. The data is  taken from Ref.~\cite{GAMBIT:2023yih, GEWMSSM:sup}. }
\end{figure}

We plot the components and the masses of the two lightest neutralinos $\tilde{\chi}_{1,2}^0$ in Fig.~\ref{fig:N1N2_CTs}, and plot the decay widths of 
 $\tilde{\chi}_2^0 ~(\tilde{\chi}_1^{\pm}) \to \tilde{\chi}_1^0+X$ 
 versus $\tilde{\chi}_2^0 ~(\tilde{\chi}_1^{\pm}) \to \tilde{G}+X$  
 in Fig.~\ref{fig:N2C2_DKs}, in which the color is coded with the mass splitting. From Fig.~\ref{fig:N1N2_CTs} one can find that when the $\tilde{\chi}_1^0$ mass is smaller than $400~{\rm GeV}$, the current experiments favor the higgsino NLSP scenario. There is also a small part of samples predicting $\tilde{\chi}_1^0$ to be bino-dominated or bino-higgsino admixture. Pure higgsinos do not decay to photons, which is verified by the fit result. The decay width of $\tilde{\chi}_2^0$ (or $\tilde{\chi}_1^\pm$) to $\tilde{\chi}_1^0$ is suppressed by the mass splitting between higgsino-dominated states. For the most compressed case, in which $M_1$ and $M_2$ are much heavier than $\mu$, the total decay width $\Gamma\left(\tilde{\chi}_2^0 \to \tilde{\chi}_1^0 + X\right)$ can reach  $10^{-20}~{\rm GeV}$, corresponding to a very small mass splitting around  $10^{-3}~{\rm GeV}$. For a mass splitting at order $1~{\rm GeV}$, the decay width of $\tilde{\chi}_2^0$ into all final states including $\tilde{\chi}_1^0$ is at the same order of $\Gamma_0$ defined in Eq.~(\ref{eq:dk-typical}).

%%%fig.3 

\subsection{\label{mod:spf}Two simplified models of light higgsino scenario}
\begin{figure}[t]
	\centering
	\includegraphics[width=0.7\linewidth]{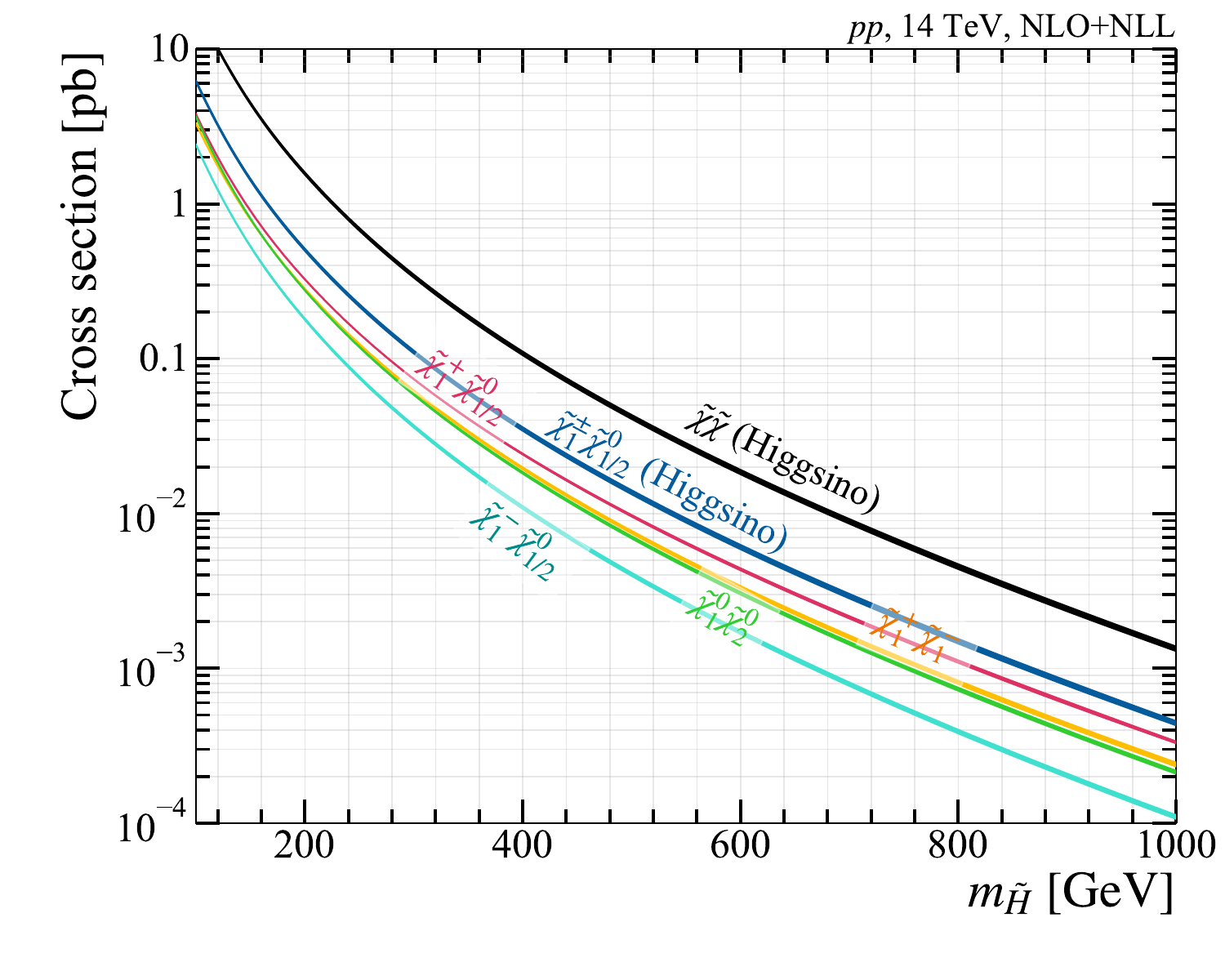}
 \vspace{-.7cm} 
	\caption{\label{fig:hino-XSect}  The cross sections for higgsino pair productions, assuming $\tilde{\chi}_1^0$, $\tilde{\chi}_2^0$ and $\tilde{\chi}_1^\pm$ with a degenerate mass $m_{\tilde{H}}$. The four production modes $\tilde{\chi}_1^+ \tilde{\chi}_1^-$, $\tilde{\chi}_1^+ \tilde{\chi}_{1/2}^0$, $\tilde{\chi}_1^- \tilde{\chi}_{1/2}^0$ and $\tilde{\chi}_1^0 \tilde{\chi}_2^0$ are shown respectively by the orange, red, cyan and green curves, while the black curve represents the sum of these modes.}
\end{figure}

We consider two simplified SUSY models inspired by gauge-mediated SUSY breaking and by the \textsf{GAMBIT} result. Both of them  contain an almost mass-degenerate higgsino system ($\tilde{\chi}_1^\pm$, $\tilde{\chi}_1^0$, $\tilde{\chi}_2^0$) and an nearly massless LSP $\tilde{G}$, while all other SUSY particles are much heavy and out of the reach of the LHC/HL-LHC.  The higgsino system offers four production processes at the LHC: $\tilde{\chi}_1^+ \tilde{\chi}_1^-$, $\tilde{\chi}_1^+ \tilde{\chi}_{1/2}^0$, $\tilde{\chi}_1^- \tilde{\chi}_{1/2}^0$ and $\tilde{\chi}_1^0 \tilde{\chi}_2^0$.  Fig.~\ref{fig:hino-XSect} shows the cross sections of chargino-neutralino pair productions for $\sqrt{s} = 14~{\rm TeV}$ at NLO plus next-leading-log (NLO-NLL) contributions using the package \textsf{Resummino}~\cite{Fiaschi:2020udf, Fiaschi:2023tkq, Fuks:2013vua} with the parton distribution functions (PDFs) given by \texttt{CTEQ6.6}~\cite{Nadolsky:2008zw}.
%(we checked that using the \texttt{MSTW2008nlo90cl} PDFs give the almost same results). 
In the calculations we assume the higgsino-like charginos and neutralinos are degenerate, denoted as $m_{\tilde{H}}$. 

%%%fig.4 

%%%fig.5 
\begin{figure}[t]
	\centering
	\makebox[\textwidth][c]{	
	\subfigure[HH channel]{\includegraphics[width=0.24\linewidth]{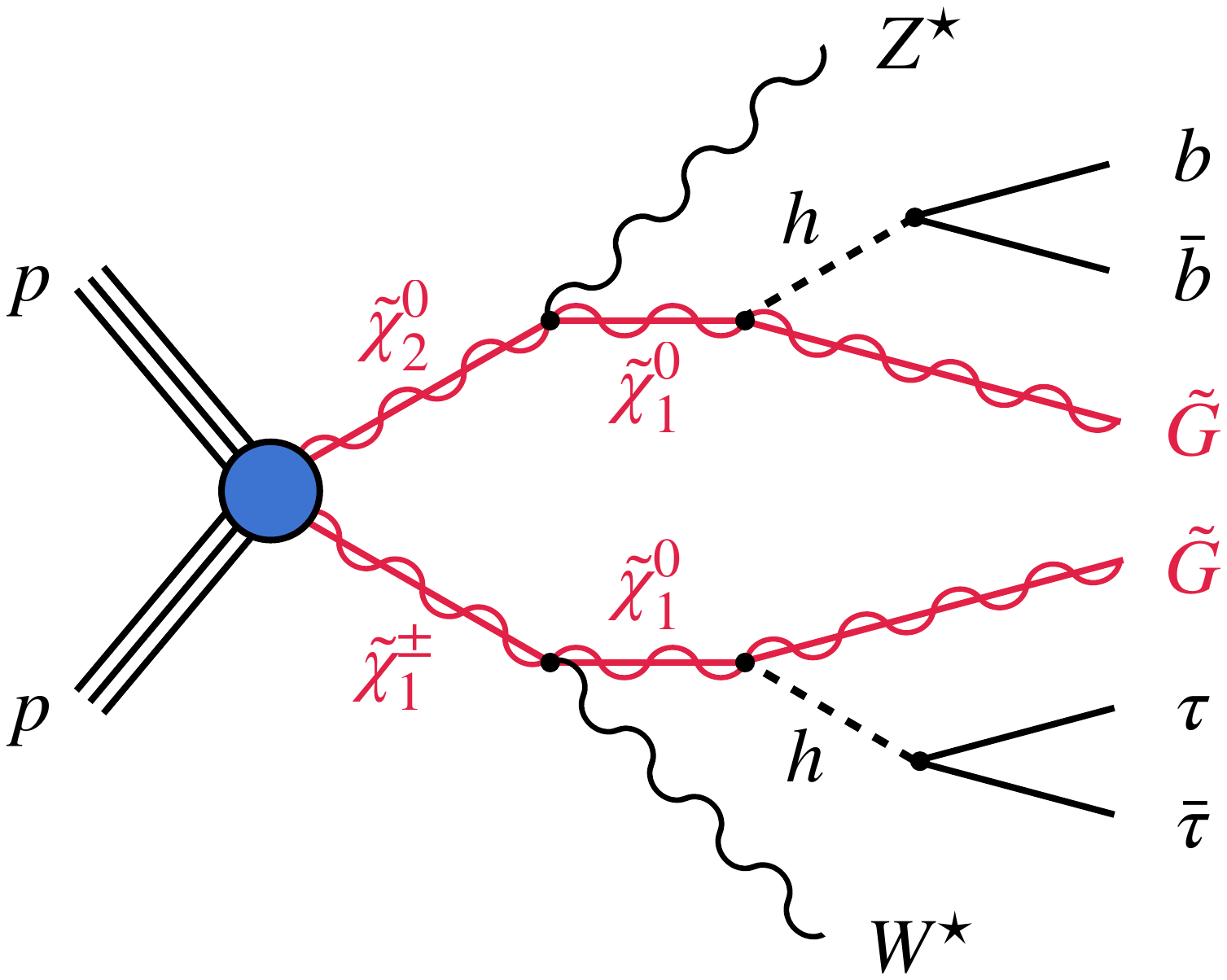}}\hspace{-0.3cm}
	\subfigure[ZH channel]{\includegraphics[width=0.24\linewidth]{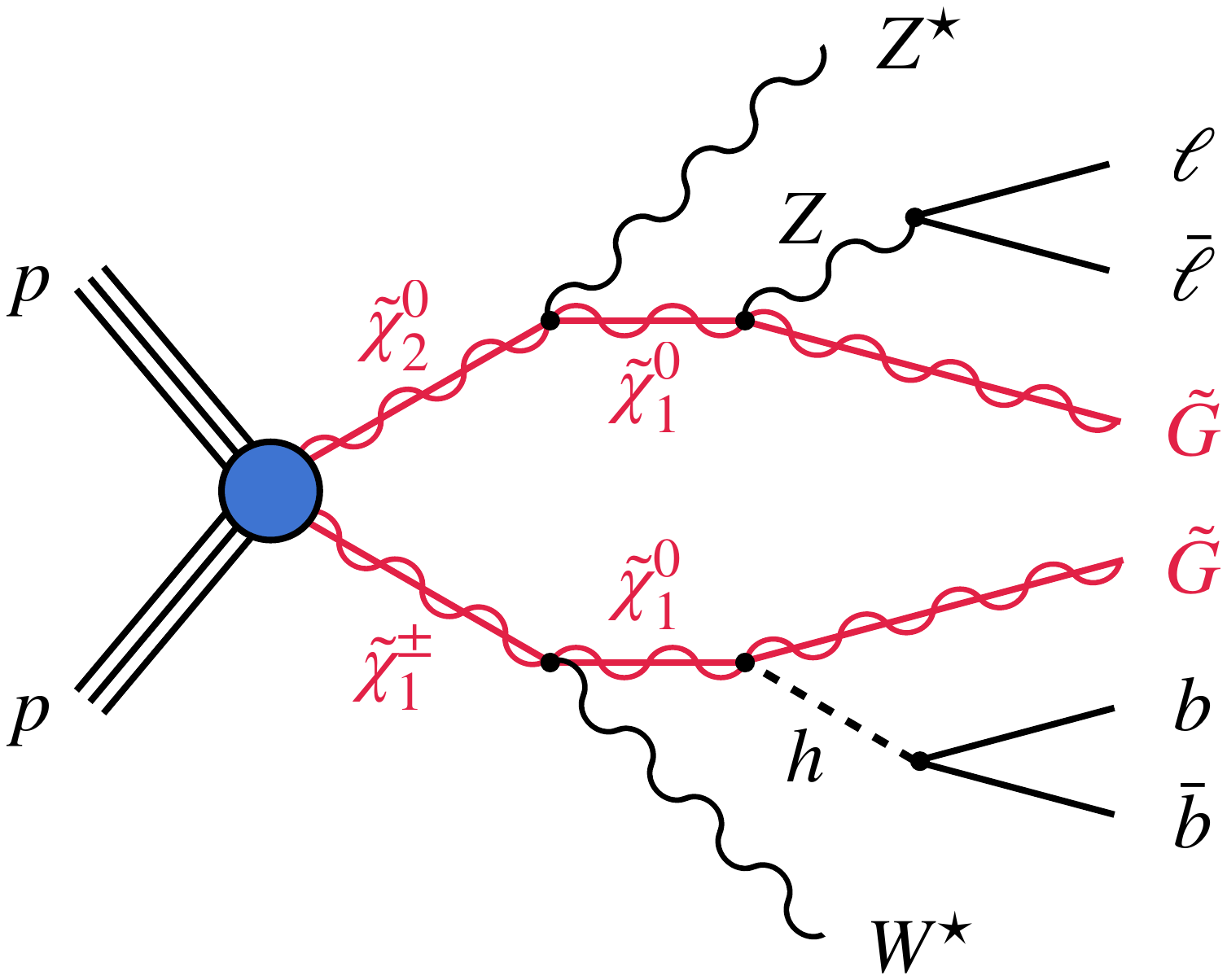}}\hspace{-0.3cm}
	\subfigure[ZZ channel]{\includegraphics[width=0.24\linewidth]{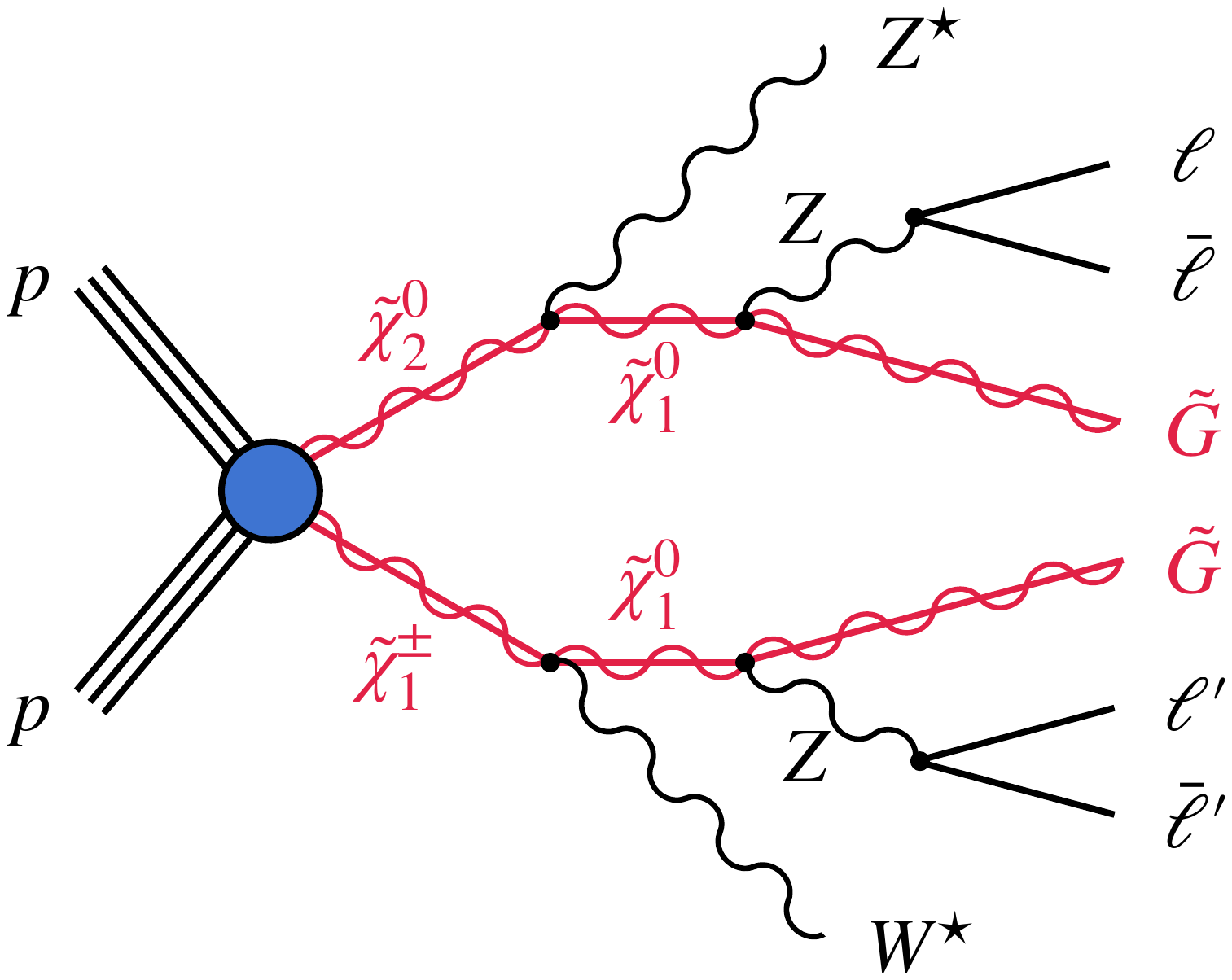}}
	\subfigure[WH channel]{\includegraphics[width=0.24\linewidth]{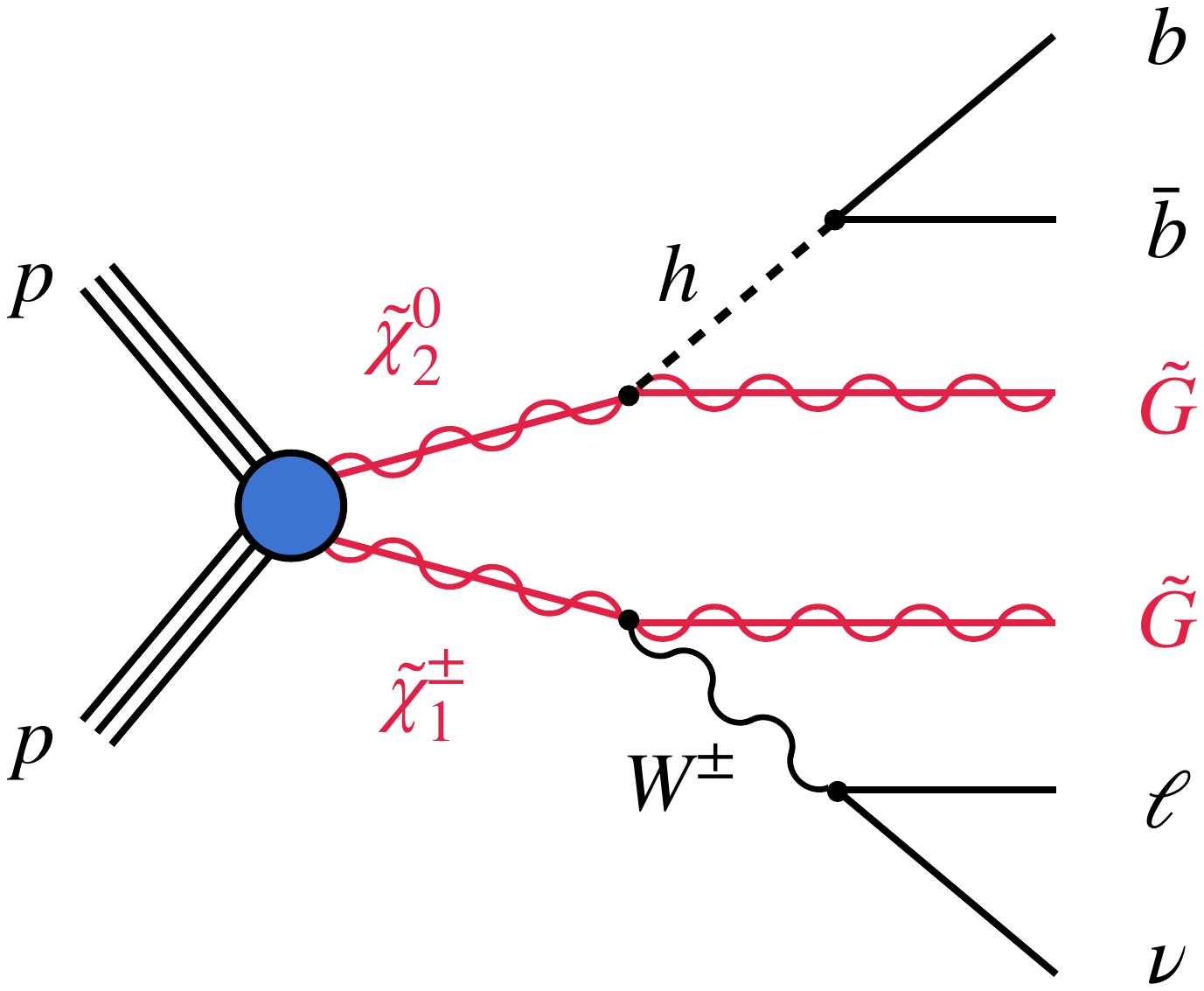}}\hspace{-0.6cm}
	\subfigure[WZ channel]{\includegraphics[width=0.24\linewidth]{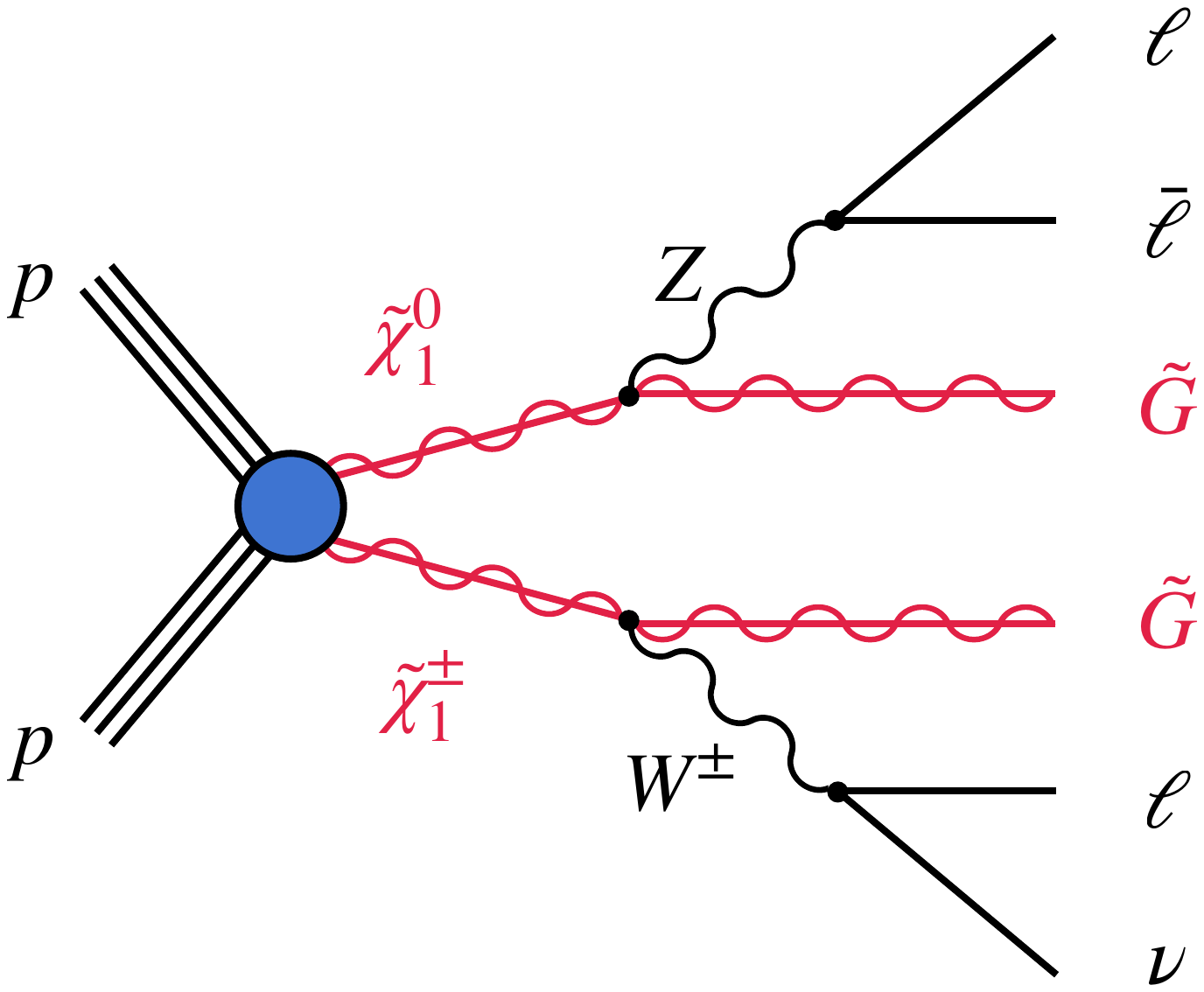}}\hspace{-0.2cm}
	}
	\caption{\label{fig:feyn-diam-ty}The typical processes predicted by the simplified models for the productions of higgsino-dominated electroweakinos with a light gravitino LSP. The neutralinos and charginos are pair produced, resulting in final states with rich leptons, $b$-jets and $E_{\rm T}^{\rm miss}$.  }
\end{figure}

\subsubsection{\label{sec:smA} Simplified model A}
In the first simplified model, the gravitino mass $m_{3/2}$ is set to $0.1~{\rm GeV}$, and the $\tilde{\chi}_1^\pm$ and $\tilde{\chi}_2^0$ masses are set to $1~{\rm GeV}$ above  $\tilde{\chi}_1^0$. According to Eq.~(\ref{eq:dk-typical}), the decay widths of electroweakinos to $\tilde{G}$ are much suppressed, so both $\tilde{\chi}_1^\pm$ and $\tilde{\chi}_2^0$ dominantly decay to $\tilde{\chi}_1^0$ via a virtual $Z/W$ boson. The  final states from the virtual $Z/W$ boson are too soft to be reconstructed, while  $\tilde{\chi}_1^0$ decays promptly to a gravitino plus a $Z$ or $h$ boson, $\tilde{\chi}_1^0 \to Z/h + \tilde{G}$. The $\tilde{\chi}_1^0$ mass and the branching ratio ${\rm BR}\left(\tilde{\chi}_1^0 \to Z \tilde{G} \right) = 1 - {\rm BR} \left(\tilde{\chi}_1^0 \to h \tilde{G} \right)$ are treated as two free parameters in our analysis for this simplified model. Therefore, there are three searching channels complementary to each other:
	\begin{enumerate}
		\item[(1)~] HH channel. This channel is targeting  the Higgs pair from the  $\tilde{\chi}_1^0$ decay. 
  As discussed in Sec.~\ref{sec:intro}, we focus on the final states of $hh$ decaying into ${b\bar{b}}\tau\bar{\tau}$, and a typical process is shown in Fig.~\ref{fig:feyn-diam-ty}~(a). 
		\item[(2)~] ZH channel. As shown in Fig.~\ref{fig:feyn-diam-ty}~(b), this channel focuses on the final state containing two $b$-jets plus one opposite-sign same-flavor (OSSF) lepton pair. 
		\item[(3)~] ZZ channel. This channel is targeting the final state of four-leptons which can form two OSSF pairs, as shown in Fig.~\ref{fig:feyn-diam-ty}~(c). 
	\end{enumerate}
	All production processes of the higgsino system (black line in Fig.~\ref{fig:hino-XSect}) are considered in this model. 
\subsubsection{\label{sec:smB} Simplified model B}
 The second simplified model is similar to the first one, except the gravitino mass $m_{3/2}$ is set to $1~{\rm eV}$. Then both $\tilde{\chi}_1^\pm$ and $\tilde{\chi}_2^0$ dominantly decay directly into $\tilde{G}$ plus a $W^{\pm}$ or $Z/h$ boson. In this simplified model, only chargino-neutralino pair productions ($\tilde{\chi}_1^\pm \tilde{\chi}_1^0$ and $\tilde{\chi}_1^\pm \tilde{\chi}_2^0$) are considered in our analysis. So this simplified model can be probed via:
	\begin{enumerate}
		\item[(4)~] WH channel. As shown in Fig.~\ref{fig:feyn-diam-ty}~(d), this channel is targeting the final state of the $W$-boson leptonic decay and the $h$-boson $b\bar{b}$ decay. 
		\item[(5)~] WZ channel. This channel is well probed in the previous LHC searches~\cite{ATLAS:2018eui, ATLAS:2019lng, ATLAS:2021moa}, and a similar search strategy focusing on $3\ell+E_{\rm T}^{\rm miss}$ final states is used in this work, which is shown in Fig.~\ref{fig:feyn-diam-ty}~(e).  
	\end{enumerate}
	Since in the light higgsino scenario we have $|\mu| \ll M_1, M_2$,  the bino and wino components of $\tilde{\chi}_1^0$ and $\tilde{\chi}_2^0$ can be neglected, $\left|N_{11}\right|^2 \sim \left|N_{12}\right|^2 \sim \left|N_{21}\right|^2 \sim \left|N_{22}\right|^2 \sim 0$, which results in a maximal mixing between $\tilde{H}_u^0$ and $\tilde{H}_d^0$ in $\tilde{\chi}_1^0$ and $\tilde{\chi}_2^0$. 
	Therefore, given the complicated model dependence on the parameters like $\tan{\beta}$ and $\alpha$, the branching ratio of the lightest neutralino $\tilde{\chi}_1^0$ is treated as a free parameter, with ${\rm BR}\left( \tilde{\chi}_1^0 \to Z \tilde{G} \right) = 1 - {\rm BR}\left( \tilde{\chi}_1^0 \to h \tilde{G} \right)$. Also, we have  the following relations approximately
	\begin{equation}\begin{split}
		{\rm BR}\left(\tilde{\chi}_1^0 \to Z \tilde{G} \right) + {\rm BR}\left( \tilde{\chi}_2^0 \to Z \tilde{G} \right) &= 1, \\
		{\rm BR}\left(\tilde{\chi}_1^0 \to h \tilde{G} \right) + {\rm BR}\left( \tilde{\chi}_2^0 \to h \tilde{G} \right) &= 1.  
	\end{split}\end{equation}
	These relations provide a good approximation according to Eq.~(\ref{eq:dkN2Hgra}) and Eq.~(\ref{eq:dkN2Zgra}) in the decoupling limit $m_A \gg m_Z$~\cite{Han:2013kza}. 
	As a result, the total branching ratio of the WH channel or WZ channel is equal to $50\%$. The corresponding cross section is shown by dark blue curve in Fig.~\ref{fig:hino-XSect}. 

\section{\label{sec:3}Detecting higgsino pair productions at HL-LHC}
Now we present our search strategies, especially the distinctive new signature from the di-Higgs events in the $bb\tau\tau$ final state. As discussed in the preceding section, enhancing this channel can significantly enhance the sensitivity compared to the leading four $b$-jets signal for the di-Higgs events. Several relevant studies have explored the electroweakino productions with a light gravitino scenario~\cite{Dai:2022isa, Nakamura:2017irk, Chang:2017niy, Kang:2015nga, Zurita:2017sfg, Biswas:2016ffy}. In this work, we conduct a comprehensive investigation for the decays of the higgsino pairs, which are subsequently combined with all other channels to provide a final assessment of the LHC sensitivity to the $\mu$ parameter.

\par We perform Monte-Carlo simulations to estimate both the SM backgrounds and the higgsino pair signals, as well as to examine the efficiency for various electroweakino productions. In our analysis, all events are generated using \textsf{Madgraph5\_aMC@NLO} event generator~\cite{Alwall:2014hca, Frederix:2018nkq} and we use \textsf{Pythia8}~\cite{Bierlich:2022pfr, Sjostrand:2014zea} for parton shower and hadronization. Next-to-leading-order (NLO) cross sections are used for background and signal normalization, calculated using \textsf{Madgraph5\_aMC@NLO}~\cite{Alwall:2014hca} and the package \textsf{Resummino}~\cite{Fiaschi:2023tkq}, respectively. The events are processed through the fast ATLAS detector simulation using the \textsf{Rivet-HEP} package~\cite{Bierlich:2019rhm} for parameterized simulation and object reconstruction with the \texttt{ATLAS\_RUN2} smearing and efficiency setting~\cite{Buckley:2019stt, ATLAS:2019kwg}.
\par We reconstruct jets using the anti-$k_T$ clustering algorithm \cite{Cacciari:2008gp} with a distance parameter of 0.4, as implemented in the \textsf{FASTJET} package~\cite{Cacciari:2011ma}. The jets are required to have $p_{\rm T} > 30~{\rm GeV}$ in the region $|\eta| < 2.47$. The $b$-jets containing $b$-hadron decays are identified from the selected jet objects with a flat efficiency of $75\%$ and the mis-tagging rates for non-$b$ jets with $\epsilon(c \to b) =0.1$ and $\epsilon(j \to b) =0.01$~\cite{ATL-PHYS-PUB-2019-005}. The reconstructed $\tau_{\rm had}$ objects are seeded by jets, with an efficiency fixed at $60\%$~\cite{Buckley:2019stt}. Light flavor lepton objects (electrons and muons) are reconstructed in the region $|\eta| < 2.47$ to have $p_{\rm T} > 15~{\rm GeV}$. The missing transverse momentum ${\bf p}_{\rm T}^{\rm miss}$, with magnitude $E_{\rm T}^{\rm miss}$, is defined as  the negative vector sum of the transverse momenta of all identified physics objects, including electrons, photons, muons, jets and an additional soft term.

\par We apply a sequential overlap removal procedure to resolve ambiguities in which multiple electron, muon or jet candidates would otherwise be reconstructed from the same detector signature. This procedure uses the definition of the angular distance, $\Delta R = \sqrt{(\Delta y)^2 + (\Delta \phi)^2}$, where $y$ is the rapidity of the object, and the following steps are applied in order:
\begin{enumerate}
	\item If a $b$-jet is within $\Delta R = 0.2$ of an electron or muon candidate, this electron or muon is rejected, as it is likely to be from a semi-leptonic $b$-hadron decay; 
	\item If the jet within $\Delta R = 0.2$ of the electron is not $b$-tagged, the jet itself is discarded, as it likely originates from an electron-induced shower;
	\item Any $\tau_{\rm had}$ object with $p_{\rm T} \leq 50~{\rm GeV}$ within $\Delta R = 0.2$ of an electron or muon is removed. Otherwise, the muon or electron is  removed;
	\item Any electron with $\Delta R = 0.2$ of a muon is removed; 
	\item Any jet within $\Delta R = 0.2$ of an electron or muon is removed; 
	\item Any jet with $\Delta R = 0.2$ of a $\tau_{\rm had}$ is removed; 
	\item Muons with $\Delta R = 0.4$ of a remaining jet are discarded to suppress muons from semi-leptonic decays of $b$-hardons.
\end{enumerate}

The detection and discovery significance $Z_A$ is evaluated by Poisson formula  
\begin{equation}
	Z_A = \left[ 2\left((s+b) \ln \left(1+\frac{s}{b} \right) -s \right) \right]^{1/2}, 
\end{equation}
where $s$ and $b$ are respectively the signal and background event numbers. 

\subsection{Searching strategy for HH channel}
\begin{figure}
	\centering
	\includegraphics[width=0.49\linewidth]{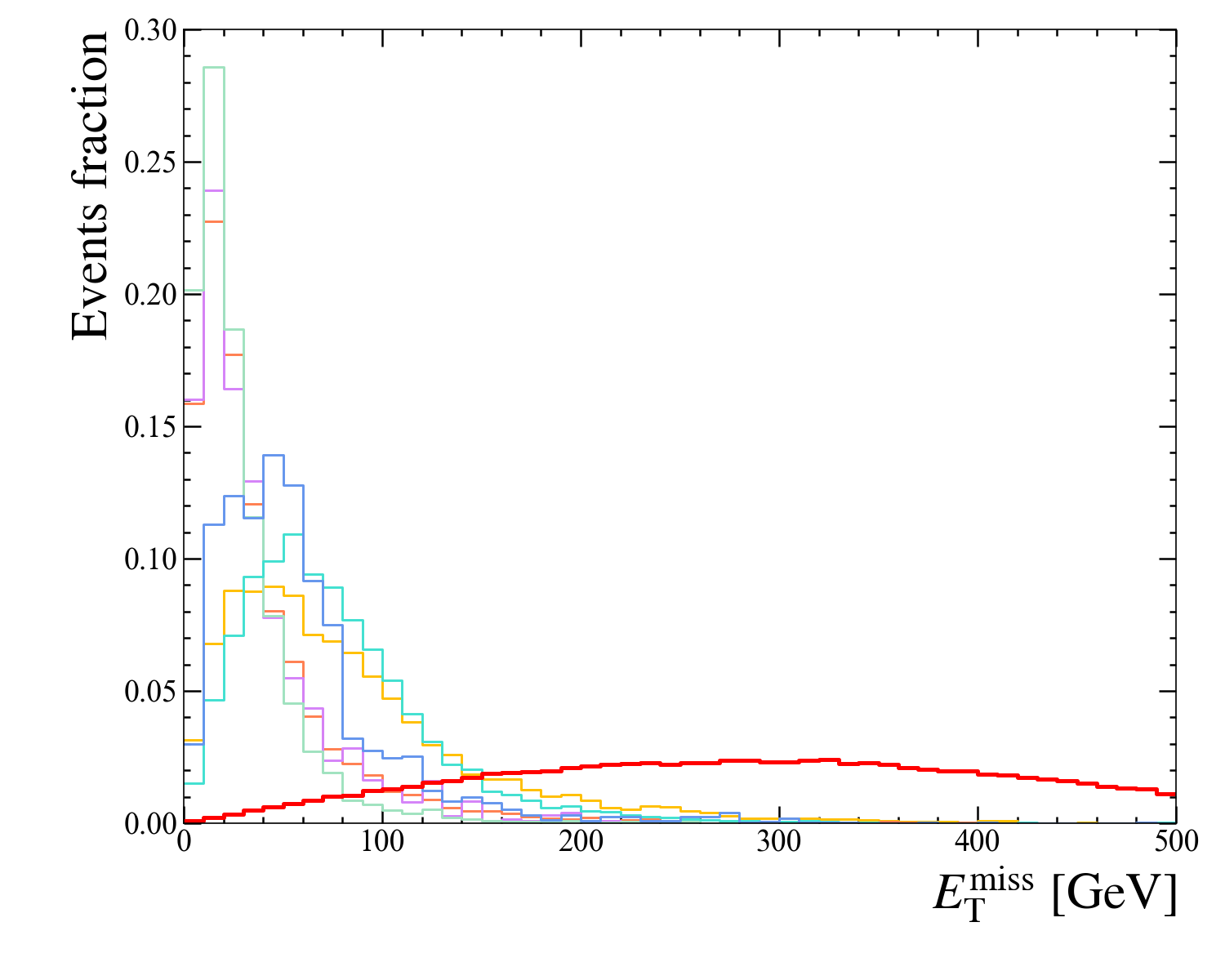}
	\includegraphics[width=0.49\linewidth]{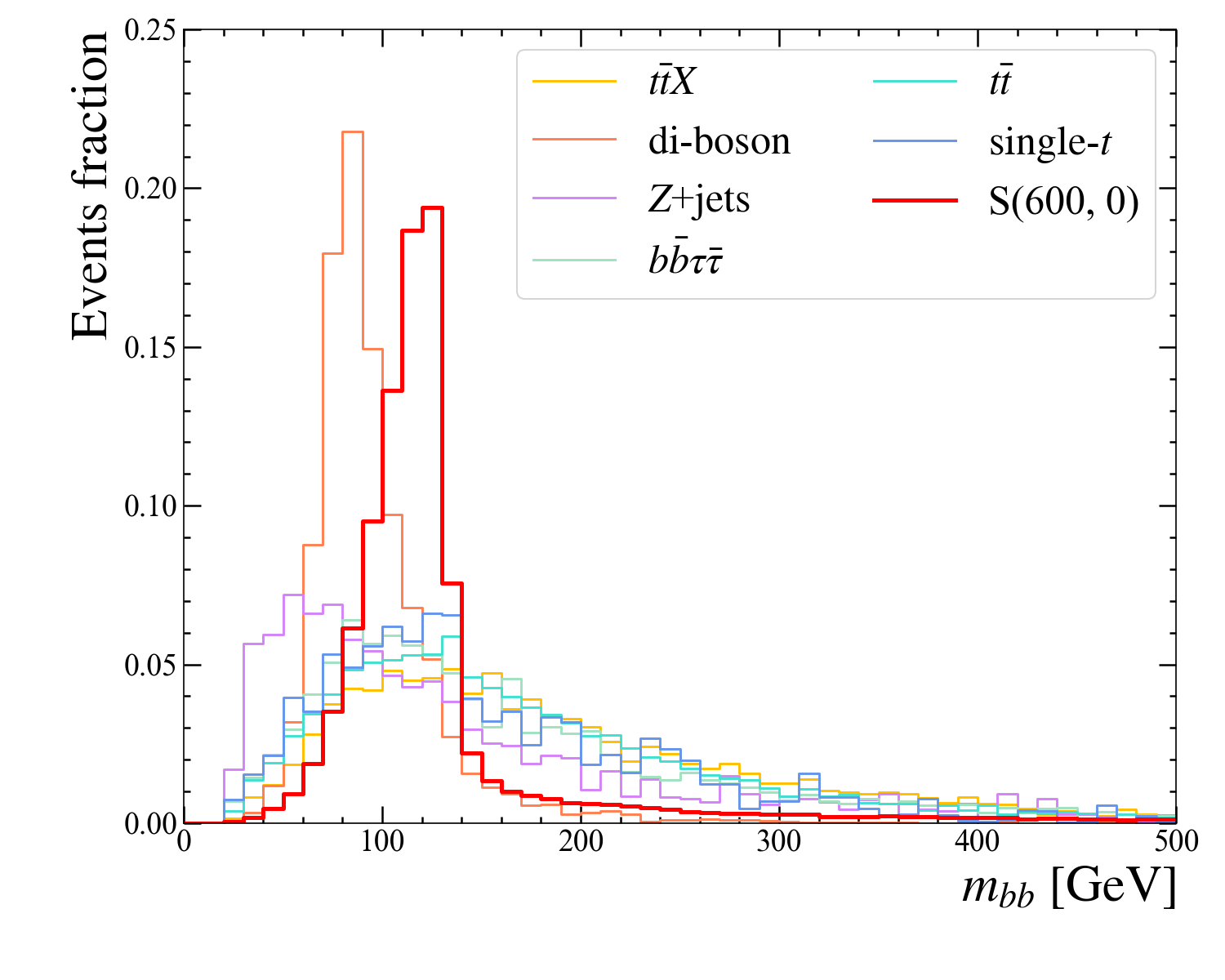} \\
	\includegraphics[width=0.49\linewidth]{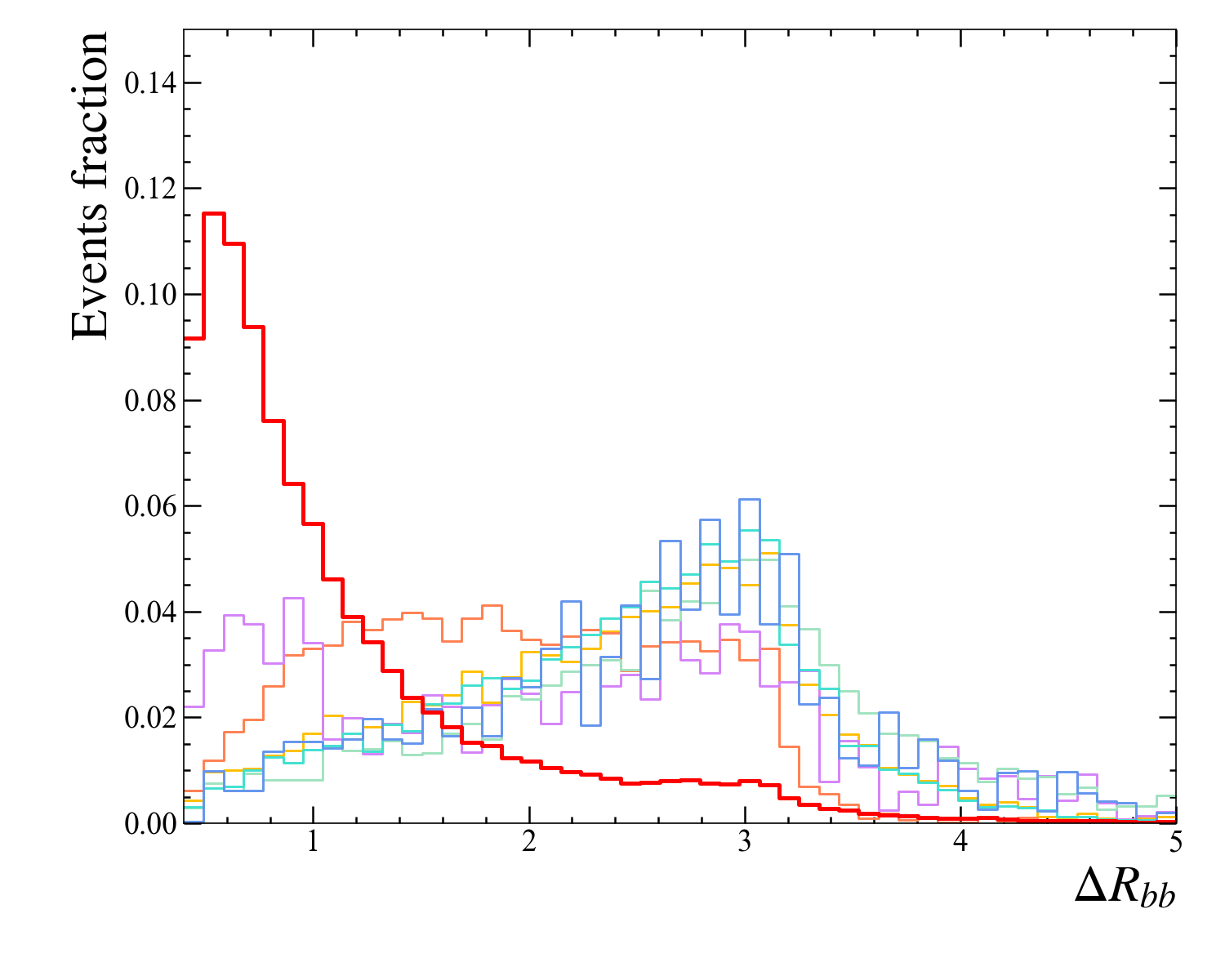}
	\includegraphics[width=0.49\linewidth]{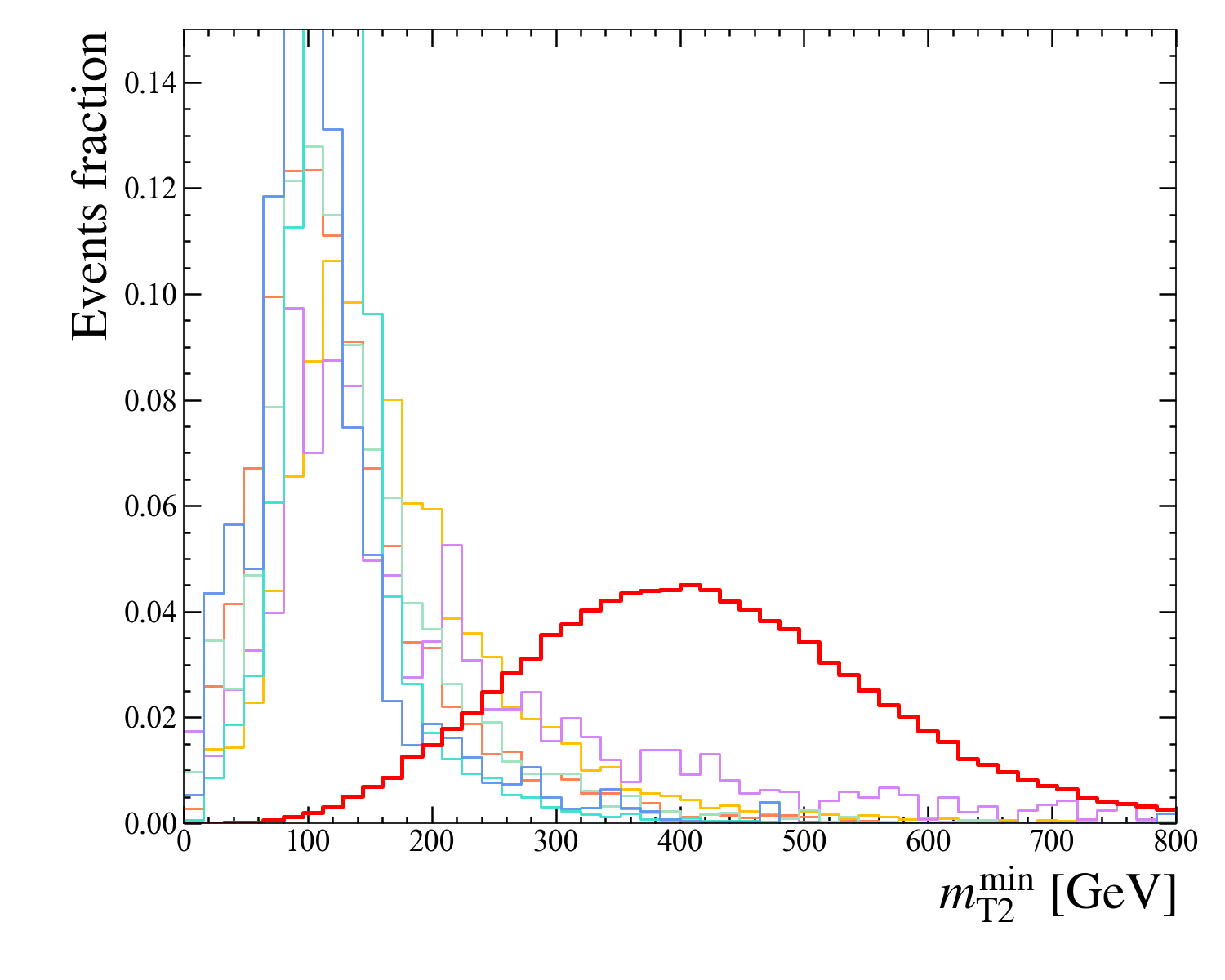}
	\caption{\label{fig:distri-HH} The $E_{\rm T}^{\rm miss}$, $\Delta R_{bb}$, $m_{bb}$ and $m_{\rm T2}^{\rm min}$ distributions of the dominated SM backgrounds and the signal with $m_{\tilde{H}}=600~{\rm GeV}$ after the HH channel preselection criteria.}
\end{figure}

\par  The HH channel targets events with two oppositely charged $\tau_{\rm had}$ candidates and two $b$-jets. The main sources of backgrounds in HH channel are $t$-quark processes, $Z$+jets, $W$+jets, di-bosons and multi-jet productions. The $t$-quark processes include the single-$t$ process, $t\bar{t}$ and $t\bar{t}X$, where $X$ stands for a $Z$ or $H$ boson. 

\par The leptonic $t\bar{t}$ background event is complicated by an additional layer of decays, with $t\to Wb$ in the first step followed by $W \to \ell\nu$ in the second. This results in an event topology with two identical branches, $t \to  b\ell\nu$, each with a visible ($b\ell$) and invisible ($\nu$) component. In order to suppress the background from the $t\bar{t}$  processes, the `stransverse mass' $m_{\rm T2}$ ~\cite{Barr:2010zj, Lester:1999tx} is an event variable used to bound the masses of an unseen pair of particles, defined as 
\begin{equation}
	m_{\rm T2} = \min_{{\bf p}_{\rm T}^{\rm miss} = {\bf p}_{\rm T}^a + {\bf p}_{\rm T}^b}
	\left[\max \left(  m_{\rm T} ({\bf p}_{\rm T}^{b^a} + {\bf p}_{\rm T}^{\tau^a}, {\bf p}_{\rm T}^a), m_{\rm T} ({\bf p}_{\rm T}^{b^b} + {\bf p}_{\rm T}^{\tau^b}, {\bf p}_{\rm T}^b)\right) \right], 
\end{equation}
where the transverse mass $m_{\rm T}$ is defined as 
\begin{equation}\label{eq:bkg_err}
	m_{\rm T}({\bf p}_{\rm T}, {\bf q}_{\rm T}) = \sqrt{2 p_{\rm T} q_{\rm T} - {\bf p}_{\rm T}\cdot {\bf q}_{\rm T}},
\end{equation}
with ${\bf p}_{\rm T}$ and ${\bf q}_{\rm T}$ being respectively the transverse momenta of the visible and invisible components, and ${p}_{\rm T}$ and ${q}_{\rm T}$ are their magnitudes correspondingly. In our analysis, for the pairing of two $b$-jets and two leptons, we select the pair that yields the smaller value for $m_{\rm T2}$, marked as $m_{\rm T2}^{\rm min}$. Several other variables used in this channel are the invariant mass of two leptons $m_{\ell\ell}$, the invariant mass $m_{bb}$ and the angular distance $\Delta R_{bb}$ of two $b$-jets. The distributions of the variables $E_{\rm T}^{\rm miss}$, $\Delta R_{bb}$, $m_{bb}$ and $m_{\rm T2}^{\rm min}$ are shown in Fig.~\ref{fig:distri-HH}.
For the signal events, the following cut sequences are summarized in Table ~\ref{tab:cut-HH}:
\begin{itemize}
	\item The basic cut requires two reconstructed $b$-jets plus two hadronic $\tau_{\rm had}$ candidates. The event containing any other jet or lepton candidate will be vetoed.  
	\item $E_{\rm T}^{\rm miss}$ trigger. This cut requires $E_{\rm T}^{\rm miss} > 100~{\rm GeV}$. 
	\item Higgs jet cut. The signal event contains a Higgs boson $H\to b\bar{b}$. Therefore, the invariant mass $m_{bb}$ and the angular distance $\Delta R_{bb}$ of the two $b$-jets requires $90~{\rm GeV}\leq m_{bb} \leq 140~{\rm GeV}$ and $\Delta R_{bb} < 2$ to suppress the non-Higgs resonance b-jets background like $t\bar{t}$. 
	\item The $m_{\rm T2}^{\rm min} > 175~{\rm GeV}$ cut. The distribution of $m_{\rm T2}$ variable has an upper endpoint at the parent particle. The cut at $175~{\rm GeV}$ here is to further suppress the background.  
\end{itemize}
After the cut-flows, we find that the significance $Z_A$ of the signal point of $m_{\tilde{H}}=600~{\rm GeV}$ can reach about $2.4\sigma$.
%%%%table 1 
\begin{table}[t]
\centering
\caption{\label{tab:cut-HH}The cut flows of HH channel at the 14 TeV HL-LHC. For the signal we choose a benchmark point $m_{\tilde{H}}=600~{\rm GeV}$, ${\rm BR}\left( \tilde{\chi}_1^0 \to Z \tilde{G} \right) = 0$, marked as S(600, 0). The event number are normalized to $3000~{\rm fb}^{-1}$. }
\resizebox{\linewidth}{!}{
\begin{tabular}{lp{.1cm}|rp{.01cm}rp{.01cm}rp{.01cm}rp{.01cm}rp{.01cm}rp{.01cm}|rp{.01cm}|c}
\hline\hline
\multirow{2}{*}{\bf Cuts} && \multicolumn{11}{c}{\bf SM backgrounds} &&{\bf signal} && \multirow{2}{*}{$Z_A$}\\ 
       && $t\bar{t}$ && $bb\tau\tau$ &&  $Z+$jets && di-boson && single-$t$ && $t\bar{t}X$     && S(600, 0) &&  \\ \hline
basic cut && 1111024.1                       && 269015.0                          && 403352.4                            && 9890.2   && 198374.4   && 21544.9 && 381.2  && 0.27       \\
$E_{\rm T}^{\rm miss} \geq 100~{\rm GeV}$          && 266068.9                        && 6480.2                            && 26653.2                             && 650.2    && 37911.4    && 6109.4  && 355.9  && 0.61       \\
$90~{\rm GeV} \leq m_{bb} \leq 140~{\rm GeV}$          && 60673.7                         && 437.8                             && 3791.3                              && 205.0    && 5848.8     && 1212.8  && 244.6  && 0.91       \\
$\Delta R_{bb} < 2 $          && 28645.9                         && 437.8                             && 2892.7                              && 184.8    && 2426.4     && 622.9   && 231.9  && 1.23        \\
$m_{\rm T2}^{\rm min} > 175~{\rm GeV}$          && 5172.2                          && 175.1                             && 2120.3                              && 141.1    && 1197.1     && 347.2   && 229.9  && 2.39        \\ \hline\hline
\end{tabular}
}
\end{table}

\subsection{Searching strategy for ZH channel}
\begin{figure}
	%\centering
\includegraphics[width=0.48\linewidth]{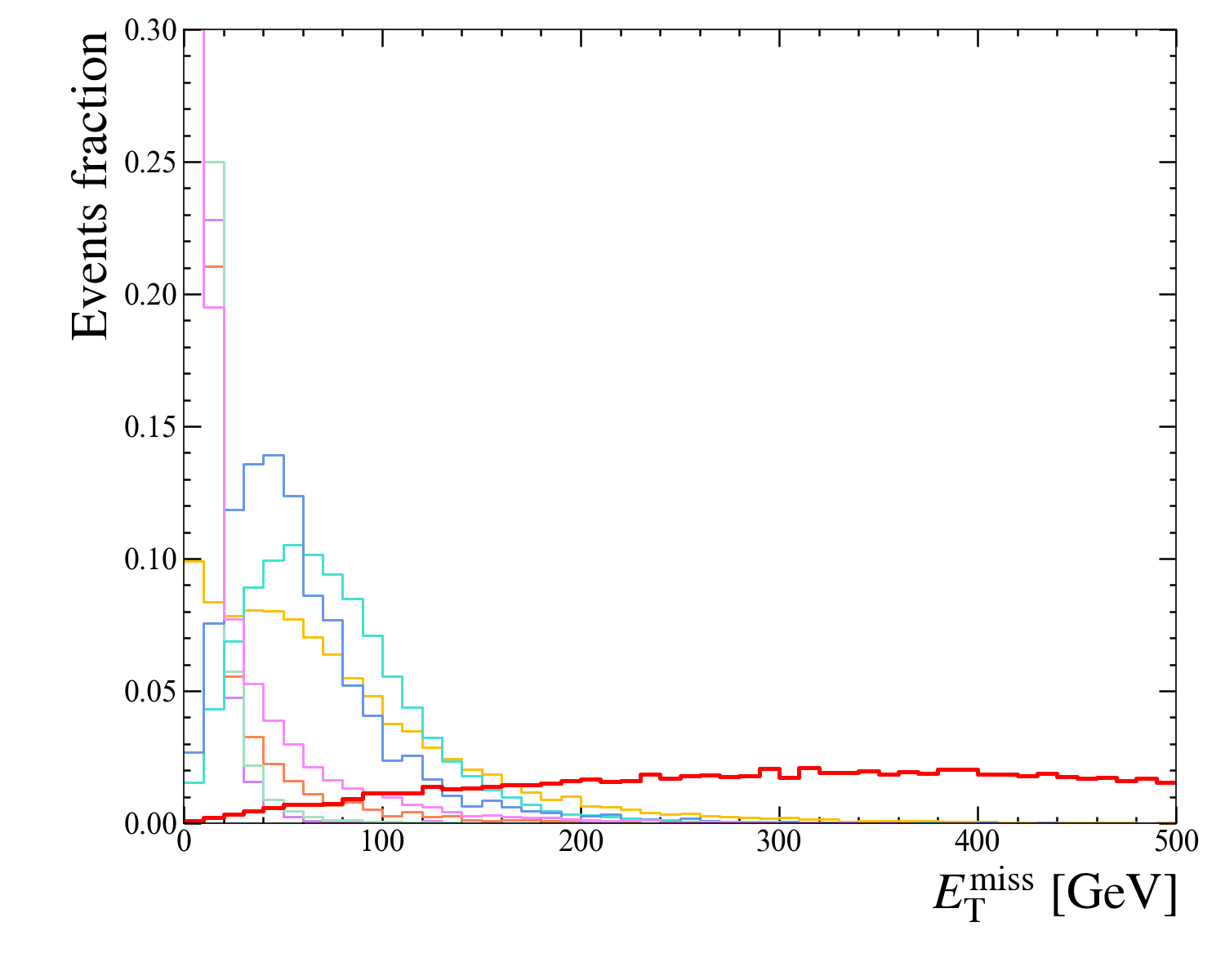}
\includegraphics[width=0.48\linewidth]{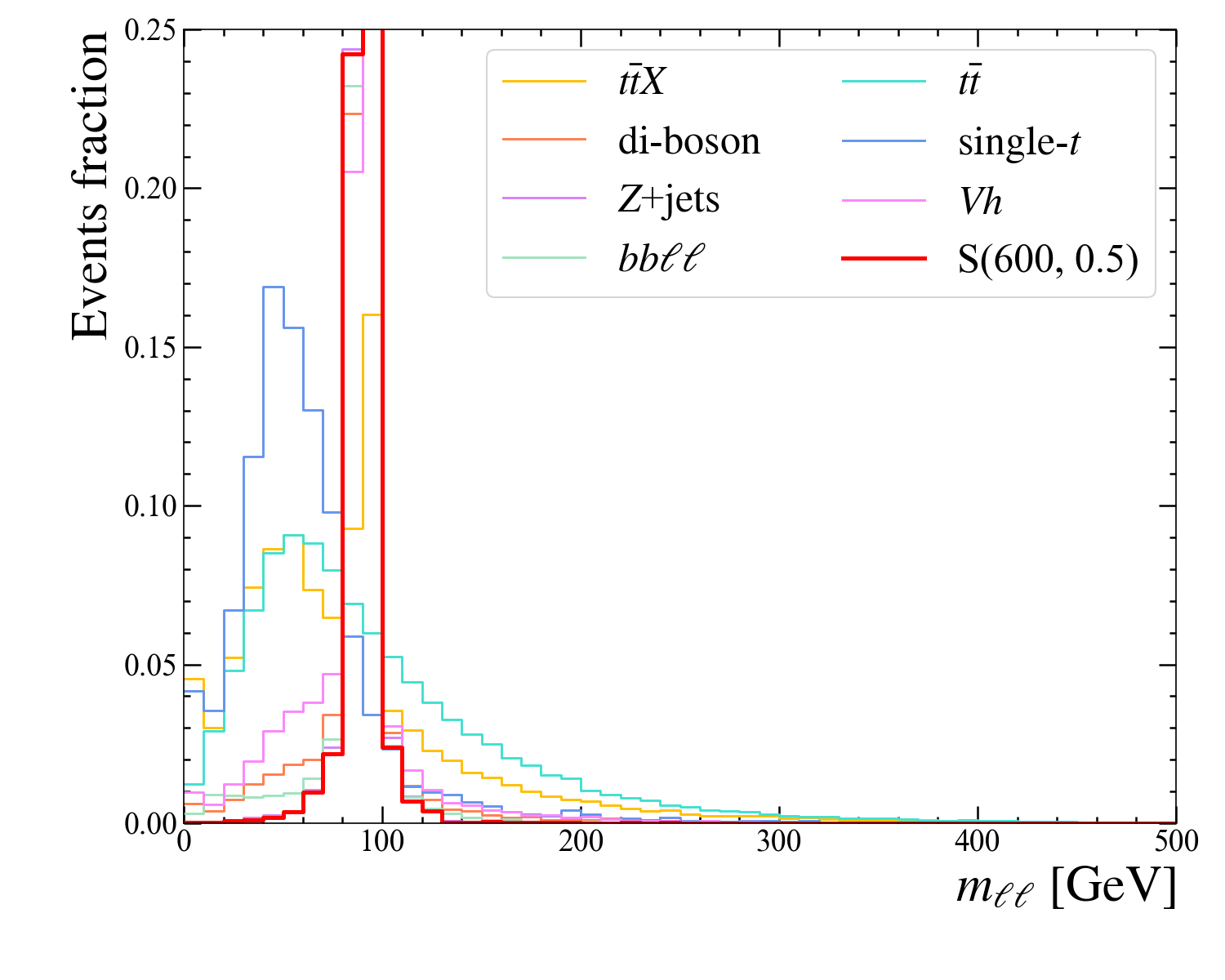} \\
\includegraphics[width=0.48\linewidth]{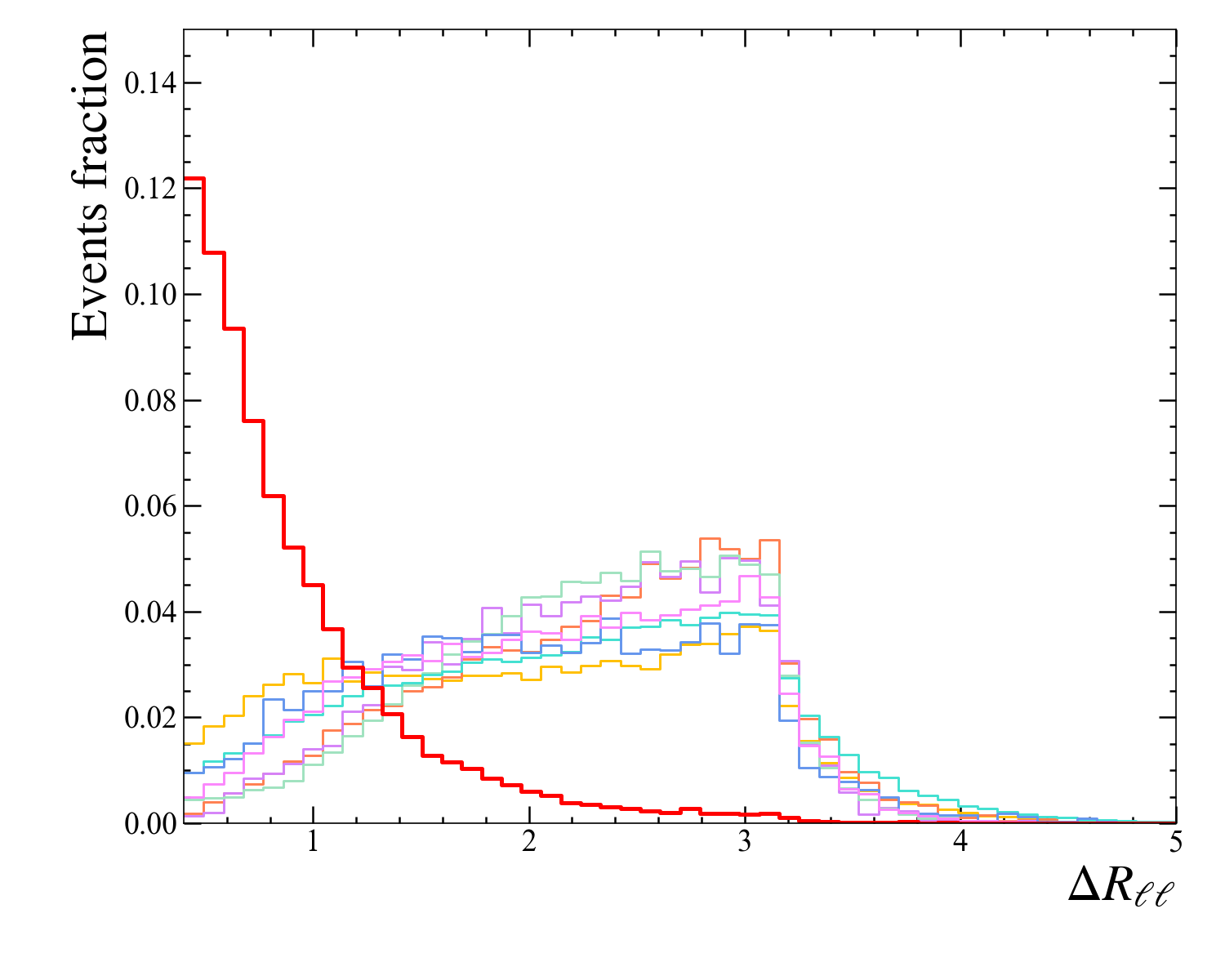}
\includegraphics[width=0.48\linewidth]{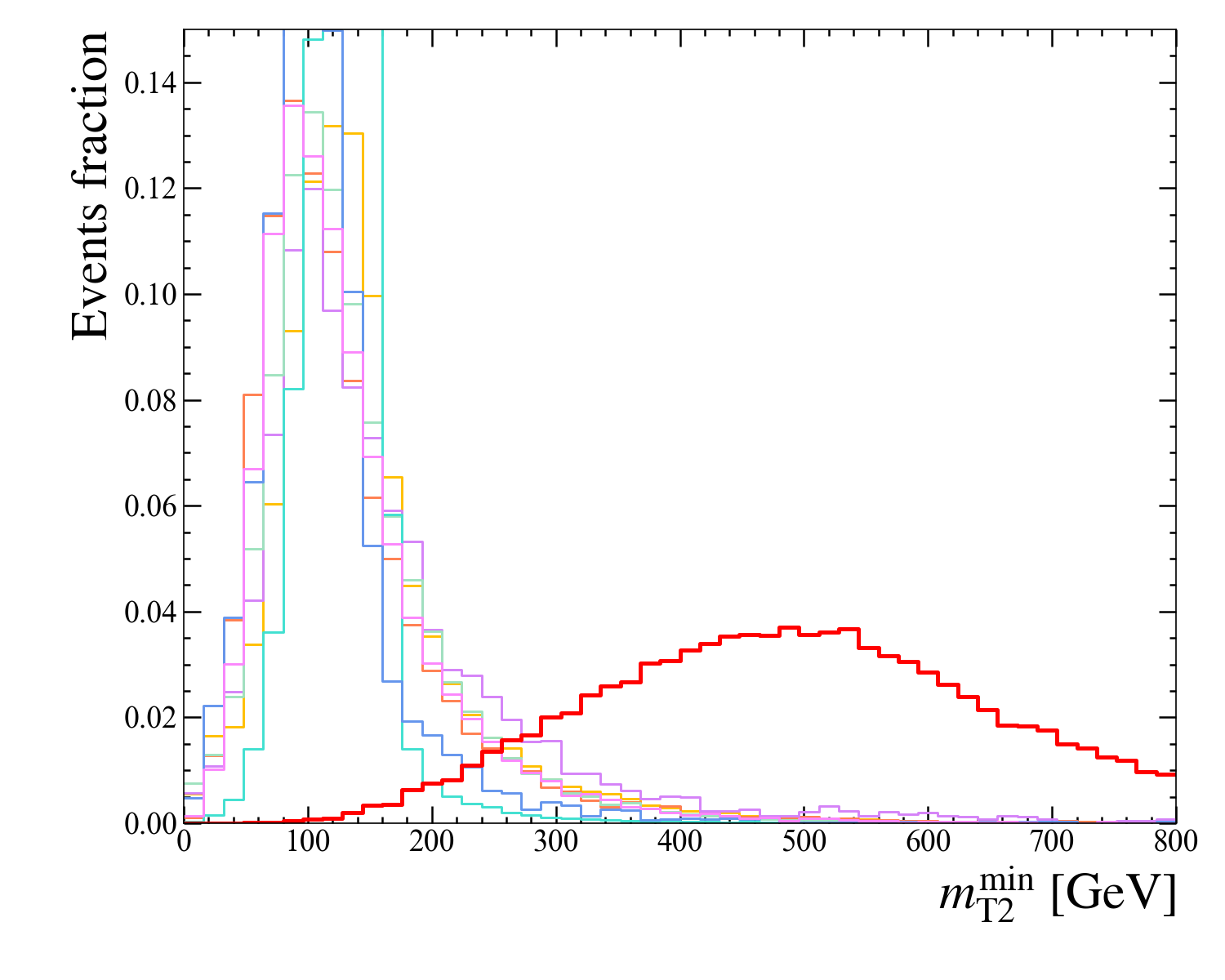}\\
 \vspace{-.4cm} 
	\caption{\label{fig:distri-ZH} The $E_{\rm T}^{\rm miss}$,  $m_{\ell \ell}$, $m_{bb}$, $\Delta R_{\ell \ell}$ and $m_{\rm T2}^{\rm min}$ distributions of the dominated SM backgrounds and the signal with $m_{\tilde{H}}=600~{\rm GeV}$ after the ZH channel preselection criteria. }
\end{figure}
The ZH channel targets the event containing exactly two $b$-jets and two leptons ($e$ or $\mu$) that can form one OSSF pair. The dominant backgrounds are $t\bar{t}$, $t\bar{t}X$, $bb\ell\ell$, VH process, $Z$+jets and di-bosons.

\par The search strategy in the ZH channel is to reconstruct $Z$ and $h$ with large $E_{\rm T}^{\rm miss}$. As summarized in Table~\ref{tab:cut-ZH} and based on the distributions of $E_{\rm T}^{\rm miss}$, $m_{\ell\ell}$, $\Delta R_{\ell\ell}$ and $m_{\rm T2}^{\rm min}$ in Fig.~\ref{fig:distri-ZH}, both the dominant background and the benchmark signal $\mu = 600~{\rm GeV}$ are selected with the following cuts:
\begin{itemize}
	\item The basic cut requires the signal event to contain two $b$-jets and an OSSF lepton pair. Events containing another jet or an additional lepton are rejected. 
	\item The $E_{\rm T}^{\rm miss}$ trigger. As shown in Fig.~\ref{fig:distri-ZH}, the background basically follows an exponentially decreasing distribution. Here $E_{\rm T}^{\rm miss}$ is required to be larger than $100~{\rm GeV}$.
	\item The $Z$ and $H$ mass peak cut. The invariant masses of the OSSF pair $m_{\ell\ell}$ and two $b$-jets of the signal event are required to located in the $Z$ peak window $80~{\rm GeV} \leq m_{\ell\ell} \leq 100~{\rm GeV}$ and in the Higgs window $90~{\rm GeV} \leq m_{bb} \leq 150~{\rm GeV}$, respectively. 
	\item The OSSF pair cone $\Delta R_{\ell\ell}$ cut. The $Z\to \ell\ell$ candidate of the signal event appears as a smaller cone compared with the SM backgrounds, so the cut $\Delta R_{\ell\ell} < 1.5$ greatly improves the statistical significance $Z_A$.    
	\item The $m_{\rm T2}$ cut. In order to further suppress the $t$-quark background, a cut of $m_{\rm T2}^{\rm min} \geq 175~{\rm GeV}$ is also applied in the ZH channel.  
	\end{itemize}
	After the cuts, the signal for $m_{\tilde{H}} = 600~{\rm GeV}$ and ${\rm BR}(\tilde{\chi}_1^0 \to Z \tilde{G}) = 50 \%$ has a statistical significance $Z_A = 2.94$.
\begin{table}[t]
\centering
\caption{\label{tab:cut-ZH} Similar to Table~\ref{tab:cut-HH}, but for ZH channel. The label `-' represent  that no event are retained after applying the corresponding cut. For the signal we choose a benchmark point $m_{\tilde{H}}=600~{\rm GeV}$, ${\rm BR}\left( \tilde{\chi}_1^0 \to Z \tilde{G} \right) = 0.5$, marked as S(600,0.5).}
\resizebox{\linewidth}{!}{
\begin{tabular}{lp{.1cm}|rp{.01cm}rp{.01cm}rp{.01cm}rp{.01cm}rp{.01cm}rp{.01cm}rp{.01cm}|rp{.01cm}|c}
\hline\hline
\multirow{2}{*}{\bf Cuts} && \multicolumn{13}{c}{\bf SM backgrounds} &&{\bf signal} && \multirow{2}{*}{$Z_A$}\\ 
       		&& $t\bar{t}$ 	&& $bb\ell\ell$ && di-boson && single-$t$ 	&& Vh 	&& $Z+$jets && $t\bar{t}X$   	&& S(600, 0.5) 			&&  \\ \hline
basic cut 	&& 11182850.0	&& 3550857.4	&& 59818.1	&& 1434990.2 && 21623.8   	&& 352670.1 && 100652.2 && 228.2  	&& 0.06      \\
$E_{\rm T}^{\rm miss} \geq 100~{\rm GeV}$          	&& 2538545.7		&& 6187.1	&& 1263.8 	&& 263070.7    && 1114.2   && 626.2  	&& 26227.6 && 214.7  && 0.13       \\
$80~{\rm GeV} \leq m_{\ell\ell} \leq 100~{\rm GeV}$ && 308938.3                         && 2237.9  && 179.2 && 29098.1   && 163.9    && 150.3  	&& 4199.9  && 198.0 && 0.34      \\
$90~{\rm GeV} \leq m_{bb} \leq 150~{\rm GeV}$   && 93895.0                         && 526.6	&& 29.1	&& 6536.2   && 102.0    	&& -  		&& 1094.5  && 153.0 && 0.48      \\
$\Delta R_{\ell\ell} < 1.5 $          	&& 3381.8                     && 394.9                            && -                              	&& 640.3    && 9.2     && -   && 	309.3 && 139.2  && 2.01        \\
$m_{\rm T2}^{\rm min} > 175~{\rm GeV}$ 				&& 1193.6  && 394.9  && -   && 334.1  && 2.3    	&& -   		&& 260.9  	&& 138.9 	&& 2.94        \\ \hline\hline
\end{tabular}}
\end{table}

\subsection{Searching strategy for ZZ channel}
\begin{figure}
	%\centering	
	 \makebox[\linewidth][c]{
 	\includegraphics[width=0.34\linewidth]{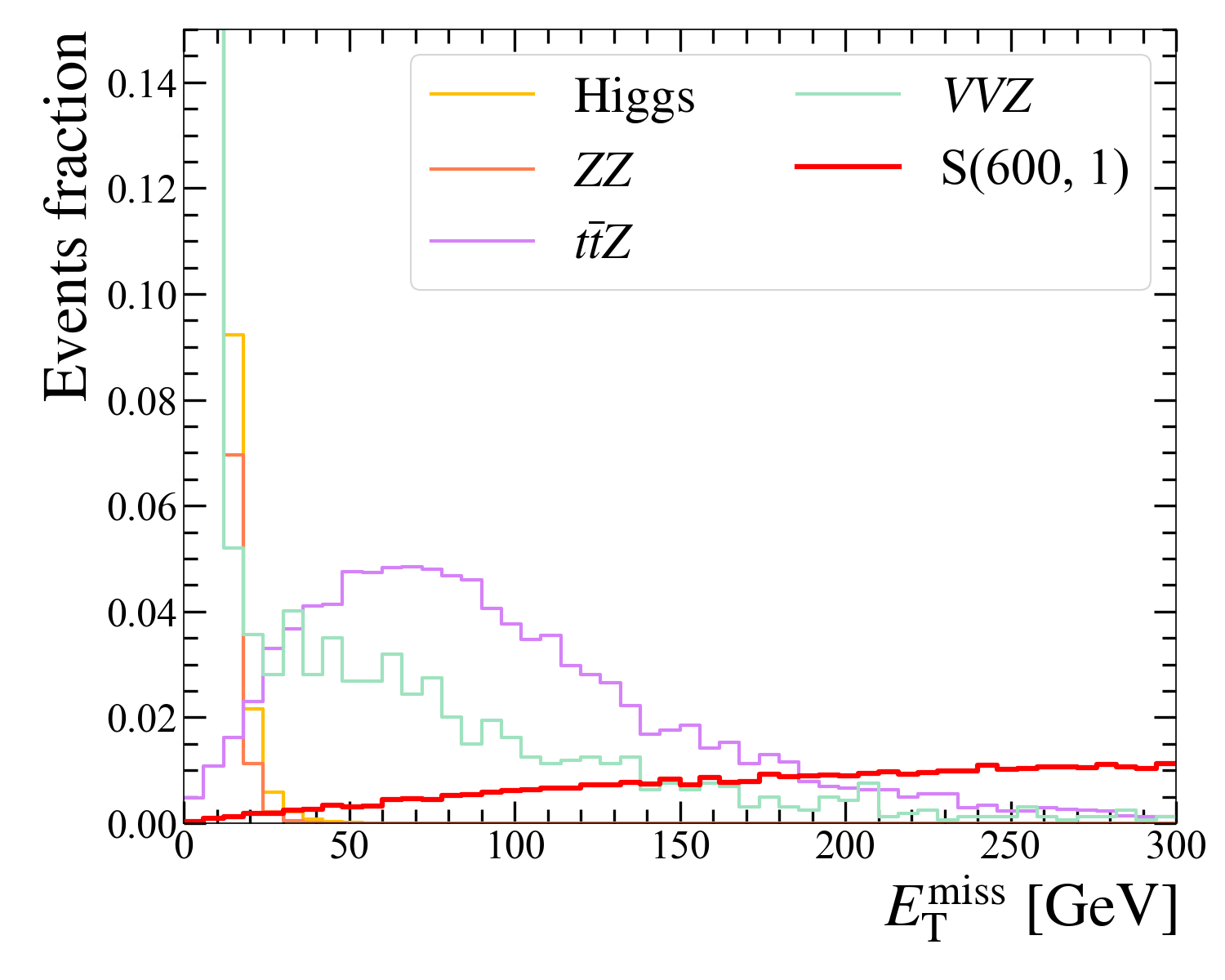}
	\includegraphics[width=0.34\linewidth]{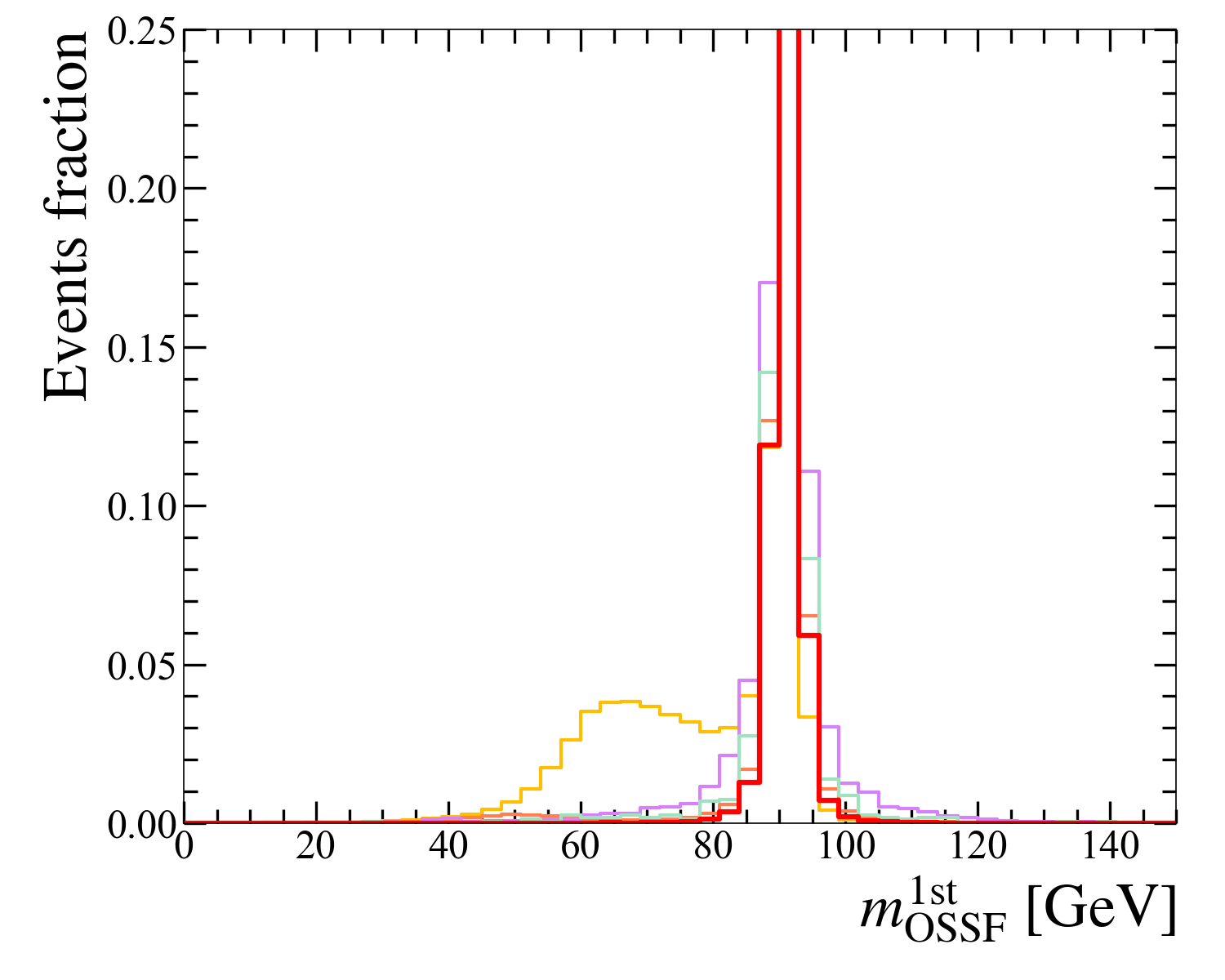}
	\includegraphics[width=0.34\linewidth]{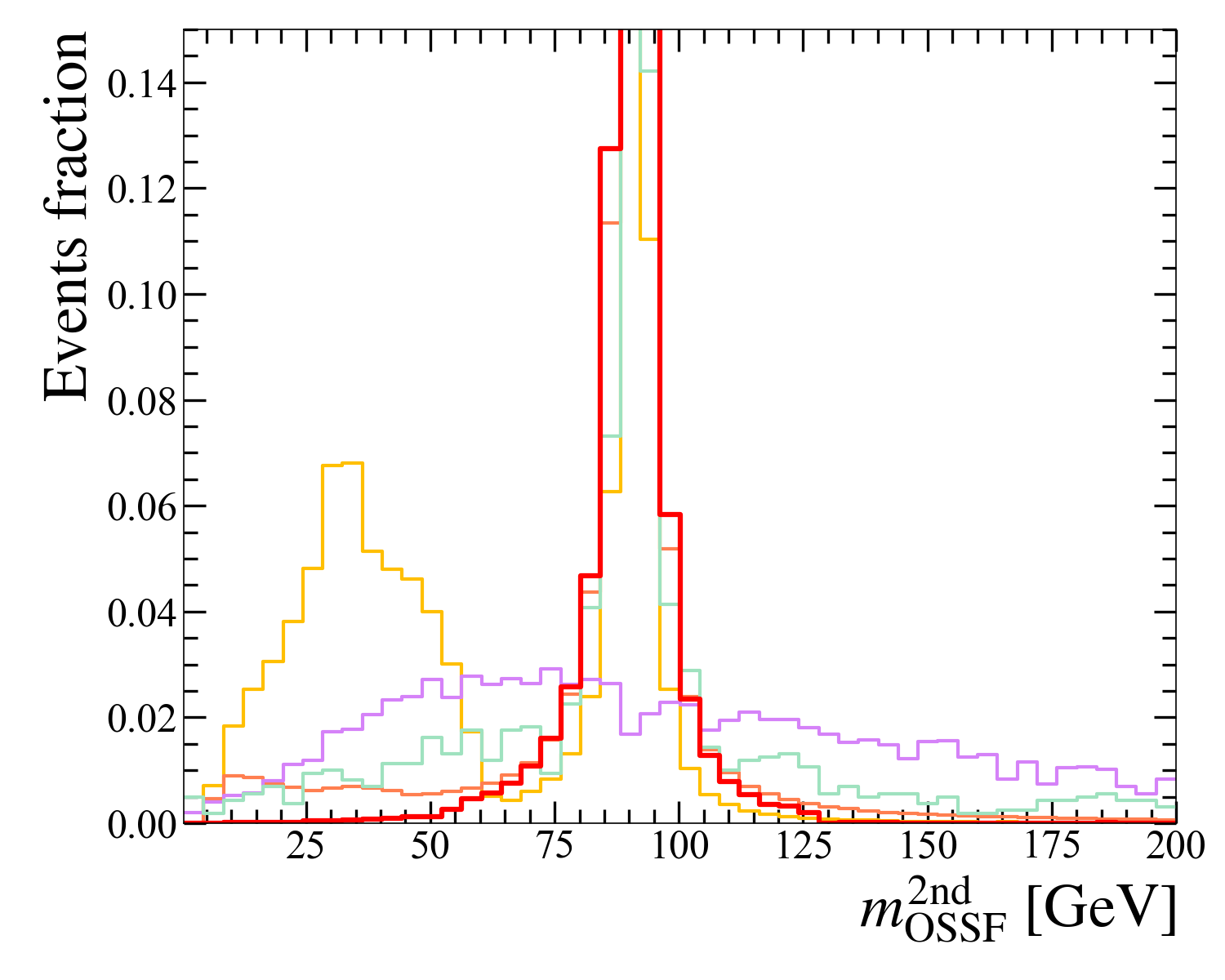}\\
  \vspace{-.4cm} 
	 }
	% \caption{\label{fig:distri-ZZ} The ``N-1'' distributions of the $m_{\rm OSSF}^{\rm 1st}$, $m_{\rm OSSF}^{\rm 2nd}$ and $E_{\rm T}^{\rm miss}$ variables after the SR requirements in ZZ channel. }	
 \caption{\label{fig:distri-ZZ} The $m_{\rm OSSF}^{\rm 1st}$, $m_{\rm OSSF}^{\rm 2nd}$ and $E_{\rm T}^{\rm miss}$  distributions of the dominated SM backgrounds and the signal with $m_{\tilde{H}}=600~{\rm GeV}$ after the ZZ channel preselection criteria. }	
\end{figure}
\begin{table}[t]
\centering
\caption{\label{tab:cut-ZZ}Similar to Table.~\ref{tab:cut-HH}, but for the ZZ channel. For the signal we choose a benchmark point $m_{\tilde{H}}=600~{\rm GeV}$, ${\rm BR}\left( \tilde{\chi}_1^0 \to Z \tilde{G} \right) = 1$, marked as S(600,1).}
\resizebox{\linewidth}{!}{
\begin{tabular}{lp{.1cm}|rp{.01cm}rp{.01cm}rp{.01cm}rp{.01cm}|rp{.01cm}|c}
\hline\hline
\multirow{2}{*}{\bf Cuts} && \multicolumn{7}{c}{\bf SM backgrounds} &&{\bf signal} && \multirow{2}{*}{$Z_A$}\\ 
       && $ZZ$ && $t\bar{t}Z$ &&  VVZ && Higgs && S(600,1) &&  \\ \hline
basic cut && 53942.8   	&& 388.3		&& 646.7	&& 516.5  	&& 197.2  	&& 0.84      \\
$80~{\rm GeV} \leq m_{\rm OSSF}^{\rm 1st} \leq 100~{\rm GeV}$		&& 51944.7                        && 356.8				&& 621.2	&& 355.8    	&& 196.2  	&& 0.85       \\
$70~{\rm GeV} \leq m_{\rm OSSF}^{\rm 2nd} \leq 110~{\rm GeV}$		&& 44124.0                         && 88.9                         && 374.3	&& 225.4    && 186.5  	&& 0.88       \\
$E_{\rm T}^{\rm miss} > 100~{\rm GeV}$ && -		&& 34.9		&& 40.9		&& -    && 176.3      && 15.92        \\
 \hline\hline
 \end{tabular}
}
\end{table}

The ZZ channel targets the final states containing two OSSF pairs of leptons in $Z$ peak plus large $E_{\rm T}^{\rm miss}$. The dominant irreducible backgrounds are the processes that can produce four charged leptons, including $ZZ$ (off-shell $Z/\gamma$ contributions are also included), $t\bar{t}Z$, $VVZ$ ($V=W/Z$), and the Higgs production via gluon-gluon fusion. 

\par As shown in Fig.~\ref{fig:distri-ZZ} and summarized in Table~\ref{tab:cut-ZZ}, for this channel we apply the following cuts sequentially:
\begin{itemize}
	\item The basic cut requires the signal event to have exactly two light flavor ($e$ or $\mu$) OSSF lepton pairs. Events with additional signal lepton candidate and any $b$-jet are discarded. 
	\item The OSSF pair with mass closer to the $Z$-boson mass is labeled as the first $Z$ candidate, while the other OSSF pair is labeled as the second $Z$ candidate. The first (second) $Z$ candidate must have an invariant mass $m_{\rm OSSF}^{\rm 1st}$ ($m_{\rm OSSF}^{\rm 2nd}$) in the range of $80-100~{\rm GeV}$ ($70-110~{\rm GeV}$). As shown in  Fig.~\ref{fig:distri-ZZ}, the narrower peak of the first $Z$ candidate is due to the ordering of the $Z$ candidates, so that widening the range of the $m_{\rm OSSF}^{\rm 2nd}$ window increases the signal acceptance and the statistical significance. 
	\item The final cut requires $E_{\rm T}^{\rm miss} \geq 100~{\rm GeV}$. Note that the distribution of $E_{\rm T}^{\rm miss}$ depends on the value of $\mu$. Although for parameter points with a relatively large  $\mu$ value, e.g. the benchmark point in Fig.~\ref{fig:distri-ZZ} with $\mu = 600~{\rm GeV}$, it seems to be possible to take a stronger cut on $E_{\rm T}^{\rm miss}$ (e.g. $E_{\rm T}^{\rm miss} \geq 150~{\rm GeV}$) to suppress backgrounds more efficiently, a stronger cut on $E_{\rm T}^{\rm miss}$ would hurt signal seriously for a lower value of $\mu$. So we take a universal soft cut here for simplicity. 
\end{itemize}
After the selection criterion, the detection sensitivity to ${\rm BR}\left(\tilde{\chi}_1^0 \to Z \tilde{G} \right)$ of $m_{\tilde{H}}=600~{\rm GeV}$ can reach to about $0.3$.

\subsection{Searching strategy for WH channel}
\begin{figure}
	\centering
	% \includegraphics[width=0.98\linewidth]{WH.png}
 % \vspace{-.8cm}
	% \caption{\label{fig:distri-WH} The ``N-1'' plots of the used variables after the \texttt{SR-WH-High} selection criteria in WH channel.}
  	\includegraphics[width=0.48\linewidth]{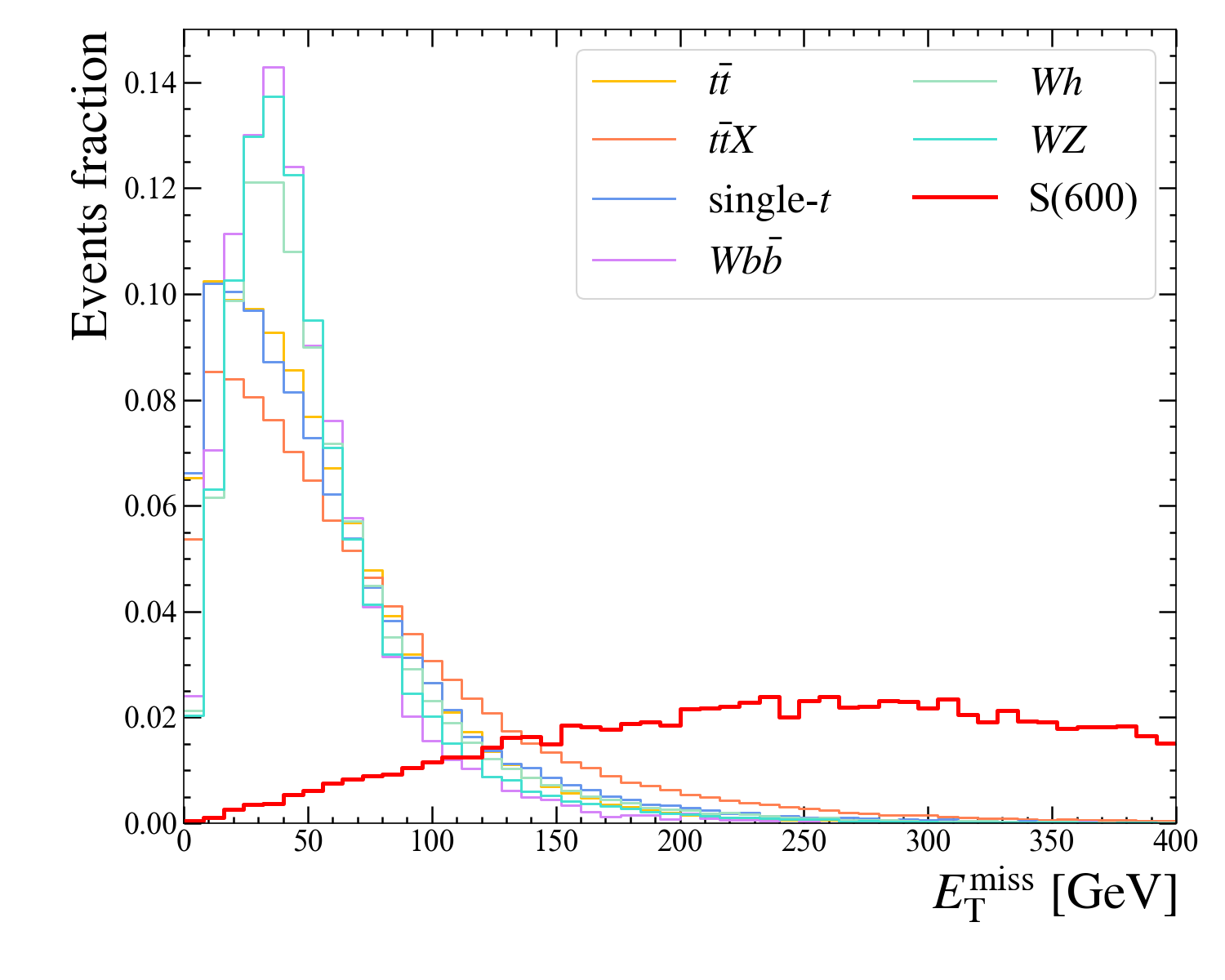}
	\includegraphics[width=0.48\linewidth]{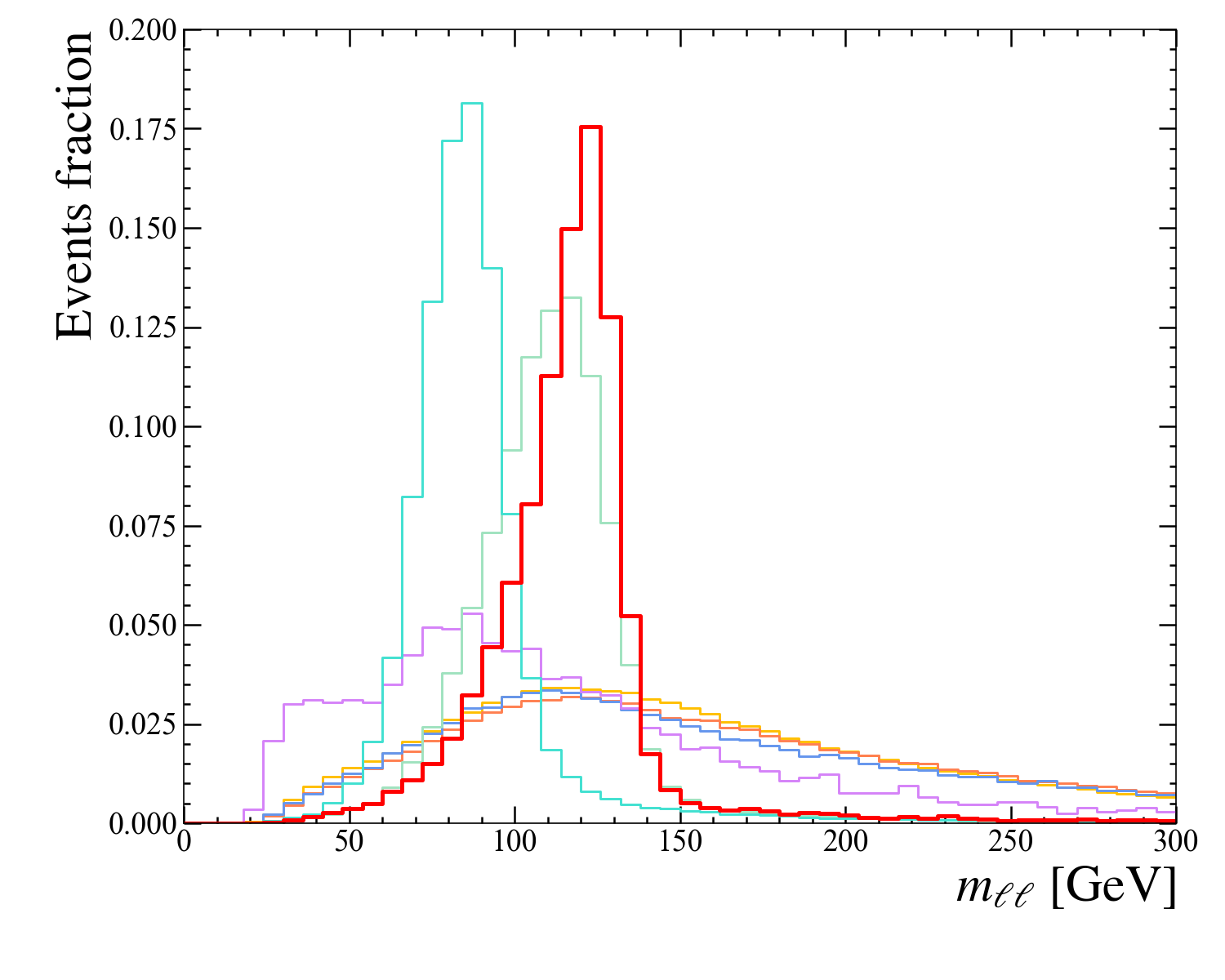}\\
	\includegraphics[width=0.48\linewidth]{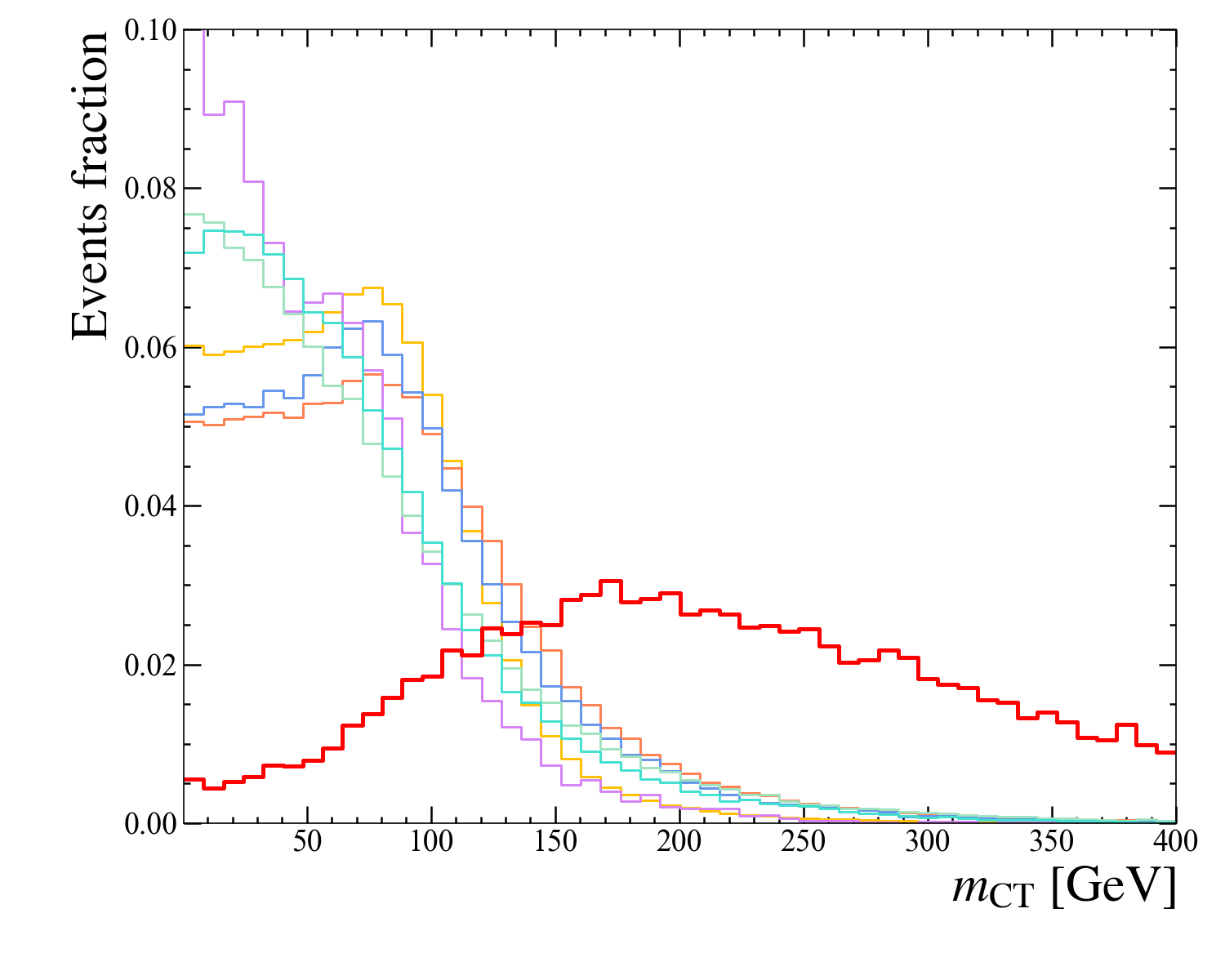}
 	\includegraphics[width=0.48\linewidth]{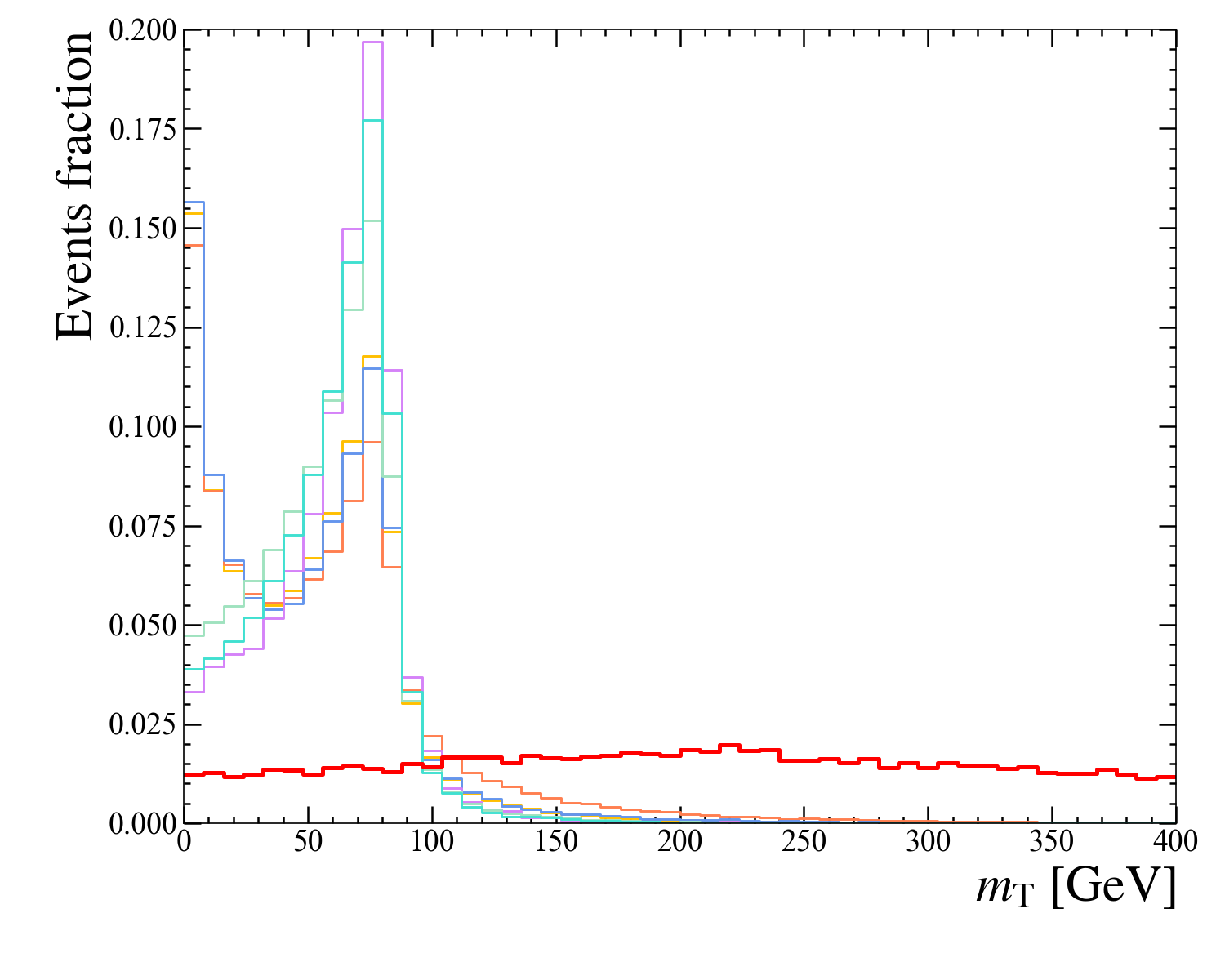}\\
  \vspace{-.4cm} 
 \caption{\label{fig:distri-WH}  The $E_{\rm T}^{\rm miss}$, $m_{bb}$, $m_{\rm CT}$ and $m_{\rm T}$  distributions of the dominated SM backgrounds and the signal with $m_{\tilde{H}}=600~{\rm GeV}$ after the WH channel preselection criteria. }	
\end{figure}
\begin{table}[t]
\centering
\caption{\label{tab:cut-WH} Similar to Table~\ref{tab:cut-HH}, but for WH channel. For the signal we choose a benchmark point $m_{\tilde{H}}=600~{\rm GeV}$, marked as S(600).}
\resizebox{\linewidth}{!}{
\begin{tabular}{lp{.1cm}|rp{.01cm}rp{.01cm}rp{.01cm}rp{.01cm}rp{.01cm}rp{.01cm}|rp{.01cm}|c}
\hline\hline
\multirow{2}{*}{\bf Cuts} && \multicolumn{11}{c}{\bf SM backgrounds} &&{\bf signal} && \multirow{2}{*}{$Z_A$}\\ 
       		&& $t\bar{t}$ 	&& $Wb\bar{b}$ && single-$t$ && $WZ$ 	&& $Wh$ 	&&  $t\bar{t}X$   	&& S(600) 			&&  \\ \hline
basic cut 	  && 49194619.2	&&2941776.0 	&&3096878.4 	&&231115.7    	&&113865.1    && 43576.2   &&641.7    	&& 0.09\\
$E_{\rm T}^{\rm miss} \geq 100~{\rm GeV}$     && 6799978.6          && 191769.0     &&429871.2     && 19070.6    	&&12726.4 	&&14647.9       && 595.3   &&0.22    \\
$90~{\rm GeV} \leq m_{bb} \leq 140~{\rm GeV}$   && 1841118.6          &&37563.0      &&105085.2     &&5634.8     	&& 10413.6	&&3410.2       &&497.6   &&0.35    \\
$m_{\rm CT} > 200~{\rm GeV}$ 	   &&35898.0           &&5931.0      &&  7273.2   &&1326.3     	&&2834.7 	&&161.9       && 310.9   &&  1.34    \\
$m_{\rm T} > 150~{\rm GeV}$ 	   && 3141.0          &&395.4      &&250.8     && 35.3    	&& 35.3	&&48.9       && 223.2   &&3.54      \\ \hline\hline
\end{tabular}
}
\end{table}

In order to search for the chargino-neutralino productions shown in Fig.~\ref{fig:feyn-diam-ty}(d), the searching strategy targets the $W$-boson decay to lepton and the Higgs boson decay to $b\bar{b}$. Then the signal event requires a single isolated electron or muon, two $b$-jets and large $E_{\rm T}^{\rm miss}$. The major backgrounds arise from SM processes containing top quarks and $W$-bosons. As shown in Fig.~\ref{fig:distri-WH}, the kinematic variables $E_{\rm T}^{\rm miss}$, the invariant mass of the two $b$-jets $m_{bb}$, the cotransverse mass variable $m_{\rm CT}$ and the $m_{\rm T}$ variable of lepton $p_{\rm T}$ show great difference between the backgrounds and signals. The \texttt{SR-WH} is defined to suppress the SM backgrounds via the following cut strategies:
\begin{itemize}
	\item The basic cut requires exactly one light flavor lepton and two $b$-jets. The primary SM processes that contribute to the preselection region are $t\bar{t}$, single top quark (mostly in the $tW$ channel), $W$+jets and $WZ$ production. Events with additional leptons or jets are rejected. 
	\item The cut $E_{\rm T}^{\rm miss} \geq 100~{\rm GeV}$ can suppress all SM backgrounds by about one order of magnitude after the basic cut. 
	\item The invariant mass of the two $b$-tagged jets is required in the range of  $90~{\rm GeV} \leq m_{bb} \leq 140~{\rm GeV}$, consistent with the mass of the SM Higgs boson.
	\item The angular distance of the two $b$-jets $\Delta R_{bb}$ is required to be smaller than 1.5, in order to tag a SM Higgs fat cone. 
	\item The cotransverse mass variable $m_{\rm CT}$ of the $b$-jet pair is defined as 
		\begin{equation}
			m_{\rm CT} = \sqrt{2 p_{\rm T}^{b1} p_{\rm T}^{b2}\left( 1+ \cos{\Delta \phi_{bb}} \right) }, 
		\end{equation}
	where $p_{\rm T}^{b1}$ and $p_{\rm T}^{b2}$ are the magnitudes of the transverse momenta of the two $b$-jets and $\Delta \phi_{bb}$ is the azimuth angle between the two $b$-jets~\cite{Tovey:2008ui}. For the $t\bar{t}$ background, the $m_{\rm CT}$ variable has a kinematic endpoint of about 150 GeV. \texttt{SR-WH} requires a harder cut $m_{\rm CT} > 200~{\rm GeV}$ to effectively suppress the $t\bar{t}$ and $tW$ backgrounds.
	\item The SM processes with one $W$-boson decaying to leptons, originating primarily from the semi-leptonic $t\bar{t}$ and $W$+jets background, are suppressed by requiring the transverse mass $m_{\rm T}$ to be greater than 150 GeV in \texttt{SR-WH}.
	\end{itemize}
As shown in Table~\ref{tab:cut-WH}, for the signal with $\mu = 600~{\rm GeV}$, the statistical significance $Z_{A}$ can reach to about $3.5\sigma$ in \texttt{SR-WH}. 

\subsection{Searching strategy for WZ channel}
\begin{figure}
	\centering
   	\includegraphics[width=0.48\linewidth]{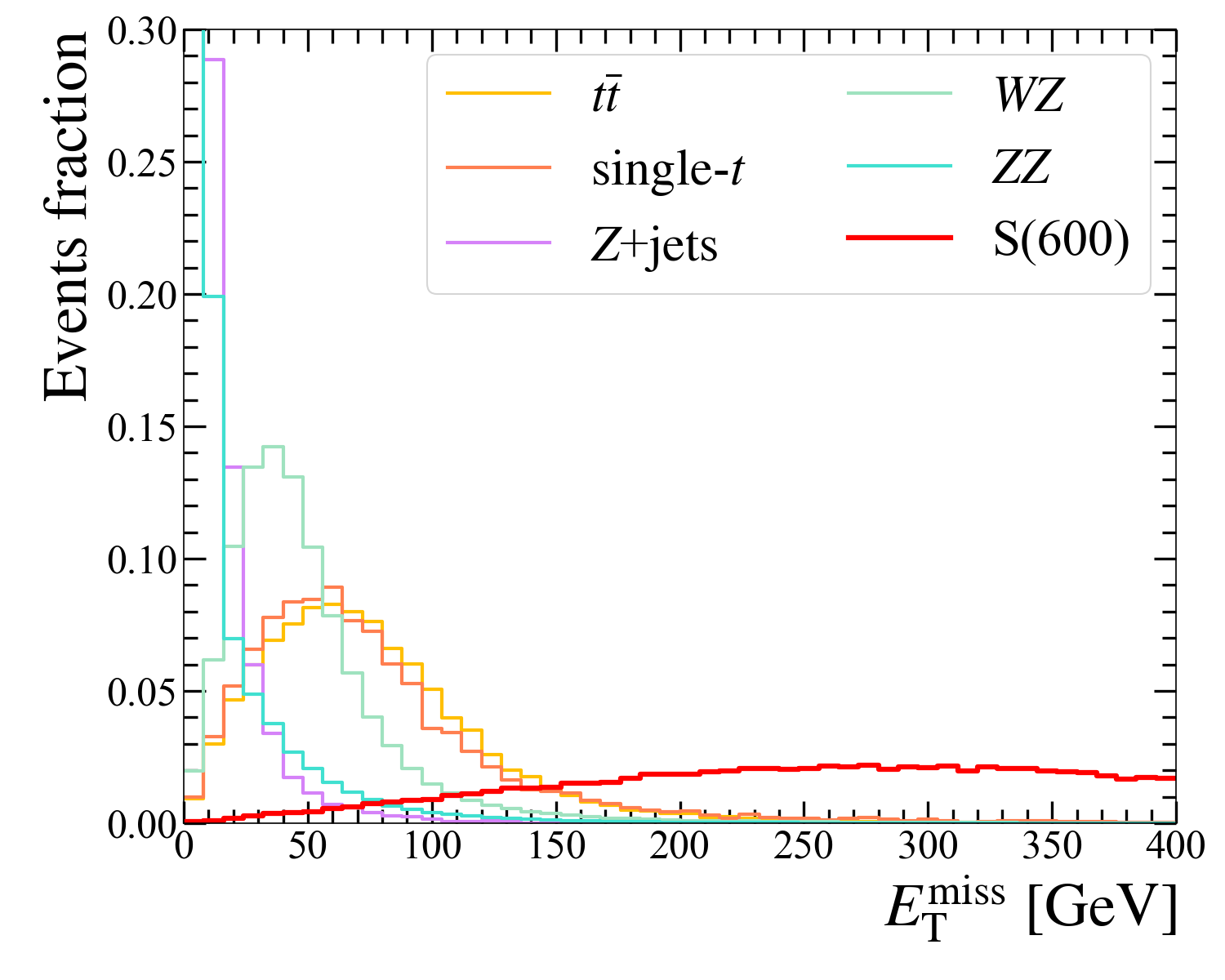}
	\includegraphics[width=0.48\linewidth]{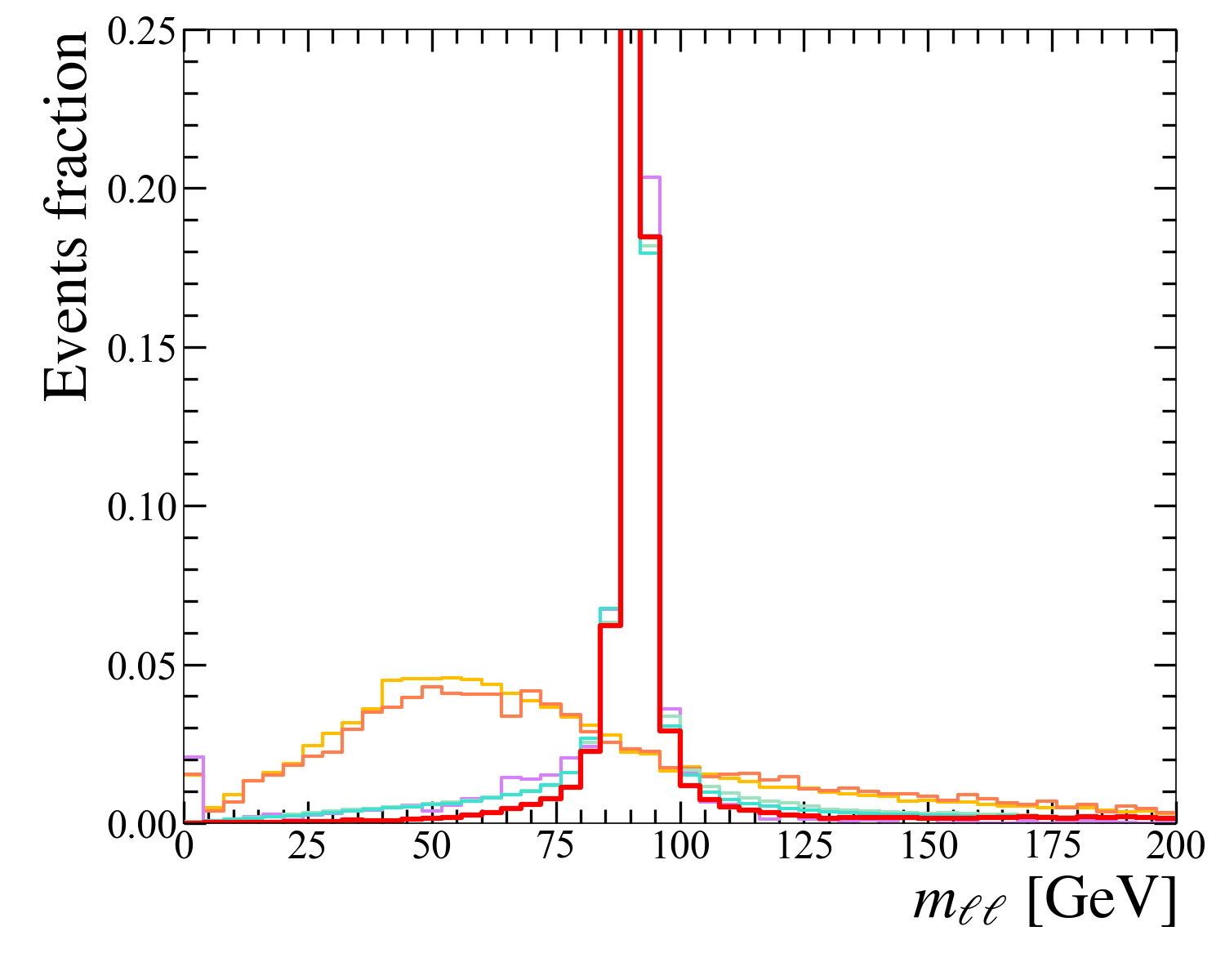}\\
	\includegraphics[width=0.48\linewidth]{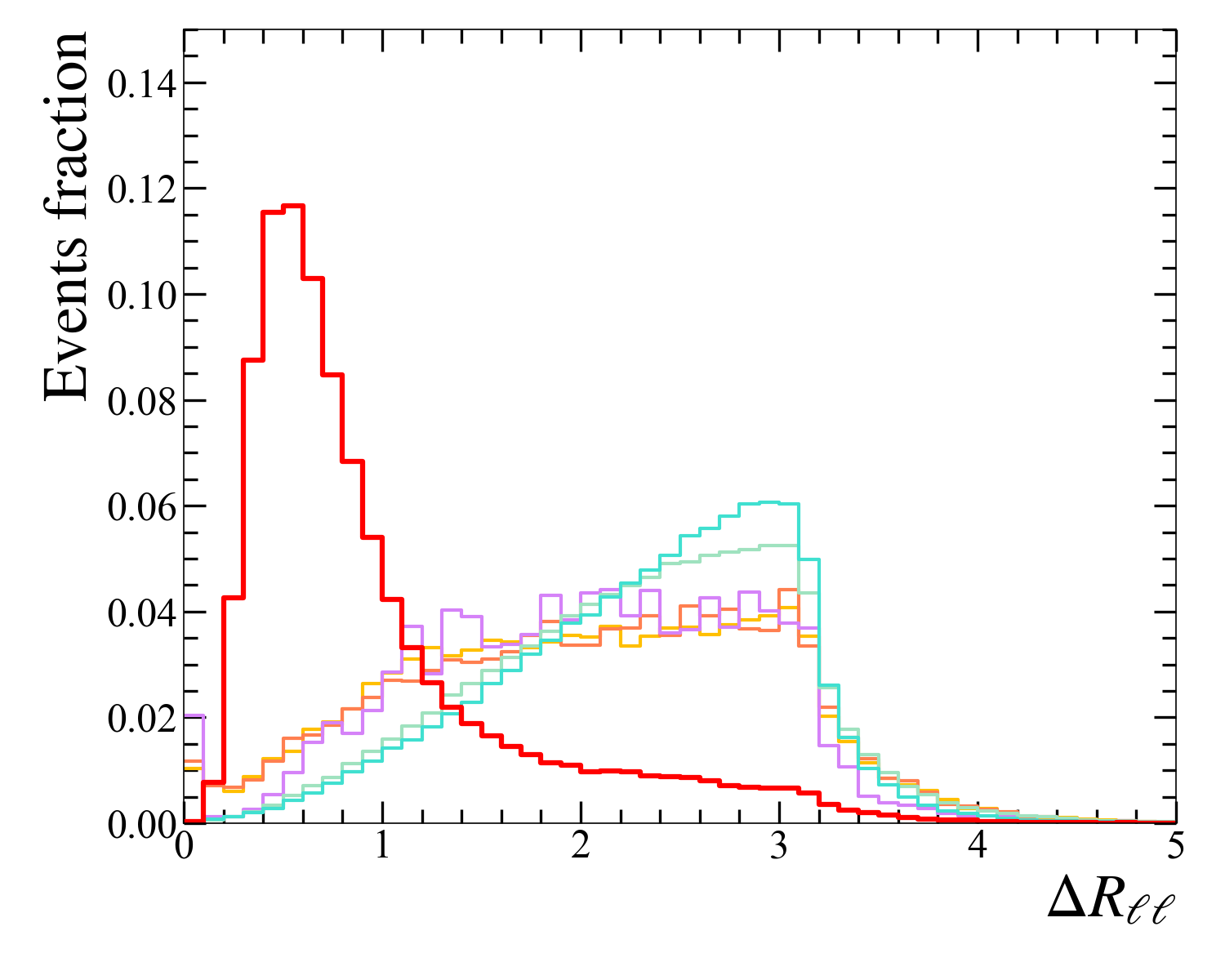}
 	\includegraphics[width=0.48\linewidth]{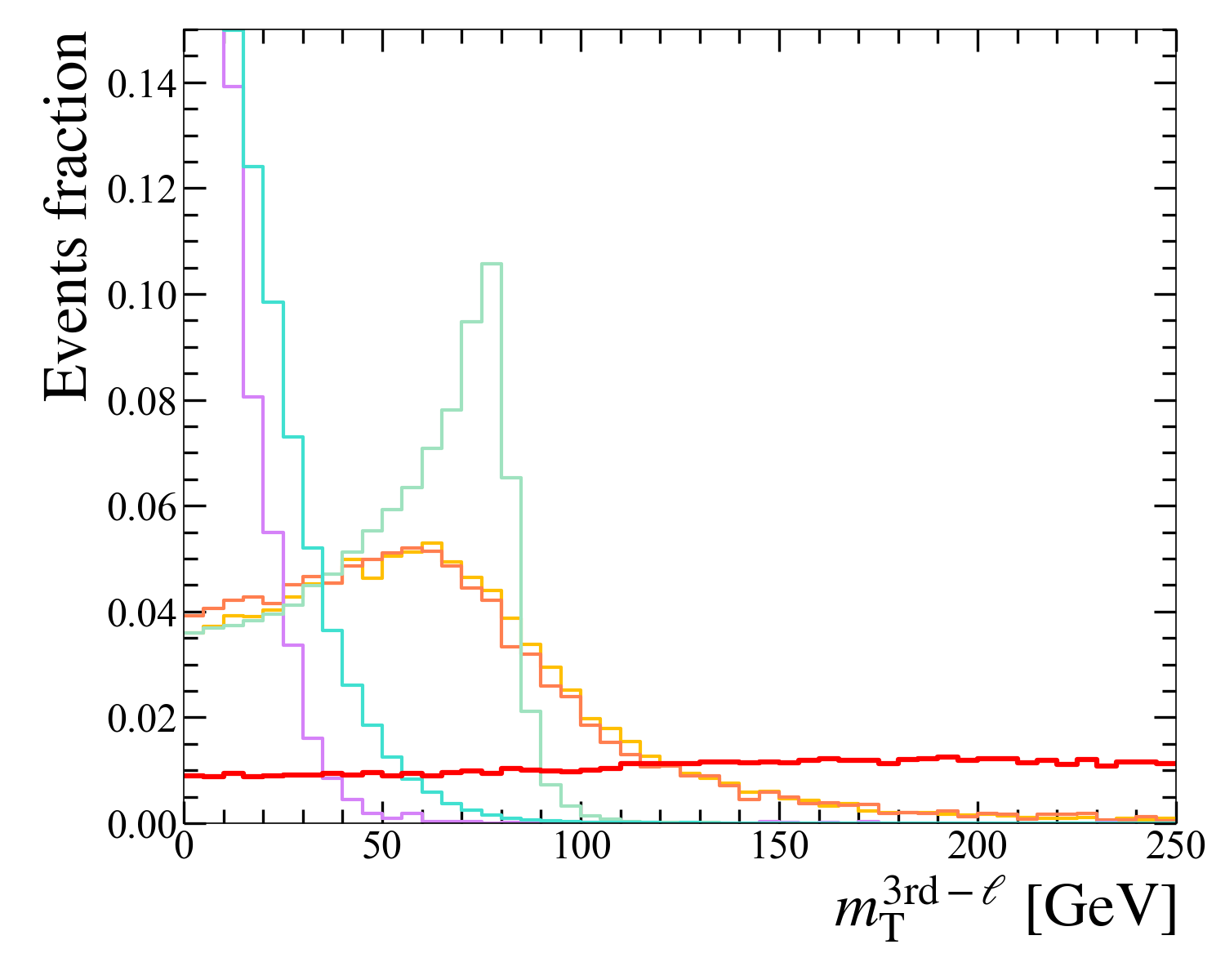}\\
  	\vspace{-.4cm} 
 	\caption{\label{fig:distri-WZ}  The $E_{\rm T}^{\rm miss}$, $m_{\ell \ell}$, $\Delta R_{\ell \ell}$ and $m_{\rm T}^{\rm 3rd-\ell}$  distributions of the dominated SM backgrounds and the signal with $m_{\tilde{H}}=600~{\rm GeV}$ after the WZ channel preselection criteria. }	
\end{figure}

\begin{table}[t]
\centering
\caption{\label{tab:cut-WZ} Similar to Table~\ref{tab:cut-HH}, but for WZ channel. The label `-' represent  that no event are retained after applying the corresponding cut. For the signal we choose a benchmark point $m_{\tilde{H}}=600~{\rm GeV}$, marked as S(600).}
\resizebox{\linewidth}{!}{
\begin{tabular}{lp{.1cm}|rp{.01cm}rp{.01cm}rp{.01cm}rp{.01cm}rp{.01cm}|rp{.01cm}|c}
\hline\hline
\multirow{2}{*}{\bf Cuts} && \multicolumn{9}{c}{\bf SM backgrounds} &&{\bf signal} && \multirow{2}{*}{$Z_A$}\\ 
       		&& $t\bar{t}$ 	&& $Z+$jets && single-$t$ && $WZ$ 	&& $ZZ$ 	   	&& S(600) 			&&  \\ \hline
basic cut 	&& 1436224.9		&& 5193973.3	&& 576197.9	&& 239646.7  	&& 65487.9 && 428.5  	&& 0.16      \\
$E_{\rm T}^{\rm miss} \geq 100~{\rm GeV}$      	&& 340170.3		&& 40790.9	&& 39442.4 	&& 53759.0    && 1385.0     	&& 403.7 		&& 0.59       \\
$80~{\rm GeV} \leq m_{\ell\ell} \leq 100~{\rm GeV}$   && 38244.3                         && 13378.2	&& 34625.4	&& 4816.3      	&& 1225.0  		&& 312.2 && 1.03      \\
$\Delta R_{\ell\ell} < 1.5$ 				&& 4426.2    && 5811.5   && 19423.0  && 612.5    	&& 824.7   		&& 295.3 	&& 1.67        \\
$m_{\rm T}^{\rm 3rd-\ell} > 110~{\rm GeV}$ 				&& 348.1    && -   && 19.7  && 111.4    	&& 1.0    	&& 235.4 	&& 10.0        \\ \hline\hline
\end{tabular}
}
\end{table}

For this channel the final state with three leptons is known as the golden channel for electroweakinos. Since in our study we tend to make a conservative evaluation for the exploration of the simplified model depicted as in Fig.~\ref{fig:feyn-diam-ty}~(e), we use  
a common approach throughout most steps of our analyses as summarized in Table~\ref{tab:cut-WZ}:
\begin{itemize}
	\item The basic cut requires exactly three leptons that can form an OSSF pair. In the cases of three leptons with same flavor, among the two combinations of the OSSF lepton pair, the combination with a smaller $m_{\rm T}$ corresponding to the third lepton is selected to reconstruct the event. Events containing any $b$-jet are discarded. 
	\item $E_{\rm T}^{\rm miss}$ is required to be greater than 100 GeV.
	\item In order to suppress the non-resonance background, such as $t\bar{t}$, the invariant mass of the OSSF lepton $m_{\ell\ell}$ is required to be in the $Z$-boson mass window $[80, 100]~{\rm GeV}$. 
	\item The angular distance between the two leptons in the OSSF lepton pair $\Delta R_{\ell\ell}$ is required to be smaller than $1.5$ to search for a boosted $Z$-boson from a heavier higgsino decay. 
	\item As shown in Fig.~\ref{fig:distri-WZ}, the $m_{\rm T}$ variable of the third lepton for the SM background has a Jacobian peak with a sharp cut-off at the $W$-boson mass, while the signals have relatively flat distributions. Therefore, the cut $m_{\rm T}^{\rm 3rd-\ell}$ being greater than $110~{\rm GeV}$ can greatly improve the statistical significance.
\end{itemize}
For $m_{\tilde{H}}=600~{\rm GeV}$, the statistical significance via this WZ channel can reach to about $10\sigma$. 

\section{\label{sec:4}Results and discussions}
\begin{figure}[th]
\centering
\subfigure[Exclusion limits for the simplified model A]{\includegraphics[width=0.49\linewidth]{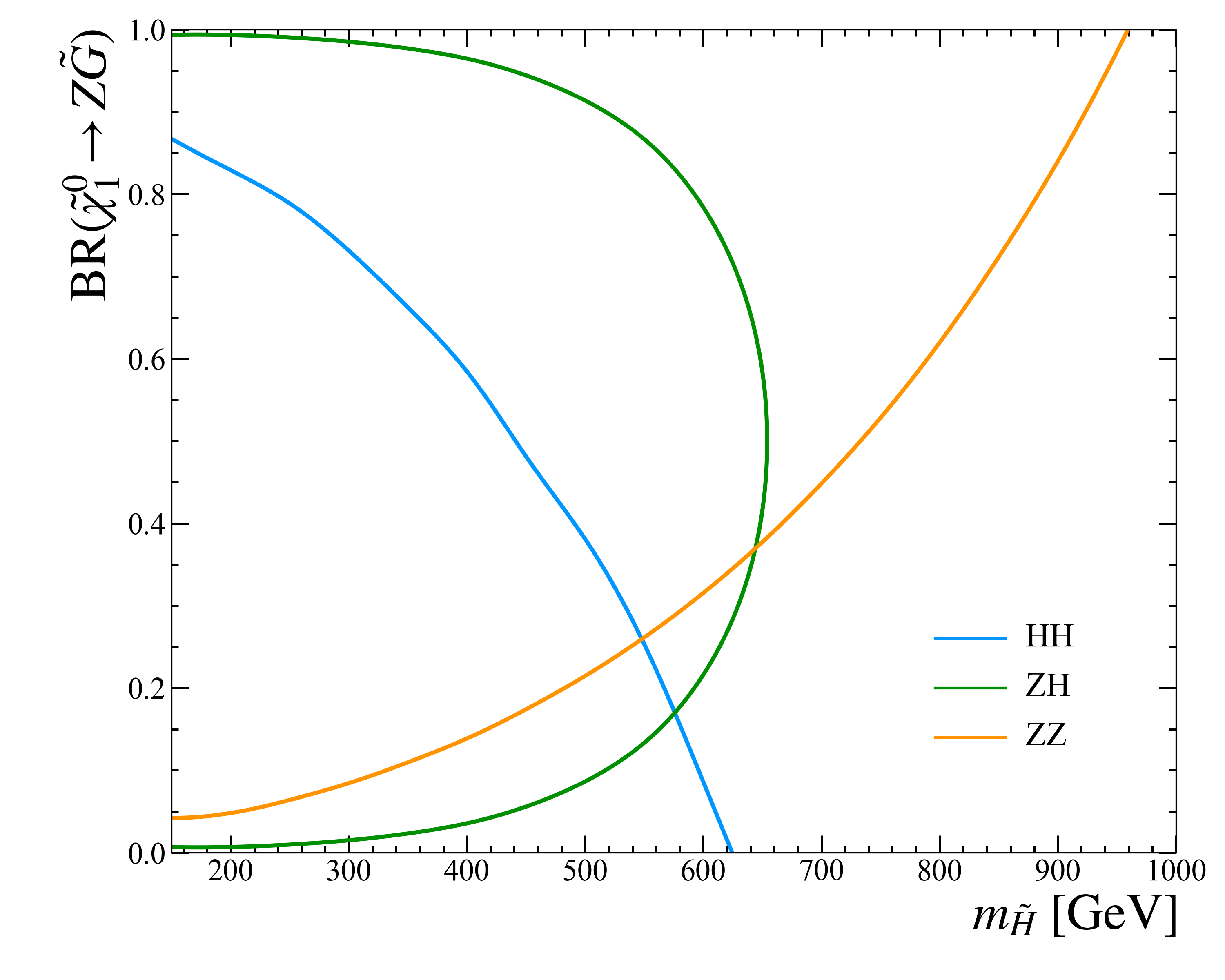}}
\subfigure[Exclusion limits for the simplified model B]{\includegraphics[width=0.49\linewidth]{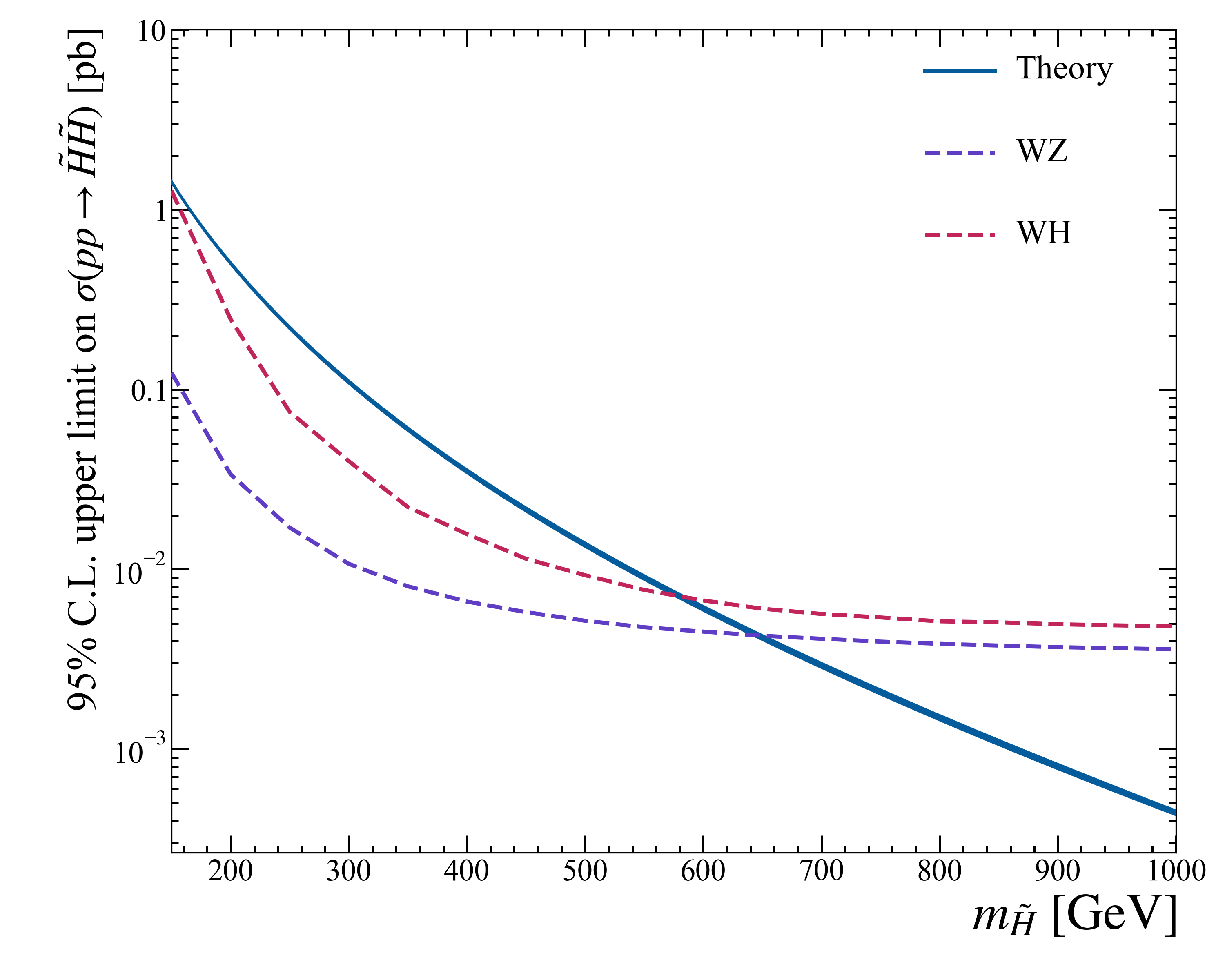}}
\caption{\label{fig:excl} Model dependent exclusion contour at $95\%$ confidence level on the higgsino with light gravitino scenario at HL-LHC. (a) Exclusion limit for the first simplified model in Sec.~\ref{sec:smA}, as a function of the lightest higgsino mass $m_{\tilde{H}}$ and the branching ratio ${\rm BR}\left( \tilde{\chi}_1^0 \to Z \tilde{G} \right) = 1 - {\rm BR}\left( \tilde{\chi}_1^0 \to H \tilde{G} \right)$. (b) The exclusion limit for the second simplified model in Sec.~\ref{sec:smB}, as a function of the signal cross sections that reproduce the expected excesses in each SR and the higgsino mass $m_{\tilde{H}}$. }
\end{figure}

All the SRs defined in this work and the corresponding SM backgrounds are summarized in Table ~\ref{tab:SRs}. 
The first three SRs, i.e., \texttt{SR-HH}, \texttt{SR-ZH} and \texttt{SR-ZZ},  target the first simplified model described in Sec.~\ref{sec:smA}, and the corresponding exclusion limits are shown in Fig.~\ref{fig:excl}~(a). The last two SRs, i.e., \texttt{SR-WH} and \texttt{SR-WZ},  target the simplified model described in Sec.~\ref{sec:smB} and the exclusion limits are shown in Fig.~\ref{fig:excl}~(b).

\par From Fig.~\ref{fig:excl}~(a), one can find that the \texttt{SR-HH}, \texttt{SR-ZH} and \texttt{SR-ZZ} channels are complementary to each other for probing a higgsino pair, and the parameter space with the higgsino mass $m_{\tilde H}$ less than $575~{\rm GeV}$ can be fully covered regardless of the branching ratio. The exclusion potential of the higgsino mass $m_{\tilde{H}}$ can reach about $960~{\rm GeV}$ using the ZZ channel with the assumption that ${\rm BR}( \tilde{\chi}_1^0 \to Z \tilde{G} ) = 1$.And in the situation that ${\rm BR}( \tilde{\chi}_1^0 \to h \tilde{G} ) = 1$, the HH channel can push $m_{\tilde{H}}$ to about $620~{\rm GeV}$. Searching di-Higgs in $bb\tau\tau$ plus $E_{\rm T}^{\rm miss}$ final state can distinguish the signal from $t\bar{t}$ background well. Meanwhile, the ZH and ZZ channel can reach the branching ratio of a higgsino into $h\tilde{G}$ to about 5\% and 1\% correspondingly.  

\par From Fig.~\ref{fig:excl}~(b) we see that the exclusion limits from the WH and WZ channel can reach to about 550 and 650 GeV, respectively. The statistical sensitivity of the WZ channel is better than the WH channel.  

\par Roughly speaking, the detection limits of the channels containing Higgs bosons are a little weaker than the channels containing $Z$-bosons. The Higgs tagging techniques, as proposed in recent works~\cite{Shen:2023ofd, Khosa:2021cyk, Guo:2020vvt, Tannenwald:2020mhq, Datta:2019ndh, Lin:2018cin, Moreno:2019neq, CMS:2020poo, Chung:2020ysf} using the machine learning methods, can improve the detection ability of the WH, HH and ZH channels. 

\begin{table}[t]
\centering
\caption{\label{tab:SRs} Selection criteria and the event numbers of all SM backgrounds simulated in signal regions (SRs). The variables have mass-dimension in $\rm GeV$. The "-" means that the corresponding background is negligible. }
\resizebox{\linewidth}{!}{
\begin{tabular}{lc|p{3cm}<{\centering}p{.01cm}p{3cm}<{\centering}p{.01cm}p{3cm}<{\centering}p{.01cm}p{3cm}<{\centering}p{.01cm} p{3cm}<{\centering}p{.01cm}}
\hline\hline 
SR &&  \texttt{SR-HH} && \texttt{SR-ZH} && \texttt{SR-ZZ}&& \texttt{SR-WH}&& \texttt{SR-WZ} \\ \hline
 $N_{b\text{-jets}}$ 				&& $=2$ && $=2$ && $=0$ && $=2$  && $=0$	\\
 $N_{\text{non-}b\text{-jets}}$ 	&& $=0$ && $=0$ && $=0$ && $\leq 2$  && $=0$\\
 $N_{\tau_{\rm had}}$ 				&& $=2$ && $=0$ && $=0$ && $=0$ && $=0$	\\
 $N_{\ell}$ 							&& $=0$ && $=2$ && $=4$ && $=1$ && $=3$	\\
 $N_{\rm OSSF}$						&& $=0$ && $=1$	&& $=2$ && $=0$ && $=1$	\\
 $E_{\rm T}^{\rm miss}$ 				&& $\geq 100$	&& $\geq 100$	&& $\geq 100$	&& $\geq 100$	&& $\geq 100$	\\ \hline
 \begin{tabular}[c]{@{}l@{}} 
	Selection \\ 
	Criteria  
 \end{tabular}
&& \begin{tabular}[l]{@{}l@{}} 
 		$m_{bb} ~\in [90, 140]$ \\
 		$\Delta R_{bb} < 2$ \\
 		$m_{\rm T2}^{\rm min} > 175$ 
 	\end{tabular}
&& \begin{tabular}[c]{@{}c@{}} 
 		$m_{\ell\ell} \in [80, 100]$ \\
 		$m_{bb} \in [90, 150]$ \\ 
 		$\Delta R_{bb} < 1.5$ \\ 
 		$m_{\rm T2}^{\rm min} > 175$  
 	\end{tabular}
&& \begin{tabular}[c]{@{}c@{}} 
 		$m_{\rm OSSF}^{\rm 1st} \in [80, 100]$\\ 
 		$m_{\rm OSSF}^{\rm 2nd} \in [70, 110]$\\ 
 	\end{tabular} 
&& \begin{tabular}[c]{@{}c@{}} 
 		$m_{bb} \in [90, 150]$ \\ 
  		$m_{\rm CT} > 200$\\ 
  		$m_{\rm T} > 150$\\ 
 	\end{tabular} 
&& \begin{tabular}[c]{@{}c@{}} 
 		$m_{\ell\ell} \in [80, 100]$\\ 
 		$\Delta R_{\ell\ell} < 1.5$\\ 
 		$m_{\rm T}^{{\rm 3rd-\ell}} > 110$\\ 
 	\end{tabular} 
\\
\hline
SM Total 	&& 9153.0 	&& 2185.8 	&& 75.8	&&4028.2  	&&480.2 \\ 
\hline
$t\bar{t}$	&& 5172.2 	&& 1193.6 	&&- 	&&3141.0 	&&348.1 \\
single-$t$	&& 1197.1  	&& ~334.1	&&- 	&& 250.8	&&19.7 \\
$t\bar{t}X$	&& ~347.2  	&& ~260.9 	&&34.9 	&&48.9 	 && -\\
$Z+$jets	&& 2120.3 	&& - 	&& - 	&& - 	&& -\\
$Wb\bar{b}$ && - 	&& - 	&& -	&&395.4 	&&- \\
di-boson	&& ~141.1  	&& -	&&- 	&&35.3 	&&112.4 \\
$VVZ$		&& - 	&& - 	&&40.9 	&&- 	&&-  \\
$Vh$		&& - 	&& ~~~2.3	&&- 	&&156.8 	&& -\\
$bb\ell\ell$	&& - 	&& ~394.9 	&&- 	&& -	&& - \\
$bb\tau\tau$	&& ~175.1  	&&- 	&&- 	&&- 	&&- \\
\hline
\end{tabular}
}
\end{table}

\begin{figure}[th]
\centering
{\includegraphics[width=0.49\linewidth]{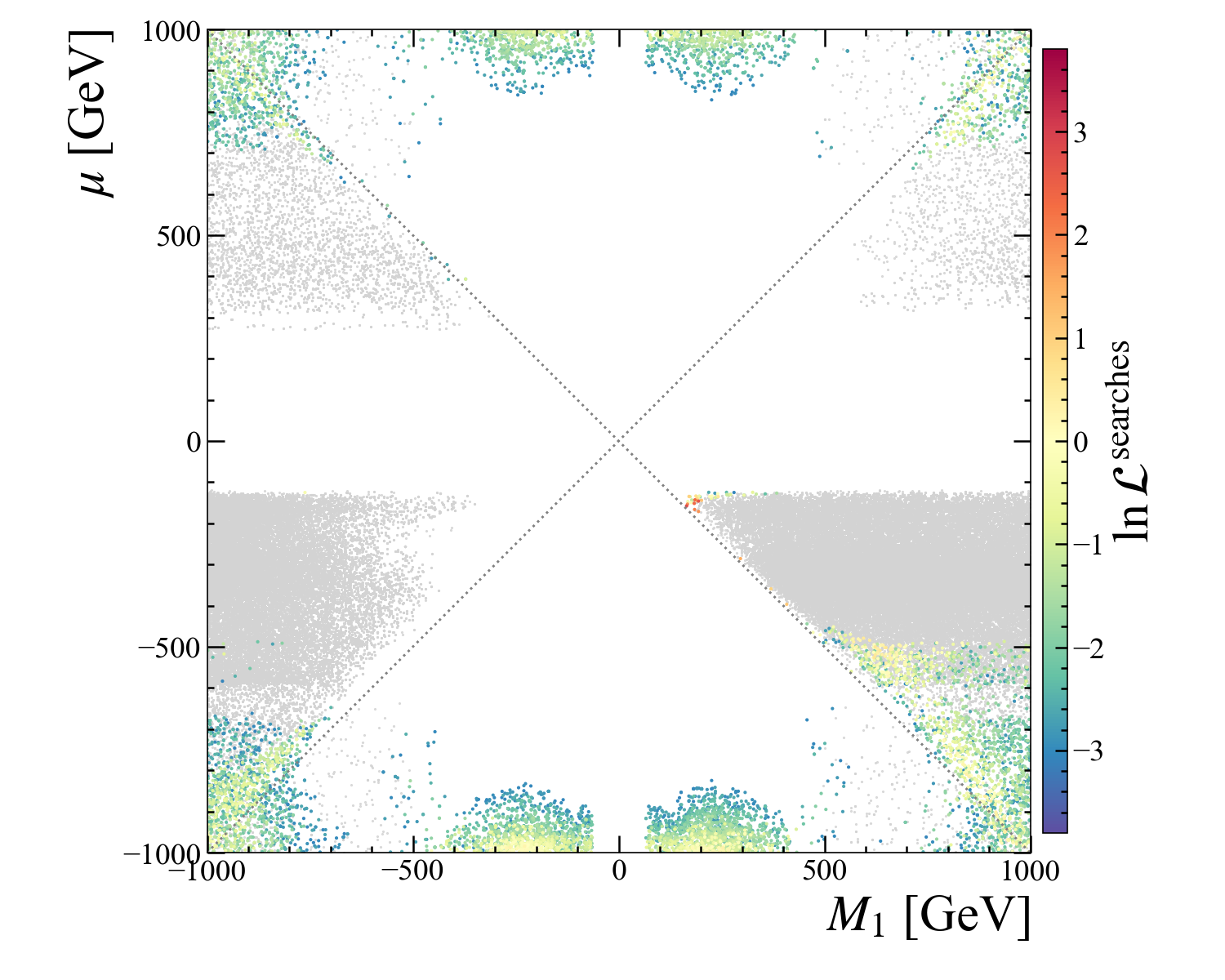}}
{\includegraphics[width=0.49\linewidth]{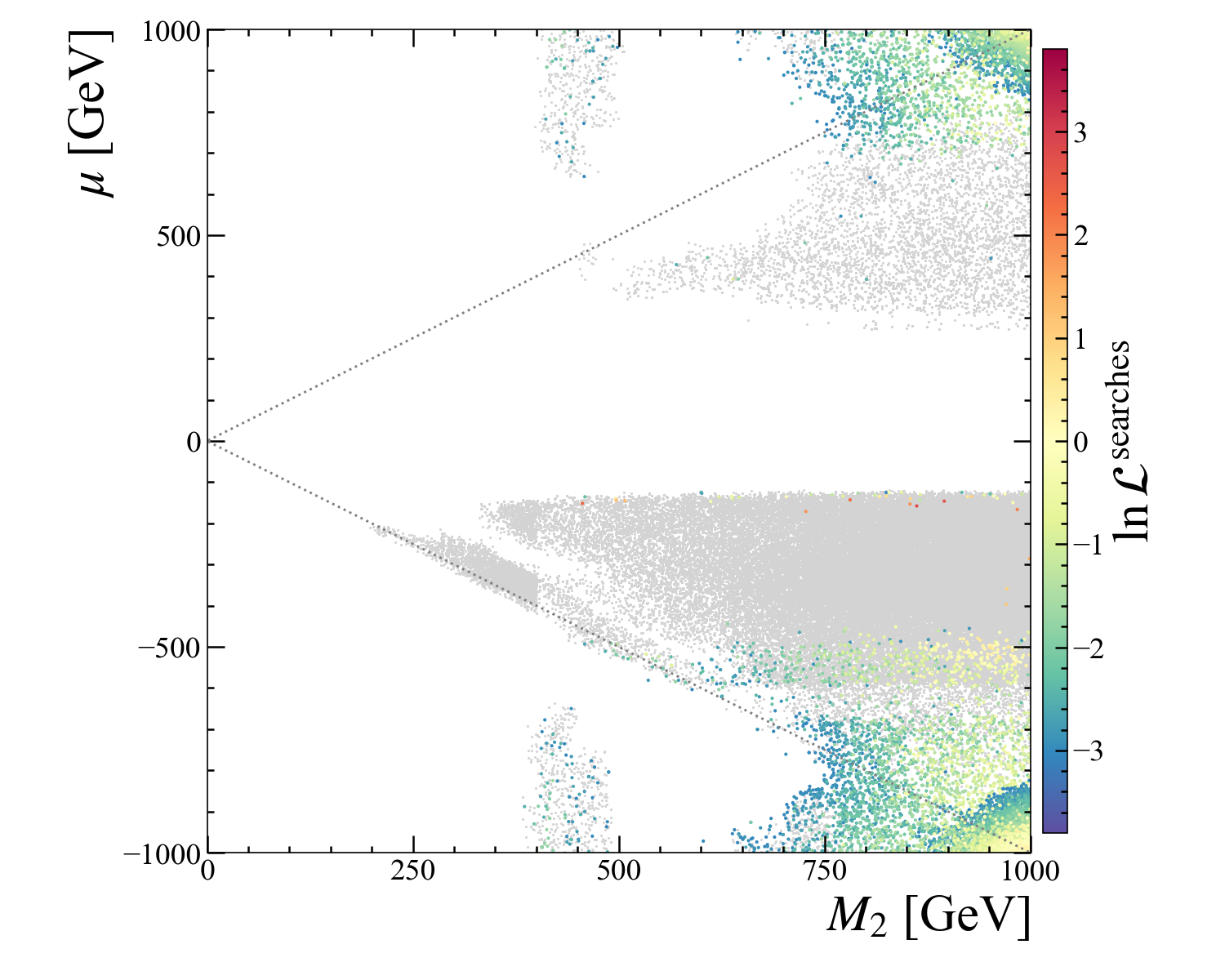}}
\caption{\label{fig:gewmssm} Same as Fig.~\ref{fig:GBfit-paras}, but the gray samples (containing almost all red samples with large likelihood in Fig.\ref{fig:gewmssm}) indicate they can be covered at the HL-LHC.}
\end{figure}

Finally, for the \textsf{GAMBIT} global fit result with the current LHC data for the $\tilde{G}$-EWMSSM, we reanalyse the samples to figure out their accessibility at the HL-LHC. As shown in Fig.~\ref{fig:gewmssm}, our search strategies can greatly probe the parameter space which survives the current LHC constraints. Comparing to Fig.~\ref{fig:GBfit-paras}, one can see that almost the whole parameter space with $|\mu| < 500~{\rm GeV}$ will be covered by the HL-LHC. The fine-tuning measure $\Delta_{\rm EW}$, defined in Eq.~(\ref{eq:intro-natrual}), is required to be larger than 100, which indicates a fine-tuning worse than 1\%. There are still a small group of samples with $-\mu \lesssim M_{1} \simeq 125~{\rm GeV}$ that are not accessible at the HL-LHC. As shown in Eq.~(\ref{eq:dkN2Hgra}) and Eq.~(\ref{eq:dkN2Zgra}), the decay widths of $\tilde{\chi}_1^0$ into $Z/h$ are greatly suppressed by a factor of $\left(1- {m_{h/Z}^2}/{m_{\tilde{\chi}_i^0}^2} \right)^4$. Other surviving samples are featured with a bino-dominated or wino-dominated $\tilde{\chi}_1^0$, and its dominant decay mode is $\tilde{\chi}_1^0 \to \gamma + \tilde{G}$, which corresponds to a decay branch ratio greater than $80\%$. Therefore, the remaining samples can be probed via the process in final states containing $2\gamma + E_{\rm T}^{\rm miss}$ at the HL-LHC.

\section{\label{sec:5}Summary}
In this paper, we studied the HL-LHC sensitivity to the production of higgsinos in the framework of the  EWMSSM assuming the LSP to be the ultra-light gravitino ($\tilde{G}$). We considered two simplified models: 
\begin{itemize}
\item In the first one, both the charged higgsinos $\tilde{\chi}^\pm_1$ and the second neutral higgsino $\tilde{\chi}^0_2$ decay to the lightest higgsino $\tilde{\chi}^0_1$ plus some invisibly soft particles. In this case we have an enhanced number of lightest higgsino pairs with each pair giving three decay channels $\tilde{\chi}^0_1 \tilde{\chi}^0_1 \to \tilde{G} \tilde{G} +ZZ/Zh/hh$. 
\item In the second model, the charged higgsino decays directly to a gravitino plus a W-boson and the second neutral higgsino decays directly to a gravitino plus a $Z/h$ boson. So the production of $\tilde{\chi}^\pm_1 \tilde{\chi}^0_2$ gives two decay channels $\tilde{\chi}^\pm_1 \tilde{\chi}^0_2\to \tilde{G} \tilde{G} + WZ/Wh$.
\end{itemize}
In each model the different channels are complementary and they were all jointly considered in our study. 
Through detailed Monte Carlo simulations for the signals and the standard model backgrounds, we found that the HL-LHC can probe the higgsinos up to about 600 GeV.
Therefore, for such a MSSM with gravitino LSP, the HL-LHC can either discover a light higgsino or 
exclude the natural SUSY scenario in case of gravitino LSP. However, we should note that for the natural SUSY scenario with higgsino LSP, a hadron collider is rather weak for exploring higgsinos, and, instead, we need a lepton collider which allows for a sensitive probe for compressed higgsinos \cite{Baer:2014yta,Yang:2022qga}. 

\par We also reinterpreted our result with $\tilde{G}$-EWMSSM global-fit data set. We found that the HL-LHC can probe almost the whole parameter space with $|\mu| \leq 500~{\rm GeV}$, and push up the electroweak fine-tuning factor $\Delta_{\rm EW}$ to 100 in case of unobservation. 

\addcontentsline{toc}{section}{Acknowledgments}  
\section*{Acknowledgments}
We thank the GAMBIT Collaboration for making the dataset from their global fit study publicly available.
This work was supported by the National Natural Science Foundation of China (NSFC) under grant Nos. 11821505, 12075300, 1233500 and 12105248, by Peng-Huan-Wu Theoretical Physics Innovation Center (12047503), by the CAS Center for Excellence in Particle Physics (CCEPP), and by the Key Research Program of the Chinese Academy of Sciences, Grant NO. XDPB15. 

\section*{Note added:}
Our work has lasted over a year due to the complicated simulations and the pandemic. When we finished our simulations and were preparing this manuscript, the ATLAS collaboration reported their work ~\cite{ATLAS:2024tqe}, which is a full run-2 version of one of the analyses in GAMBIT.  They searched the HH channel requiring both Higgs bosons to decay to $b\bar{b}$ and used the BDT (Boosted Decision Trees) approach to enhance the LHC sensitivity, while our study used the traditional cut-flow method for the $bb\tau\tau$ final state and gave a relatively conservative sensitivity. 
Absolutely, one can expect better constraints from the HL-LHC when using some advanced machine-learning approaches, which leaves to our future work.

\addcontentsline{toc}{section}{References}
\bibliography{references.bib}
\bibliographystyle{CitationStyle}

\end{document}